\documentclass[%
 reprint,
 amsmath,amssymb,
 aps,
]{revtex4-2}

\usepackage{url}
\usepackage{graphicx}
\usepackage{dcolumn}
\usepackage{bm}
\usepackage{subcaption}
\usepackage{bigints}

\begin{document}

\preprint{APS/123-QED}

\title{Non-ideal stability analysis of differentially rotating plasmas with global curvature effects}

\author{Alexander Haywood}
 \email{ahaywood@princeton.edu}
\affiliation{%
 Department of Electrical and Computer Engineering, Princeton University, Princeton, New Jersey 08544}%
\author{Fatima Ebrahimi}
\affiliation{Princeton Plasma Physics Laboratory, Princeton University, Princeton, New Jersey 08543}%
\affiliation{Department of Astrophysical Sciences, Princeton University, Princeton, New Jersey 08544}%

\date{\today}

\begin{abstract}
The linear stability of global non-axisymmetric modes in differentially rotating, magnetized, non-ideal plasma is crucial for understanding turbulence and transport phenomena. We investigate the competition between the local Magneto-Rotational Instability (MRI) and the Magneto-Curvature Instability (MCI)--a distinct non-axisymmetric low-frequency curvature-driven global branch--by developing and applying a non-ideal global spectral method, validated against NIMROD code simulations, and an extended effective potential formalism. Our analysis reveals that the global, low-frequency MCI persists at low magnetic Reynolds numbers (Rm), whereas the localized, high-frequency MRI is stabilized by diffusive broadening of its structure around its Alfvénic resonances. Consequently, we identify the global MCI as the primary onset mechanism for magnetohydrodynamic instability in systems with finite curvature, e.g., astrophysical rotators. We establish distinct parameter regimes for mode dominance: MCI prevails in geometrically moderate-thickness disks with high curvature and intermediate radial gaps, while MRI dominates in thin, low-curvature disks with large radial gaps. Mode competition is also highly sensitive to the flow profile, particularly vorticity and its gradient, with non-uniform shear profiles exhibiting more robust instability due to flow-curvature and shear contributions. A key outcome is the development of ``spectral diagrams" derived from the global spectral method. These diagrams comprehensively map dominant instabilities and their characteristics, offering a predictive tool for critical onset parameters (i.e., flow curvature, magnetic field, and Rm) and facilitating the interpretation of experimental and simulation results. Notably, these diagrams demonstrate that the global MCI is generally the sole unstable mode at the initial onset of instability.
\end{abstract}

\maketitle
\section{Introduction}

Observations \citep{shakura_black_1973} and experiments have established that differentially rotating plasmas in the presence of magnetic fields exhibit turbulence-enhanced momentum transport. Compact objects, such as M87, are well theorized to exhibit large differential flow, global curvature, and strong magnetic fields \cite{salas_resolution_2024}. Such observations have emphasized the need to study the global drivers of turbulence, especially in the strong-field limit. 

Given a weak magnetic field, such systems are unstable to magnetohydrodynamic (MHD) perturbations separately from the traditional hydrodynamic (HD) instability limit-- defined as $q(r) =-\frac{r\Omega'}{\Omega} < 2$ for the 
rotation profile $\Omega(r)$. In the context of accretion flows, the magnetorotational instability (MRI) \cite{velikhov_stability_1959, chandrasekhar_hydrodynamic_1981, balbus_powerful_1991} has long been theorized to be the primary driver of this turbulence in hydrodynamically stable flows. In this paper, we extend the stability analysis of differentially rotating systems to account for global spatial and (magnetic and velocity) field curvature effects, which are particularly significant in the context of astrophysical rotators such as accretion disks ~\cite{asada_m87_2024} and stellar interiors, as well as for interpreting laboratory experiments ~\citep{goodman_magnetorotational_2002,sisan_experimental_2004,wei_numerical_2016,wang_identification_2022}

Both global axisymmetric and non-axisymmetric MRI have been investigated thoroughly locally ~\citep{hawley_local_1995,stone_three-dimensional_1996, fromang_mhd_2007, pessah_angular_2007, bodo_aspect_2008}, globally ~\citep{machida_global_2000, hawley_global_2000, goodman_magnetorotational_2002}, and verified experimentally ~\citep{goodman_magnetorotational_2002,sisan_experimental_2004,wei_numerical_2016,wang_identification_2022, mishra_helical_2022}. Recent studies \citep{ebrahimi_nonlocal_2022,wang_identification_2022, wang_observation_2024} have emphasized the importance of studying non-axisymmetric perturbations, with simulations uncovering the onset of a low-frequency global non-axisymmetric mode occurring at magnetic Reynolds number lower than expected for ideal Taylor-Couette flows \citep{wang_observation_2024}. A global mode with similar characteristics to that observed was recently uncovered in the ideal limit, the magneto-curvature instability (MCI), and was shown to be driven by both spatial and (flow and magnetic) mean-field curvature \cite{ebrahimi_nonlocal_2022, ebrahimi_generalized_2025}. Non-axisymmetric global modes were shown to be mostly confined between Alfvénic resonant points (where the Doppler shifted mode frequency is equal to the Alfvén frequency) ~\citep{matsumoto_magnetic_1995, ogilvie_non-axisymmetric_1996}, and could persist at stronger magnetic fields (where axisymmetric MRI modes are stable) \cite{ebrahimi_nonlocal_2022, ebrahimi_generalized_2025}.  We will demonstrate that the global structure of this mode makes it persistent upon introducing diffusive effects and argue that non-axisymmetric MHD onset will generally occur at this mode.

This study investigates the stability of rotating plasma disks containing vertical or azimuthal magnetic fields. We employ a combination of techniques: local (WKB) analysis, global spectral analysis, and direct linear initial-value simulations. Our investigation yields the following key findings:
\begin{enumerate}
    \item
    The Magneto-Curvature Instability (MCI) and Magneto-Rotational Instability (MRI) retain distinct modal characteristics, each exhibiting unique mode structure and frequency, under identical magnetic field strength ($B_0 = V_A/V_0$), wave numbers (m, k), domain, and flow configurations. Although they could partially overlap over some parameter range, each mode defines a separate instability boundary.
    \item MRI's localized structure is selectively damped upon introducing diffusive effects. Consequently, the general onset of instability (at critical Rm)  is expected to occur at the MCI mode. Using \textit{spectral diagrams}, we also show that at finite Rm the system loses its time-reversal symmetry. 
    \item Larger vorticity (and thus shear) will generally lead to the onset of instability at a lower magnetic Reynolds number (Rm), as it destabilizes both MRI and MCI. Vorticity gradients have a more complicated relationship with both modes, and damping/destabilization intrinsically depends on mode frequency and field strength. Flow curvature (vorticity gradient and non-uniform q(r)) leads to increased instability of the MCI modes.   
    \item Regimes with either a turning point in the vorticity profile or a sufficiently deep vorticity well exhibit hydrodynamic (HD) instability, and at finite field, these profiles exhibit hybrid MCI-HD modes, as the MCI mode bifurcates into an HD branch at zero field. These MCI-HD hybrid modes are resonance localized at low magnetic fields. The onset of MHD instability in these regimes can be viewed as a bifurcation process of the Alfvénic resonances about the Corotation point in the flow.
    \item MCI dominates in disks with moderate relative thickness ($\Delta z /\Delta r \sim 1$), large spatial curvature $(r_2/r_1\rightarrow 1$), and moderate aspect ratio (radial gap) $r_1/(r_2-r_1) \sim 1$. Similarly, MRI dominates in thin disks ($\Delta z/\Delta r \ll 1$), low spatial curvature ($r_2/r_1\rightarrow \infty$), and small aspect ratio (large radial gap) $(r_1/(r_2-r_1) \rightarrow 0)$.
\end{enumerate}

We extend the effective potential formalism and conclude that introducing (fluid and magnetic) diffusive effects causes all modes to become broader with respect to their Alfvénic resonances. This, in turn, leads to an increased stabilizing effect for resonance-localized modes, such as MRI and low-field MCI, unlike global MCI, which exhibits only a single resonance in the domain. We utilize this observation to conclude that the general onset of MHD instability will occur at the global MCI mode and introduce a method to provide a coarse onset boundary and set of critical onset parameters. In addition, this method provides coarse boundaries of where each unstable mode will be dominant. These boundaries offer rough, experimental domains of where to search for MRI modes and predict that the onset of MRI can be observed as a discontinuous transition in the mode's frequency from low to high as the Magnetic Reynolds number (Rm) is increased. 

The paper is structured as follows: Section \ref{sec:model} describes the physical model and governing equations we utilize. Section \ref{sec:local} details our first method, a local WKB approximation, which we utilize to obtain a non-ideal dispersion relation describing instability growth rates; we use this method to explore instability characteristics under perturbation of the domain, flow configuration, and the introduction of diffusive effects. Section \ref{sec:methods} introduces our global stability analysis method, in which we derive and solve an ordinary differential equation (ODE) that captures the large-scale structure of the instabilities across the disk. This ODE is an extension of the global method derived in \cite{ebrahimi_nonlocal_2022},  which we extend by employing an approximation on the diffusive contributions' structure and validating findings against simulations. We utilize this extended ODE to form an effective potential formalism for the confinement of ideal and non-ideal modes and explore how confining potentials vary with diffusive contributions. 

Section \ref{sec:vf-res} presents results from direct initial-value NIMROD simulations for different magnetic field configurations and compares them with solutions to the global ODE to verify our methodologies. Here, we introduce ``spectral diagrams" that map out the instability domain for every unstable mode and provide the tools to classify the conditions of MHD onset coarsely. We then examine how the variation of flow profile (specifically, the shear parameter $q(r)$, vorticity, and its gradient) affects MRI and MCI, demonstrating that both remain MHD unstable. We also discuss the conditions under which purely hydrodynamic instability arises and how it interacts with the MCI. We establish that in hydrodynamically unstable configurations, the MCI bifurcates into a Corotation localized hydrodynamic mode in the absence of a magnetic field. We similarly demonstrate that hydrodynamic instability can occur for $q(r) < 2$ given sufficiently deep vorticity wells and show that these configurations also exhibit resonance-localized MCI modes. Section \ref{sec:AR} then explores the impact of changing the disk's overall curvature, shape (aspect ratio), and thickness, confirming that current experimental setups likely operate in a regime where the MCI is dominant. Finally, Section \ref{sec:conc} summarizes our key findings and suggests potential experimental strategies for isolating and observing these different types of instabilities. 

\begin{figure}
    \centering
\includegraphics[width=.8\linewidth]{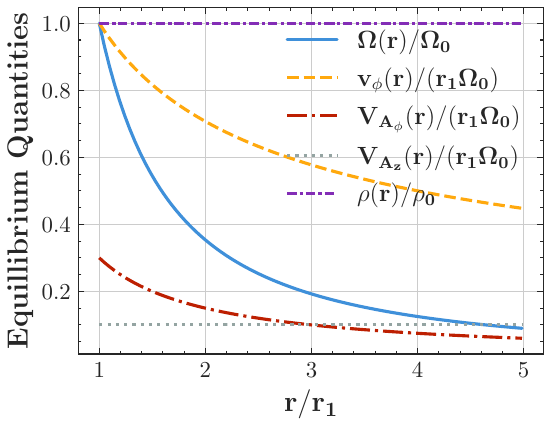}
    \caption{Variation of normalized equilibrium quantities with radius for an unstratified, currentless Keplerian disk. The magnetic field orientation is chosen such that $ V_{A_\phi}(r=0)/(r_1\Omega_0) = 0.3$ and $ \left. V_{A_z} \right(r=0)/(r_1\Omega_0) = 0.1$, which is consistent with past analysis \cite{ebrahimi_nonlocal_2022}.}
    \label{fig:Dimlessqtys}
\end{figure}

\section{Model}  \label{sec:model}

This study focuses on modeling the dynamics of cylindrical plasma disks threaded with vertical or azimuthal magnetic fields. Our investigation utilizes three main approaches: local WKB analysis, global spectral analysis, and linear initial-value simulations. All methods are based on the incompressible, resistive Magnetohydrodynamics (MHD) equations,

\begin{equation}\label{eqn:mom}
    \rho\left(\frac{\partial }{\partial t} + \mathbf{v} \cdot \nabla \right)\mathbf{v} = \frac{(\mathbf{B}\cdot\nabla)\mathbf{B}}{\mu_0} - \nabla\left(p + \frac{B^2}{2\mu_0}\right) + \rho\nu\nabla^2\mathbf{v}
\end{equation}
\begin{equation}\label{eqn:ind}
    \frac{\partial \mathbf{B}}{\partial t} =  \nabla \times (\mathbf{v} \times \mathbf{B}) - \eta \nabla^2\mathbf{B}
\end{equation}
where $\mathbf{v}, \;\mathbf{B}, \;p, \mu_0, \rho, \;\eta,$ and $\nu$ are the fluid velocity, magnetic field, pressure, magnetic permeability, density, magnetic, and viscous diffusivities, respectively. 

In this paper, we utilize three sets of models with increasing hierarchy. First, in Section \ref{sec:local}, we begin by performing a local WKB expansion on mean fields $\mathbf{v}_0 =r\Omega(r)\hat{\phi}$, $\mathbf{B}_0 = B_\phi\hat{\phi}+B_z\hat{z}$ and obtain a complete diffusive dispersion relation. 

Second,  we perform a global linear perturbative analysis and derive a second-order ordinary differential equation (ODE) governing the dynamics of the instabilities. We utilize shooting to obtain both eigenmodes and eigenvalues of the instabilities. We recast the ODE via an integrating factor transform to obtain an extended effective potential formalism \citep{ebrahimi_generalized_2025}, utilizing eigenvalues from shooting to obtain confining potentials. We explore scaling and structure of these modes/potentials with resistivity, flow, and domain configuration in Sections \ref{sec:vf-res}, \ref{sec:flow}, \ref{sec:AR} respectively. Our analyses consider ten explicit flow configurations and three aspect ratios. We perturb flow configurations to classify the scaling of MHD modes with different relative vorticity, vorticity gradients, and the presence of hydrodynamic instability. All linear simulations are ran with the Keplerian profile, however, we simulate all explicit aspect ratios. In these domains, the relative ``shape" of the disk is kept constant ($\Delta z/ \Delta r = 1/2$). This is in contrast to the short analysis done by \cite{ebrahimi_nonlocal_2022}, where aspect ratio was varied by moving the inner wall while keeping the $z$ boundaries fixed--changing the disk's relative ``thickness" ($\Delta z/\Delta r)$. 

Third, we perform direct linear initial-value simulations using the NIMROD code (Non-Ideal Magnetohydrodynamics with Rotation, an Open Discussion project) \citep{sovinec_nonlinear_2004}. NIMROD solves the time-dependent MHD equations numerically using a high-order finite element mesh for the poloidal plane ($r$-$z$). Meanwhile, it is pseudo-spectral in the periodic ($\phi$) direction with Fast Fourier Transforms (FFTs) (quantized by $m$). Although NIMROD can simulate complex non-linear physics, here we use it to model only the growth of the non-axisymmetric ($m = 1$) modes from small random initial noise. Simulations begin from an equilibrium state (e.g., Keplerian flow with uniform vertical or azimuthal field) and track the evolution of the single mode (m=1) non-axisymmetric perturbations. The radial boundaries are treated as solid, conducting walls, while the vertical and azimuthal directions are periodic. The numerical accuracy of these simulations is verified by varying mesh resolution and simulation time step. The parameters and geometries tested are detailed in Table \ref{table:AR0} and Figure \ref{fig:Ar-Config}. Comparing the simulation results with the global ODE methods allows us to validate our models and classify the stability in different configurations. However, understanding the combined effect of multiple coexisting modes on accretion would require future non-linear simulations.

\section{Local Stability Methods} \label{sec:local}

\begin{figure*}
    \centering
    \begin{minipage}{0.19\textwidth}
        \centering
        \includegraphics[width=\textwidth]{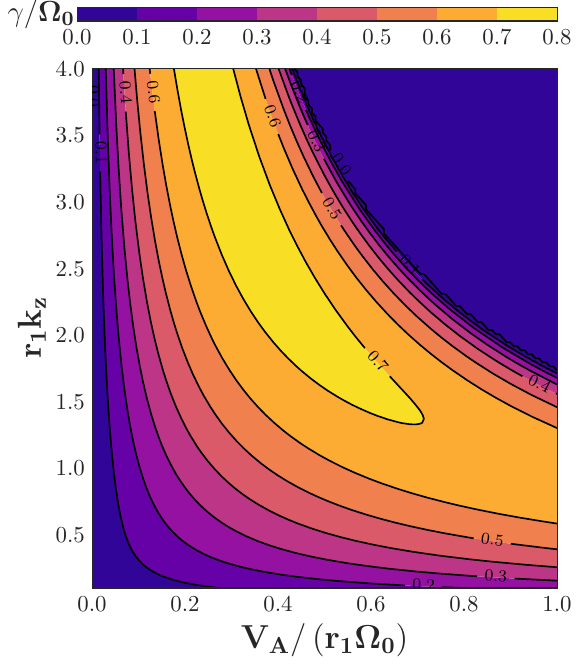}
        \subcaption{Ideal, $r_1k_r = 0$, $m = 1$, $\delta_c = 1$}  
    \end{minipage} \hfill
    \begin{minipage}{0.19\textwidth}
        \centering
        \includegraphics[width=\textwidth]{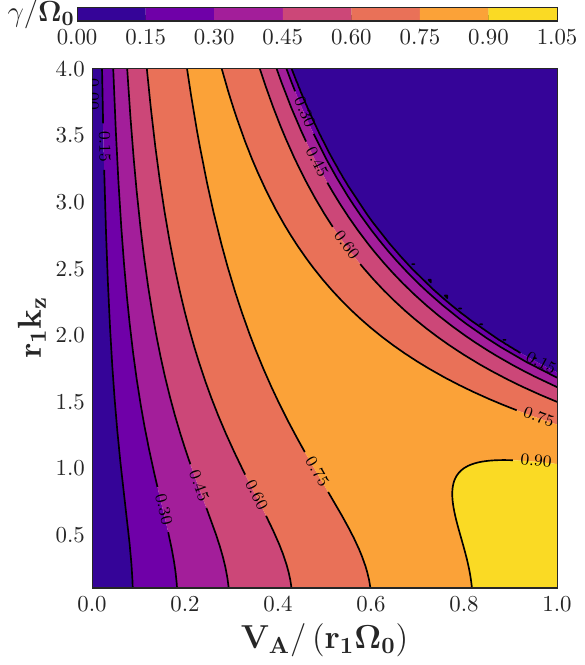}
        \subcaption{Ideal, $r_1k_r = 0$, $m = 1$, $\delta_c = 0$}  
    \end{minipage} \hfill
    \begin{minipage}{0.19\textwidth}
        \centering
        \includegraphics[width=\textwidth]{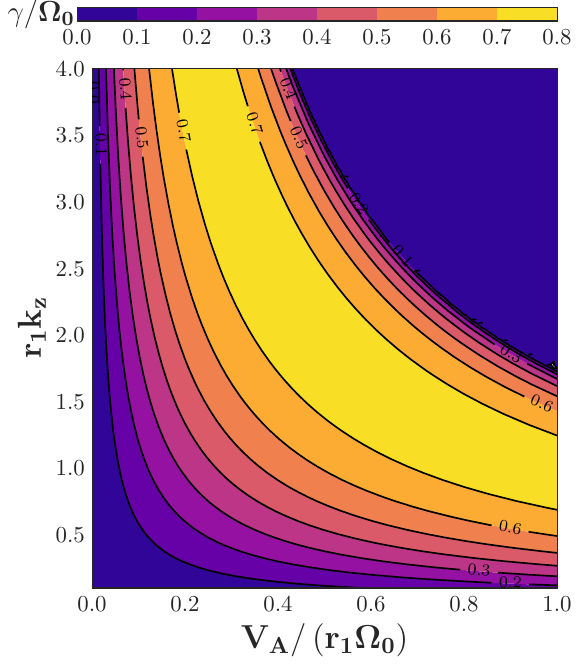}
        \subcaption{Ideal,  $r_1k_r = 0$, $m = 0$, $\delta_c = 0$}  
    \end{minipage} \hfill
    \begin{minipage}{0.19\textwidth}
        \centering
        \includegraphics[width=\textwidth]{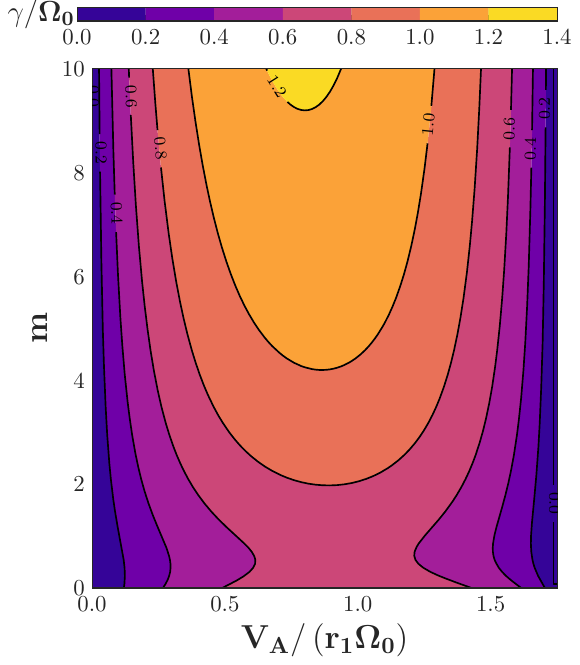}
        \subcaption{Ideal, $r_1k_r = 0$, $r_1k_z = 1$, $\delta_c = 1$}  
    \end{minipage} \hfill
    \begin{minipage}{0.19\textwidth}
        \centering
        \includegraphics[width=\textwidth]{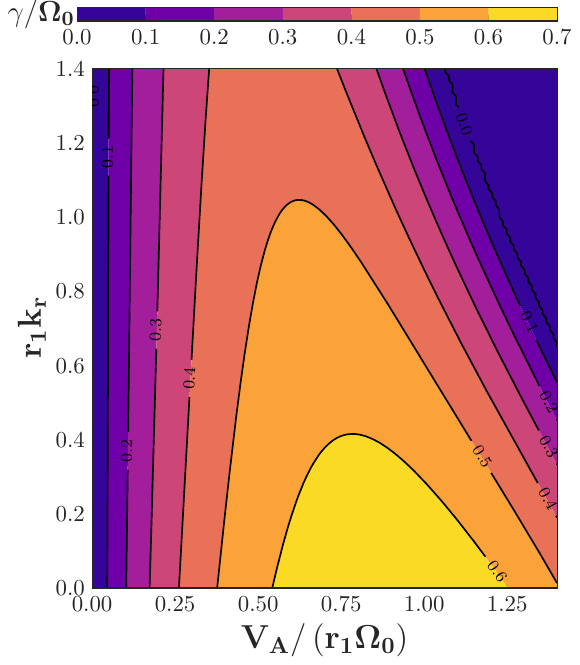}
        \subcaption{Ideal, $m = 1$, $r_1k_z = 1$, $\delta_c = 1$}  
    \end{minipage}

    \vspace{1em} 

    \begin{minipage}{0.19\textwidth}
        \centering
        \includegraphics[width=\textwidth]{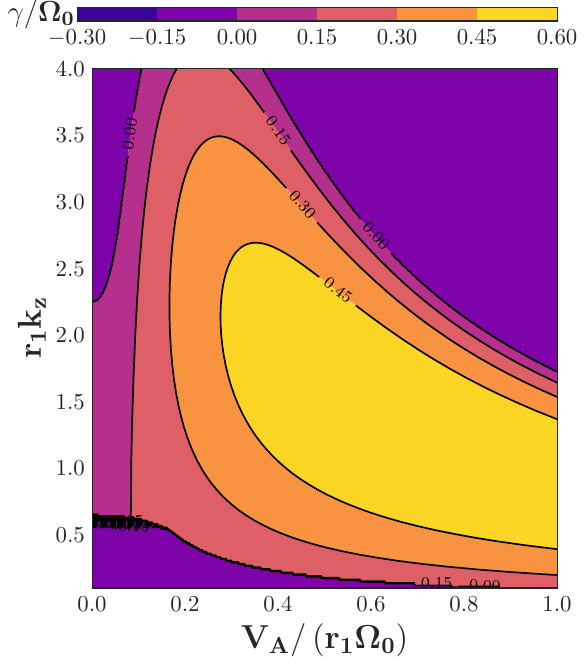}
        \subcaption{Non-ideal, $r_1k_r = 0$, $m = 1$, $\delta_c = 1$}  
    \end{minipage} \hfill
    \begin{minipage}{0.19\textwidth}
        \centering
        \includegraphics[width=\textwidth]{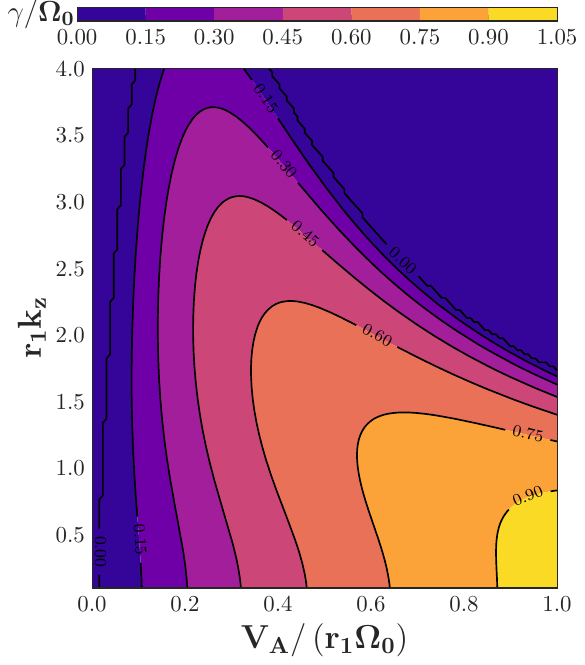}
        \subcaption{Non-ideal, $r_1k_r = 0$, $m = 1$, $\delta_c = 0$}  
    \end{minipage} \hfill
    \begin{minipage}{0.19\textwidth}
        \centering
        \includegraphics[width=\textwidth]{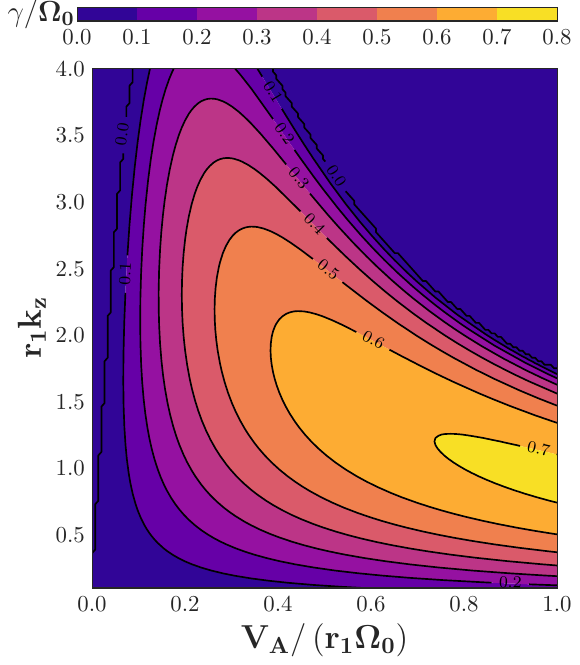}
        \subcaption{Non-ideal,  $r_1k_r = 0$, $m = 0$, $\delta_c = 0$}  
    \end{minipage} \hfill
    \begin{minipage}{0.19\textwidth}
        \centering
        \includegraphics[width=\textwidth]{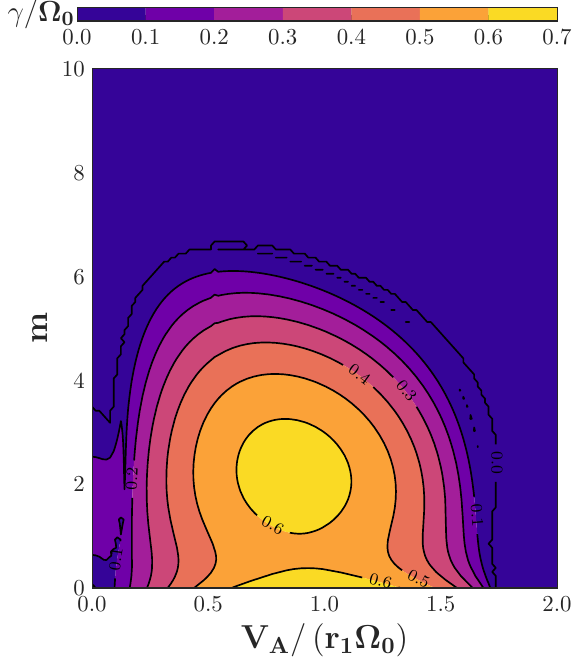}
        \subcaption{Non-ideal, $r_1k_r = 0$, $r_1k_z = 1$, $\delta_c = 1$}  
    \end{minipage} \hfill
    \begin{minipage}{0.19\textwidth}
        \centering
        \includegraphics[width=\textwidth]{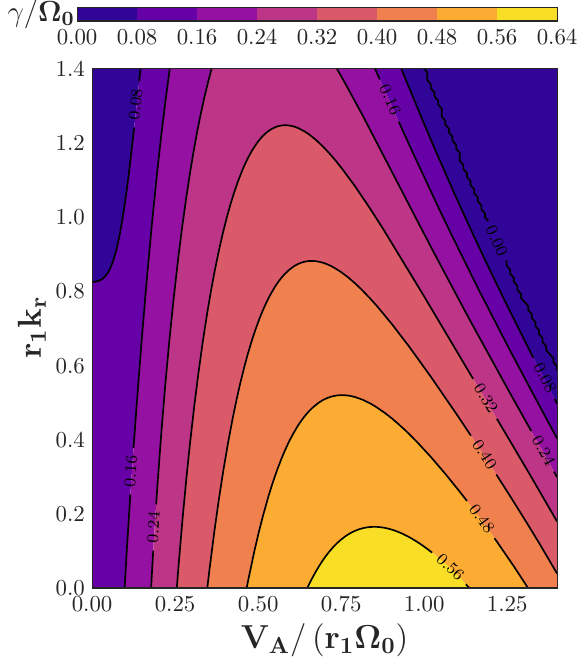}
        \subcaption{Non-ideal, $m = 1$, $r_1k_z = 1$, $\delta_c = 1$}  
    \end{minipage}

         \caption{Vertical field Keplerian WKB method growth rates from local dispersion relation (Eq. \ref{eqn:local-disp}). (a), (b), (c), (d), and (e) are ideal dispersion relations ($\mathrm{Pm} = 0, \; \mathrm{Rm} = \infty$). Meanwhile, (f), (g), (h), (i), and (j) are non-ideal with $\mathrm{Pm} = 1$ and $\mathrm{Rm} = 31$. (a), (b), (d), (f), (g), and (h) consider non-axisymmetric ($m = 1$) modes, while (c), (h) consider axisymmetric ($m = 0$) modes and (d), (i) explore variation with $m$. For (d), (e), (i), and (j), $r_1k_z = 1$. Meanwhile, for (a), (b), (c), (e), (f), (g), (h), and (I), we have set $r_1k_r = 0$. Moreover, (a), (f), have finite spatial curvature ($\delta_c = 1$), while (b), (c), (g), and (h) have zero spatial curvature ($\delta_c = 0$). (a) reproduces the ideal non-axisymmetric finite spatial curvature dispersion relation and (c) reproduces the ideal Cartesian limit dispersion relation depicted by \cite{ebrahimi_nonlocal_2022}. Irrespective of spatial curvature, it is evident that increased non-idealities lead to the need to study low $k_z$ global (low $k_r$) non-axisymmetric modes ($m \neq 0$). Finally, all radial quantities are evaluated at the inner boundary.}  \label{fig:WKB-Kep} 
\end{figure*}

\begin{figure*}
    \centering
    \begin{minipage}{0.24\textwidth}
        \centering
        \includegraphics[width=\textwidth]{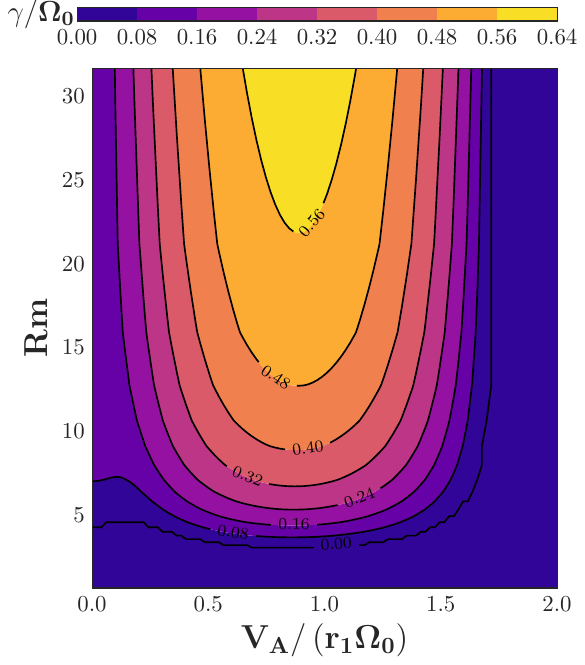}
        \subcaption{Kep ($q = 3/2$), $r_1k_r = 0$, $m = 1$, $r_1k_z = 1$, $\delta_c = 1$}  
    \end{minipage} \hfill
    \begin{minipage}{0.24\textwidth}
        \centering
        \includegraphics[width=\textwidth]{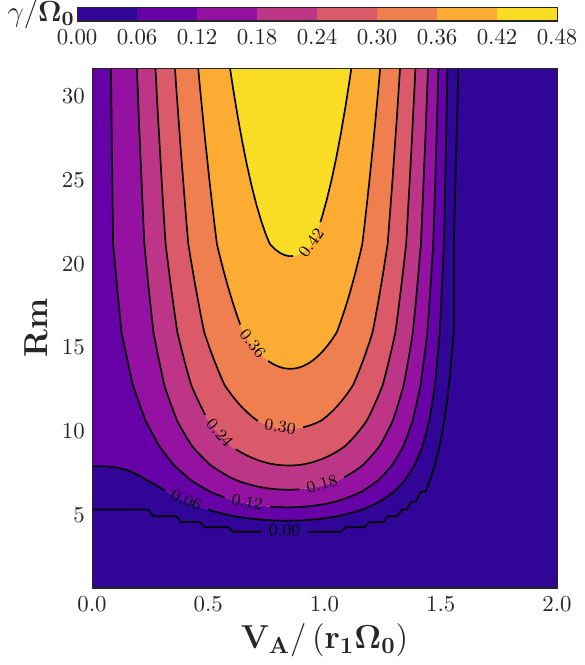}
        \subcaption{$q = 5/4$, $r_1k_r = 0$, $m = 1$, $r_1k_z = 1$, $\delta_c = 1$}  
    \end{minipage} \hfill
    \begin{minipage}{0.24\textwidth}
        \centering
        \includegraphics[width=\textwidth]{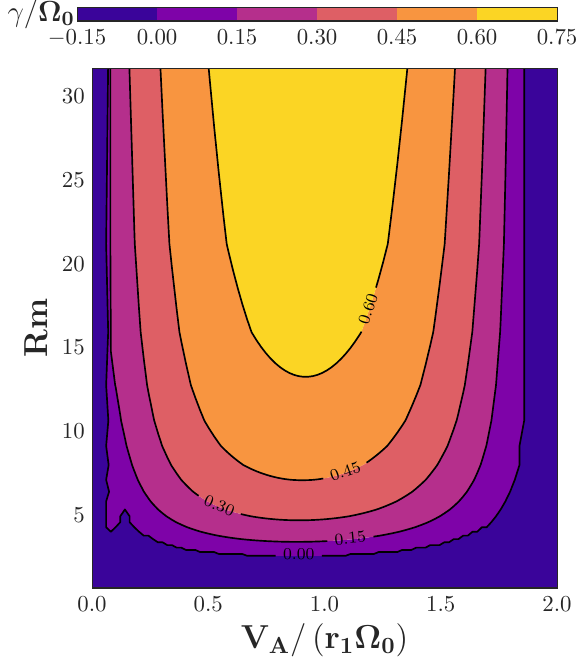}
        \subcaption{$q = 7/4$, $r_1k_r = 0$, $m = 1$, $r_1k_z = 1$, $\delta_c = 1$}  
    \end{minipage} \hfill
    \begin{minipage}{0.24\textwidth}
        \centering
        \includegraphics[width=\textwidth]{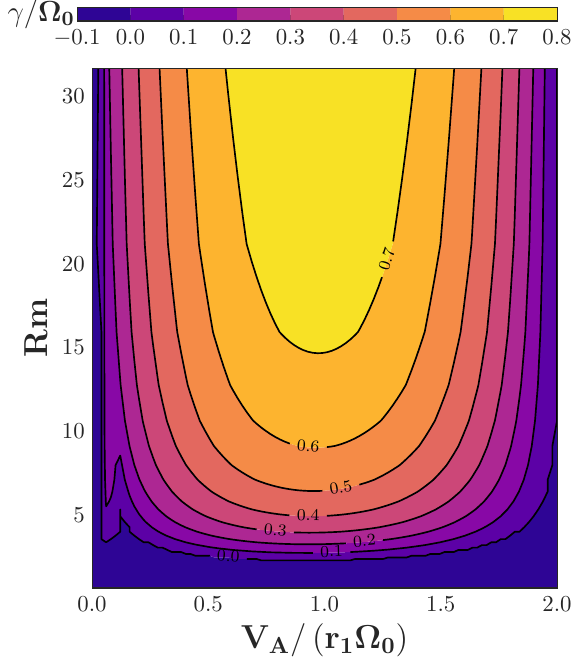}
        \subcaption{$q = 2$, $r_1k_r = 0$, $m = 1$, $r_1k_z = 1$, $\delta_c = 1$}  
    \end{minipage} \hfill

         \caption{Vertical field WKB method growth rates from local dispersion relation (Eq. \ref{eqn:local-disp}) for Kep. and power law profiles $\Omega(r)/\Omega_0 = 1/r^q$ for $q = 5/4$ (b), $q = 7/4$ (c), and $q = 2$ (d) for $\mathrm{Pm} = 1$. All plots are studying the non-axisymmetric ($m = 1$) modes with $r_1k_r = 0$, $r_1k_z = 1$, and finite curvature $\delta_c = 1$. Flow configuration changes the domain over which the modes are unstable at finite Rm and the critical Rm and magnetic field for the onset of instability. All radial quantities were evaluated at the inner boundary.}  \label{fig:WKB-DiffFlow-Rm} 
\end{figure*}

We begin by employing a WKB approximation to study the local dynamics of linear perturbations about mean fields  $\mathbf{v}_0 = r\Omega(r)\hat{\phi}$ and $\mathbf{B}_0 = B_\phi(r)\hat{\phi} + B_z\hat{z}$. We apply this approximation to the linearized MHD equations (Eqns. \ref{eqn:mom}, \ref{eqn:ind}) assuming perturbations take a wave-like form $\pmb{\xi} = \pmb{\xi}_0 e^{i(\mathbf{k} \cdot \mathbf{r}  - \omega t)}$. Here $\mathbf{k}$ is the wave-vector with components $k_r$, $k_\phi =m/r$, and $k_z$ representing variations in the radial, azimuthal, and vertical directions, respectively (with $m$ being the azimuthal mode number). Meanwhile, the complex frequency $\omega = \omega_r + i\gamma$ contains the mode's frequency ($\omega_r$) and growth rate ($\gamma$). By further assuming the plasma is incompressible, these perturbed quantities can be represented in the matrix equation $\mathbf{M} \pmb{\chi} = 0$ (Eq. \ref{eqn:local-disp-matrix}), where $\pmb{\chi}$ contains the perturbed velocity and magnetic field components.

\begin{widetext}
\begin{equation} \label{eqn:local-disp-matrix}
    \pmb{\mathrm{M}} \boldsymbol{\chi} = 
\begin{bmatrix}
i\bar{\omega} -\eta\bar{k}^2 & -\frac{2ik_\phi\eta\delta_c}{r} & iF & 0 \\
\frac{\partial \Omega}{\partial \mathrm{ln}r} + \frac{2ik_\phi\eta\delta_c}{r} & i\bar{\omega} -\eta \bar{k}^2 & \frac{2B_\phi}{r}\delta_c & iF \\
\frac{iF}{\mu_0 \rho}\left(1 - \frac{\bar{k}_rk_r}{k_z^2} \right) & \frac{1}{\mu_0\rho} \left(iF \frac{k_rk_\phi}{k_z^2} -\frac{B_\phi}{r}\delta_c \right)& i\bar{\omega}\left( 1 - \frac{\bar{k}_rk_r}{k_z^2}\right) - \nu\bar{k}^2 & 2\Omega + \frac{i\bar{\omega}k_rk_\phi}{k_z^2}-\frac{2ik_\phi\nu\delta_c}{r}\\
 \frac{iF}{\mu_0 \rho}\frac{\bar{k}_rk_\phi}{k_z^2} & \frac{iF}{\mu_0 \rho} \left(1 + \frac{k_\phi^2}{k_z^2}\right) & i\bar{\omega}\frac{\bar{k}_rk_\phi}{k_z^2} - \frac{\kappa^2}{2\Omega} + \frac{2ik_\phi\nu\delta_c }{r} & i\bar{\omega}\left( 1 + \frac{k_\phi^2}{k_z^2}\right) - \nu\bar{k}^2
\end{bmatrix} \begin{bmatrix} \tilde{B}_r \\ \tilde{B}_\phi \\ \tilde{V}_r \\ \tilde{V}_\phi \end{bmatrix} = 0 
\end{equation}
\end{widetext}

where $F = \mathbf{k}\cdot\mathbf{B}_0 = k_\phi B_\phi + k_zB_z$, $k_\phi = m/r$. Meanwhile, $\bar{\omega} = \omega - m\Omega$ and $\kappa^2 = 4\Omega^2 + \frac{\mathrm{d}\Omega^2}{\mathrm{dln}r}$ are the Doppler shifted and epicyclic frequencies respectively. Meanwhile, we define $k^2 = \bar{k}_rk_{r}+k_\phi^2+k_z^2$, with $\bar{k}_r = k_r - \frac{i\delta_c}{r}$ and $\bar{k}^2 = k^2 + \frac{\delta_c}{r^2}$, with $\delta_c$ arising due to the cylindrical geometry's finite curvature. Finally, the current-free condition mandates that $\frac{\partial B_\phi}{\partial r} = -B_\phi/r$. 

The requirement for a non-trivial solution (i.e., for an instability to exist) is that the determinant of the matrix $\mathbf{M}$ must be zero. This condition yields a fourth-order polynomial equation (Eq. \ref{eqn:local-disp}) known as the local dispersion relation.
\begin{equation}\label{eqn:local-disp}
    \bar{\omega}^4 + C_3\bar{\omega}^3 + C_2 \bar{\omega}^2 + C_1 \bar{\omega} + C_0 = 0
\end{equation}
with coefficients $C_3, C_2, C_1,$ and $C_0$ defined as,
\begin{widetext}
\begin{equation}
    C_3 = 2i\Omega\frac{\bar{k}_rk_\phi}{k^2} - \frac{i\kappa^2}{2\Omega }\frac{k_rk_\phi}{k^2} + \frac{i\bar{k}^2}{k^2}\Biggl[\left(2\eta + \nu\right)k^2 + \nu k_z^2\Biggr] +\frac{2\nu k_\phi^2\delta_c}{r^2k^2}
\end{equation}
\begin{equation}
\begin{split}
    C_2 = \frac{i\bar{k}_rk_\phi}{2k^2}\omega_A\omega_c\delta_c - \frac{1}{2}\left(\frac{k_z^2+k_\phi^2}{k^2}\right)\omega_c^2\delta_c- \kappa^2 - 2\omega_A^2  +\frac{i k_z^2k_\phi\nu\delta_c}{rk^2}\left(\frac{\kappa^2}{\Omega} + 4\Omega\right) - \frac{\bar{k}^2k_\phi\eta}{k^2}\left(4\Omega\bar{k}_r-\frac{\kappa^2}{\Omega}k_r\right) + \frac{2ik_\phi\eta\delta_c}{r}\frac{\partial \Omega}{\partial\mathrm{ln}r} \\
    +\left(\frac{4k_\phi^2\delta_c}{r^2}-\bar{k^4}\right)\left(\frac{k^2\eta^2+k_z^2\nu^2}{k^2}\right)-\frac{2\eta\nu}{k^2}\Biggl[\bar{k}^4\big\{k^2+k_z^2\bigr\} -\frac{2i\bar{k}^2k_\phi^2\delta_c}{r^2}\Biggr]
\end{split}
\end{equation}
\begin{equation}
\begin{split}
        C_1 =  \frac{1}{k^2}\Biggl[\frac{ik_rk_\phi\kappa^2}{2\Omega} -2i\bar{k}_rk_\phi-\frac{2\nu k_\phi^2\delta_c}{r^2}-i\bar{k}^2\biggl\{ \left(2\eta + \nu\right)\left(\bar{k}_rk_r+k_\phi^2\right) + 2\left(\eta + \nu \right)k_z^2 \biggr\} \Biggr]\omega_A^2 \\
        + \frac{1}{k^2}\Biggl[\frac{1}{2}\biggl\{ \frac{\partial\Omega}{\partial\mathrm{ln}r}-4\Omega\biggr\}\bigl\{k_z^2+k_\phi^2\bigr\} -k_z^2\frac{\kappa^2}{4\Omega} + \frac{ik_\phi}{r}\biggl\{ 2\eta\bar{k}_rk_r + \bigl(2\nu + 3\eta\bigr)k_\phi^2 + 3\bigl( \nu + \eta\bigr )k_z^2 + ir\bar{k}^2\biggl( \frac{\eta\bar{k}_r}{2} + \nu k_r\biggr) \biggr\} \Biggr]\omega_A\omega_c\delta_c 
        \\ 
        -\frac{i\bar{k}^2}{2k^2}\Biggl[\eta k_\phi^2 + \bigl\{\eta+\nu)\bigr\}k_z^2 \Biggr]\omega_c^2\delta_c
        -\frac{i}{k^2}\Biggl[ 2\kappa^2\bar{k}^2k_z^2 + \biggl\{ 2\Omega\bar{k}_rk_\phi - \frac{\kappa^2}{2\Omega}k_rk_\phi\biggr\} \biggl\{ \bar{k}^4\eta + \frac{2ik_\phi\delta_c}{r}\left(\frac{\partial\Omega}{\partial\mathrm{ln}r} + \frac{2ik_\phi\eta}{r} \right) \biggr\} \Biggr] 
        \\
        + \frac{\eta\nu}{k^2}\Biggl[\frac{4k_\phi^3\delta_c}{r^3}\biggl\{ \frac{2\eta k_\phi}{r} - i\frac{\partial\Omega}{\partial\mathrm{ln}r} \biggr\} -\frac{2\eta\bar{k}^4k_\phi^2\delta_c}{r^2}
        -i\bar{k}^6\biggl\{ \eta\bar{k}_rk_r+\eta k_\phi^2 + 2\left(\eta + \nu\right)k_\phi^2 \biggr\} 
        + \frac{2\bar{k}^2k_\phi\delta_c}{r}\biggl\{ \left(\bar{k}_rk_r + k_\phi^2\right) \left(\frac{\partial\Omega}{\partial\mathrm{ln}r} + \frac{2i\eta k_\phi}{r}\right) \\
        + 2k_z^2\left(\frac{\partial\Omega}{\partial\mathrm{ln}r} + \frac{2i \left(\eta + \nu \right)k_\phi}{r} - 2\Omega - \frac{\kappa^2}{2\Omega}  \right)\biggr\}\Biggr]
\end{split}
\end{equation}

\begin{equation}
\begin{split}
        C_0 = \Biggl[\omega_A^2 + \biggl\{ \frac{k_\phi^2+k_z^2}{k^2} \biggr\} \frac{\partial\Omega^2}{\partial\mathrm{ln}r} + \frac{\bar{k}^2}{k^2}\biggl\{2\eta\Omega\bar{k}_rk_\phi - \biggl(\eta\frac{\kappa^2}{2\Omega} -  \nu\frac{\partial\Omega}{\partial\mathrm{ln}r} \biggr)k_rk_\phi \biggr\} 
        \\ + \frac{ik_\phi\delta_c}{rk^2} \biggl\{ \eta\left(\bar{k}_rk_r+k_z^2\right)\frac{\kappa^2}{\Omega} + \left(4\eta\Omega - 2\nu\frac{\partial\Omega}{\partial\mathrm{ln}r} \right)\left(k_\phi^2 + k_z^2\right) \biggr\} + 
    \frac{\eta \nu}{k^2} \biggl\{ \left(k^2+k_z^2\right)\biggl(\bar{k}^4 + \frac{4k_\phi^2\delta_c}{r^2}\biggr)-\frac{4i\bar{k}^2k_\phi^2\delta_c}{r^2} \biggr\} \Biggr]\omega_A^2 \\
    + \Biggl[ \frac{4\eta\bar{k}_rk_\phi^2}{rk^2} + \frac{i\bar{k}^2}{2k^2} \biggl\{ \eta\frac{\kappa^2}{2\Omega} + 4\eta\Omega\left(k_\phi^2+k_z^2\right)  - \nu\frac{\partial\Omega}{\partial\mathrm{ln}r} \biggr\} + \frac{\eta \nu}{k^2}   \biggl\{i\bar{k}^4k_rk_\phi + \frac{2i\bar{k}^2k_\phi}{r}\left(k^2+2k_z^2\right)  - \frac{4i\bar{k}_rk_\phi^3}{r^2}
 \biggr\}
 \Biggr]\omega_A\omega_c\delta_c \\ 
 +  \frac{i\bar{k}_rk_\phi}{2k^2}\omega_A^3\omega_c\delta_c + \frac{\eta\nu \bar{k}^4}{2}
\omega_c^2 \delta_c
 + \frac{2i\eta k_\phi\delta_c \kappa^2}{r} \frac{\partial\Omega}{\partial\mathrm{ln}r} + \kappa^2\eta^2\Biggl[\bar{k}^4 - \frac{4k_\phi^2\delta_c}{r^2}\Biggr]
 + \eta\nu\Biggl[\eta\bar{k}^4 + \frac{2ik_\phi\delta_c}{r}\frac{\partial\Omega}{\partial\mathrm{ln}r} - \frac{4\eta k_\phi^2\delta_c}{r^2} \Biggr] \\ \times \Biggl[\nu\bar{k}^4-\frac{ik_\phi\delta_c}{r}\left\{\frac{\kappa^2}{\Omega} + 4\Omega\right\} + \frac{4\nu k_\phi^2\delta_c}{r^2}\Biggr]
\end{split}
\end{equation}
\end{widetext}

where $\omega_A = \frac{F}{\sqrt{\mu_0\rho}}$ and $\omega_c = \frac{2B_\phi}{r\sqrt{\mu_0\rho}}$. Upon substitution of $k_r \rightarrow -ik_r$, this matches the previously known local dispersion relation for non-axisymmetric modes \citep{ebrahimi_nonlocal_2022}. In the zero-curvature axisymmetric limit, $k_\phi, \delta_c \rightarrow 0$ and $\bar{k}_r\rightarrow k_r$, $\bar{k}^2\rightarrow k^2$, this becomes the local dispersion relation for the axisymmetric MRI mode \citep{balbus_powerful_1991}. 

In a real disk, $k_z$ and $m$ are quantized by the disk's boundaries. We take $k_r = 2\pi l/(r_2-r_1)$ and $k_z = 2\pi n/(z_2-z_1)$ where $n,l$ are positive integers. To compare with later results, we use the inner radius ($r_1$) and the rotation frequency $(\Omega_0)$ to define dimensionless numbers: the magnetic Prandtl number $\mathrm{Pm} = \nu/\eta$, the magnetic Reynolds number $\mathrm{Rm} = r_1^2\Omega_0/\eta$, and the Lehnert number $B_0 = V_A/(r_1\Omega_0)$. All radially varying quantities ($\Omega, \omega_A,$ etc) are evaluated at the inner boundary for this local analysis. 

We first study the dispersion relation with a standard Keplerian profile ($\Omega(r) \propto 1/r^{3/2}$) and a purely vertical magnetic field, with results shown in Figure \ref{fig:WKB-Kep}.
In the ideal case (no diffusion, panels a-e), the dominant instabilities (largest $\gamma$) exhibit small vertical wavelengths (high $k_z$), large radial wavelengths (low $k_r$), and are strongly non-axisymmetric. However, in the presence of significant diffusion (high $\eta, \nu$, corresponding to low Rm given Pm = 1 in panels f-j), the dominant mode shifts towards large vertical and radial wavelengths (low $k_z$ and $k_r$) yet remains distinctly non-axisymmetric. Thus, introducing diffusivity leads to increased globality in the structure of the dominant instabilities. Consequently, motivated by recent simulations and experiments that observed such modes at low Rm, this paper will focus primarily on the $m = 1$ non-axisymmetric modes.

We also examined how the flow configuration affects instabilities by considering three additional power-law profiles ($\Omega(r) \propto 1/r^q$ with $q = 5/4, 7/4, 2$) shown in Figure \ref{fig:WKB-DiffFlow-Rm}. Across all profiles, low Rm dominant instabilities are characterized by large-scale global (low $k_r$, $k_z$) non-axisymmetric ($m = 1$) structure. Figure \ref{fig:WKB-DiffFlow-Rm} demonstrates explicitly how the growth rate depends on the magnetic field strength ($V_A$) and magnetic Reynolds number (Rm) for the most global modes ($m = 1$, $r_1k_z = 1$, $k_r = 0$) among different flow configurations. A key finding is that the conditions required for the onset of instability (the minimum field strength and Rm) strongly depend on the flow profile. Specifically, flow configurations with steeper rotation profiles (larger $q$) become unstable at lower Rm (cf. Figs. \ref{fig:WKB-DiffFlow-Rm}a,b,c,d) We will demonstrate a similar relationship in the global approach by studying the free energy contributions (see Sec. \ref{sec:vorticity}). We will demonstrate that $q(r)$ can similarly be thought of as the vorticity of the profile, and show that vorticity directly provides free energy to instabilities.

However, the WKB method has a significant limitation: it assumes that background configurations are uniform locally, whereas in a real disk, properties like the velocity ($v_\phi = r\Omega(r)$) and the magnetic field strength are intrinsically radially-dependent (as shown in Figure \ref{fig:Dimlessqtys}). This radial dependence makes the definition of a radial wave number ill-posed. Therefore, the WKB results should indicate scaling behavior rather than being quantitatively precise for the whole disk. To perform a more accurate and comprehensive analysis that accounts for these radial variations, we need to employ global methods that explicitly solve for the instability structure across the entire radial extent of the disk.

\section{Non-ideal global stability methods} \label{sec:methods}

Moving beyond the local WKB analysis, we develop a global model that accounts for radial variations across the disk. Our approach extends the ideal MHD framework developed by \cite{ebrahimi_nonlocal_2022} to include the effects of finite resistivity and viscosity (corresponding to finite and non-zero Rm and/or Pm). Core to this development is the derivation of a one-dimensional Ordinary Differential Equation (ODE) that governs the radial structure of linear instabilities. To keep the ODE tractable and non-coupled, we employ an approximation targeting only the diffusive contributions (the last two terms on the RHS of Eqns. \ref{eqn:mom}, \ref{eqn:ind}), an approach first proposed by \cite{zou_analysis_2020}. This method requires the definition of a pseudo-radial wave-vector for the diffusive contributions, and we propose two candidate approximations. The approximations we term the ``MWKB" (modified WKB-like) and ``TWKB" (traditional WKB-like) and test by comparing to direct numerical simulations using NIMROD (details in Sec. \ref{sec:qr}).

Once the extended ODE is established, we solve it numerically using the shooting method to find the instability properties: the eigenvalues ($\omega$), which represent the complex frequencies (frequency Re$(\omega) = \omega_r$ and growth rate Im$(\omega) = \gamma$), and the corresponding eigenfunctions, which describe the radial structure of the instability. This approximation global model allows us to:
\begin{enumerate}
    \item Identify unstable modes and map how different families (or ``branches") of solutions change as flow/field/disk configuration and Rm/Pm/$B_0$ are varied (Sec. \ref{sec:scan}).
    \item Define locations of Alfvénic resonances in the presence of diffusion -- points in the regime where the Doppler shifted frequency equals the Alfvén frequency. These resonances are critical to describing instabilities' mode structure and confinement mechanisms (Sec. \ref{sec:res-resonance}).
    \item Extend the effective potential formalism, previously used in the ideal MHD regime \citep{ebrahimi_generalized_2025}, to the non-ideal regime to analyze how diffusion affects the confinement and stability of non-axisymmetric modes (Sec. \ref{sec:res-pot}).
\end{enumerate}
Our global analysis method consider small linear perturbations about the same equilibrium state as before: a differentially rotating flow $\mathbf{v}_0 = r\Omega(r)\hat{\phi}$ and a magnetic field $\mathbf{B}_0 = B_\phi(r)\hat{\phi} + B_z\hat{z}$. Unlike the WKB method, we now allow the radial structure of perturbations to vary arbitrarily across the disk. We assume these perturbations are wave-like in the azimuthal ($\phi$) and vertical ($z$) directions, oscillating with complex frequency $\omega = \omega_r + i\gamma$. Thus, any perturbed quantity (like velocity $\tilde{\mathbf{v}}$, magnetic field $\tilde{\mathbf{B}}$ or pressure $\tilde{P}$) takes the form $a(r)\mathrm{exp}\left(i\left( m\phi + k_zz -\omega t\right) \right)$. The radial profile $a(r)$ is determined by the dynamics for the given quantity, with the only constraint being that the perturbations vanish at the inner and outer radial boundaries. It's worth noting in the ideal limit ($\eta, \nu = 0$), the displacement vector ($\boldsymbol{\xi}$)  \citep{chandrasekhar_stability_1960} can be introduced to decouple $\tilde{\mathbf{v}}$ and $\tilde{\mathbf{B}}$ as the magnetic field is ``frozen" in the moving fluid. Given finite viscosity and resistivity, such a method cannot be taken, as the magnetic field lines can now diffuse across fluid elements. 

Including the full effects of resistivity and viscosity directly into the global perturbation equations leads to a very complex, high-order differential equation system, making it difficult to solve and interpret physically. To simplify this while retaining the essential physics of diffusion, we adopt an approximation strategy proposed by \cite{zou_analysis_2020} that targets solely the diffusive terms (viscosity $\propto \nabla^2\mathbf{v}$ and resistivity $\propto \nabla^2\mathbf{B}$). If these diffusive contributions are assumed to be wave-like with radial wavenumber $k_r(r)$, which we need to specify (see Sec. \ref{sec:qr}, then leading order diffusive contributions in the short-wave limit can be simplified by the expression $-\nabla^2 \approx k_r^2(r) + m^2/r^2 +k_z^2$. This short-wave condition can be expressed via the radial wavenumber as $\left<k_r(r)\right>_r|(r_2-r_1) \gg 1$ for $r_1,r_2$ the boundaries of the domain and $\left<.\right>_r$ the radial average. Considering global variation of perturbed quantities, we utilize $r_2-r_1$ as the characteristic length instead of $r_1$. Crucially, this approximation only modifies the diffusive contributions, leaving the ideal dynamics unchanged, and does not presuppose the actual global structure of the instability. Although the approximation was derived in the short-wave limit, we will demonstrate that it remains accurate even for large-scale (global, long-wavelength) modes. This simplification allows us to reduce the governing equations to a second-order one-dimensional ODE.

Applying this diffusive approximation to the viscous and resistive terms in the linearized MHD equations (Eqns. \ref{eqn:mom}-\ref{eqn:ind}) leads to a system that can be written in matrix form as $\mathbf{M}\tilde{\pmb{\xi}} = 0$ for $\tilde{\boldsymbol{\xi}} = \left(\tilde{v}_r, \tilde{v}_\phi, \tilde{v}_z, \tilde{B}_r, \tilde{B}_\phi, \tilde{B}_z, \tilde{p}\right)$ and $\mathbf{M}$,
\begin{equation}
\begin{split}
    \mathrm{\textbf{M}} = \\ 
    \begin{pmatrix}
    -i\bar{\omega}_\nu & - 2\Omega & 0 & -\frac{iF}{\rho\mu_0} & \frac{2B_\phi}{r\rho\mu_0} & 0 &  \frac{1}{\rho}\frac{d}{dr} \\
    \frac{1}{r}\frac{d}{dr}\left(r^2\Omega\right) & -i\bar{\omega}_\nu & 0 & 0 & -\frac{iF}{\rho\mu_0} & 0 & \frac{im}{r\rho} \\
    0 & 0 & -i\bar{\omega}_\nu & 0 & 0 & -\frac{iF}{\rho\mu_0} & \frac{ik_z}{\rho} \\
    -iF & 0 & 0 & -i\bar{\omega}_\eta & 0 & 0 & 0 \\ 
    -\frac{2B_\phi}{r} & -iF & 0 & -r\frac{d\Omega}{dr} & -i\bar{\omega}_\eta & 0 & 0 \\
    0 & 0 & -iF & 0 & 0 & -i\bar{\omega}_\eta & 0 \\ 
    \frac{1}{r} + \frac{d}{dr} & \frac{im}{r} & ik_z & 0 & 0 & 0 & 0 \\
    0 & 0 & 0 & \frac{1}{r} + \frac{d}{dr} & \frac{im}{r} & {ik_z} & 0
    \end{pmatrix}
\end{split}
\end{equation}
where $F = \boldsymbol{k}\cdot\boldsymbol{B}_0 = \frac{mB_\phi}{r} + k_zB_z$, and $\bar{\omega}_{\eta,\nu} = \bar{\omega} - i(\eta,\nu)\cdot\left(k_r^2(r)+\frac{m^2}{r^2} + k_z^2\right)$. Moreover, the current free condition has been imposed ($-\frac{\partial B_\phi}{\partial r} = B_\phi/r$). This system can be algebraically reduced to a single second-order ODE for the radial structure of instabilities, which we write in terms of the variable $\xi_r = -r\tilde{v}_r/(i\bar{\omega}_\eta)$:
    \begin{equation} \label{eqn:shooting}
        \frac{d}{dr}\left(f\frac{d\xi_r}{dr}\right) + s\frac{d\xi_r}{dr} - g\xi_r = 0
    \end{equation}
   where the coefficients $f$, $s$, and $g$ are defined as,
\begin{equation}
\begin{split}
    f =\frac{r\left(\omega_A^2 - \bar{\omega}_\nu\bar{\omega}_\eta\right)}{k_z^2r^2 + m^2}, \; \; \; s = \frac{m(\bar{\omega}_\nu - \bar{\omega}_\eta)}{k_z^2r^2 + m^2}r\Omega', \\
    g = \frac{\mathrm{d}}{\mathrm{dr}}\Biggl\{\frac{m}{\left(k_z^2r^2 + m^2\right)}\Biggl[\left(\bar{\omega}_\eta-\bar{\omega}_\nu\right)r\Omega' + \left(2\Omega\bar{\omega}_\eta+\omega_A\omega_c\right)\Biggr]\Biggr\} \\
    + \frac{\left(\omega_A^2 + \omega_c^2- \bar{\omega}_\nu\bar{\omega}_\eta\right)}{r} + \frac{\omega_s^2}{r}\Biggl(\frac{\omega_A^2 - \frac{k_z^2r^2\bar{\omega}_\eta^2 + m^2\bar{\omega}_\eta\bar{\omega}_\nu}{k_z^2r^2+m^2}}{\omega_A^2 - \bar{\omega}_\nu\bar{\omega}_\eta}\Biggr) \\ 
    - \frac{k_z^2r}{k_z^2r^2 + m^2 }\frac{\left(2\Omega\bar{\omega}_\eta + \omega_A\omega_c\right)^2}{\omega_A^2 - \bar{\omega}_\nu\bar{\omega}_\eta}
     + \frac{k_z^2r^2\left(\bar{\omega}_\eta-\bar{\omega}_\nu\right)}{k_z^2r^2+m^2}\frac{\omega_A\omega_c\Omega'}{\omega_A^2 - \bar{\omega}_\nu\bar{\omega}_\eta}
\end{split}
\end{equation}
and the coefficients $\omega_A$ and $\omega_c$ are defined as,
\begin{equation}
    \omega_A = \frac{F}{\sqrt{\rho\mu_0}}= \frac{\mathbf{k}\cdot \mathbf{B}_0}{\sqrt{\rho\mu_0}}, \; \omega_c = \frac{2B_\phi}{r\sqrt{\rho\mu_0}}
\end{equation}

\begin{figure}
\centering
 \begin{subfigure}[b]{0.45\columnwidth}
     \includegraphics[width=\textwidth]{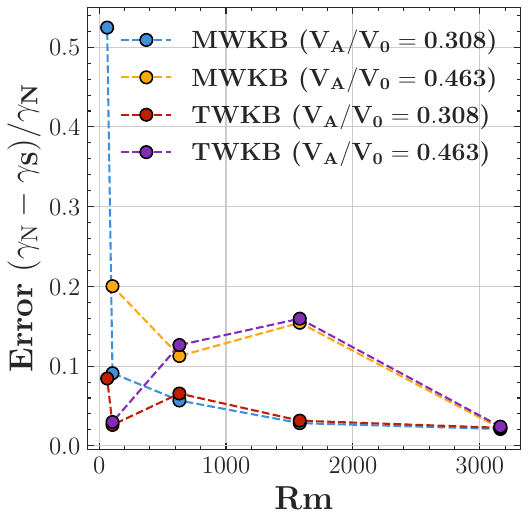}
     \caption{}
     \label{fig:c}
 \end{subfigure}
 \hfill
 \begin{subfigure}[b]{0.45\columnwidth}
     \includegraphics[width=\textwidth]{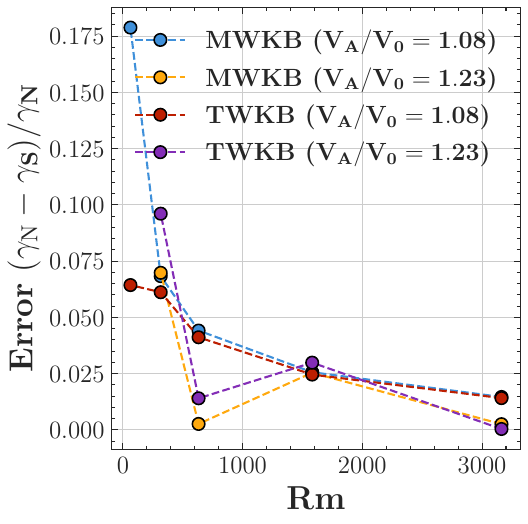}
     \caption{}
     \label{fig:d}
 \end{subfigure}
  \caption{MCI $k_z = 1k_1$ ($k_1 = \pi/4$) error between resistive global shooting method growth rates ($\gamma_S$) and NIMROD simulations ($\gamma_N$) for vertical (a) and azimuthal (b) magnetic field configuration between ``MWKB" approximation and ``TWKB" on $k_r^2(r)$.}
  \label{fig:NIMROD-error}
\end{figure}

This ODE matches the results of \cite{zou_analysis_2020} upon transform of variables, and matches previously derived ideal MHD results \cite{ebrahimi_nonlocal_2022} (see Sec. \ref{sec:conv}). In the following section, (Sec. \ref{sec:qr}), we will discuss how effective wave-vectors are utilized to model this ODE's diffusive contributions. Then, in Section \ref{sec:scan}, we will discuss how we obtain global solutions to this ODE and analyze these solutions compared to NIMROD simulations in Section \ref{sec:vf-res}.

\subsection{Choosing $k_r^2(r)$} \label{sec:qr}
A crucial input to the non-ideal ODE (Eq. \ref{eqn:shooting}) is the choice of the effective radial wavenumber $k_r(r)$, used in the diffusive approximation. This choice dictates how the radial scale of diffusion is represented. Since we are modeling global modes, we expect this parameter to depend on the radii of the domain ($r_1$, $r_2$) and/or the aspect ratio $r_1/(r_2-r_1)$. We test two primary candidates:
\begin{enumerate}
    \item MWKB (Modified WKB-like): $k_r^2(r)= \left(\frac{r_1}{r_2-r_1}\right) \times \frac{4\pi^2}{r^2}$. This form incorporates a linear dependence on the aspect ratio and power law envelope. We developed this method based on the previous suggestion by \cite{zou_analysis_2020} (scaling with aspect ratio squared), which did not accurately capture the growth rates across different simulated geometries (AR0, AR1, AR2). We found, however, that a method that scales linearly with aspect ratio, and is multiplied by a WKB-like contribution, was able to reproduce this scaling, with this WKB-like contribution taking the form $(2\pi/r)^2$, instead of $(2\pi/(r_2-r_1))^2$.
    \item TWKB (Traditional WKB-like): $k_r^2(r) = \left(\frac{\pi}{r_2-r_1}\right)^2 - i\left(\frac{\pi}{r_2-r_1} \right)/r$. This form arises from applying a standard WKB method to the Laplacian operator in cylindrical coordinates, representing the lowest radial mode number. Importantly, this form introduces an imaginary component related to the curvature of the domain.
\end{enumerate}

We benchmarked these two approximations against direct linear simulations using the NIMROD code, comparing the spectral growth rates ($\gamma_S$ from shooting vs. $\gamma_N$ from NIMROD) as shown in Figure \ref{fig:NIMROD-error}. Generally, both methods perform better for purely azimuthal magnetic fields, often yielding errors below 20\% even at low Rm (see Fig. \ref{fig:NIMROD-error}b). For vertical fields, the MWKB approximation shows significantly lower error across much of the test regime. However, the TWKB approximation becomes more accurate as Rm $\rightarrow 0$.
We begin by benchmarking both approximations relative to NIMROD simulations in Figure \ref{fig:NIMROD-error}. Generally, both approximations are more accurate with the azimuthal magnetic field configuration, typically having less than $20\%$ error even at low Rm. For a vertical magnetic field, there is a clear distinction in the error offered through both methods, with the MWKB method having nearly half the error for the entire regime. Nonetheless, both methods can reproduce the general scaling under the introduction of diffusive contributions.

The domains of accuracy of each approximation can be quantified by the short-wave condition ($\bigl|\left<k_r\right>_r\bigr|\left(r_2-r_1\right) \gg 1$). The TWKB approximation, rooted in the WKB limit, is expected to be most accurate in the limit $r_2/r_1 \rightarrow 1$, and we find this satisfies the short-wave condition. Conversely, we find that the MWKB approximation is better suited for large disks ($r_2/r_1\gg 1$), and thus both approximations are essential to consider. While finding an optimal $k_r(r)$ that can represent the scaling of diffusion contributions in both limits remains an open question, both MWKB and TWKB approximations provide inexpensive methods to reproduce mode properties (frequencies, growth rates) and structures that correlate well with simulation observables. Further systematic simulations varying $r_2/r_1$ and $z_2/r_1$ could help refine the optimal form of $k_r(r)$. This remains a path of future research.

\subsection{Global solutions at finite Rm}\label{sec:scan}

\begin{figure}
\centering
 \begin{subfigure}[b]{.85\columnwidth}
    \includegraphics[width=\textwidth]{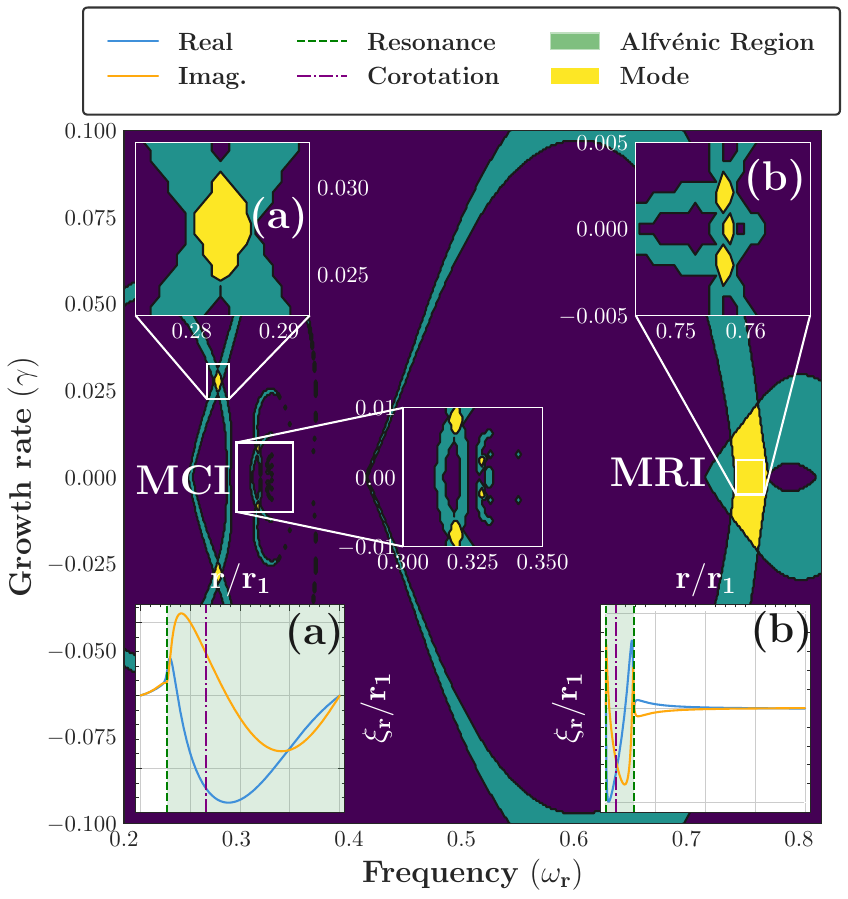}
     \caption{}
     \label{fig:c}
 \end{subfigure}
 \hfill
 \begin{subfigure}[b]{.85\columnwidth}
    \includegraphics[width=\textwidth]{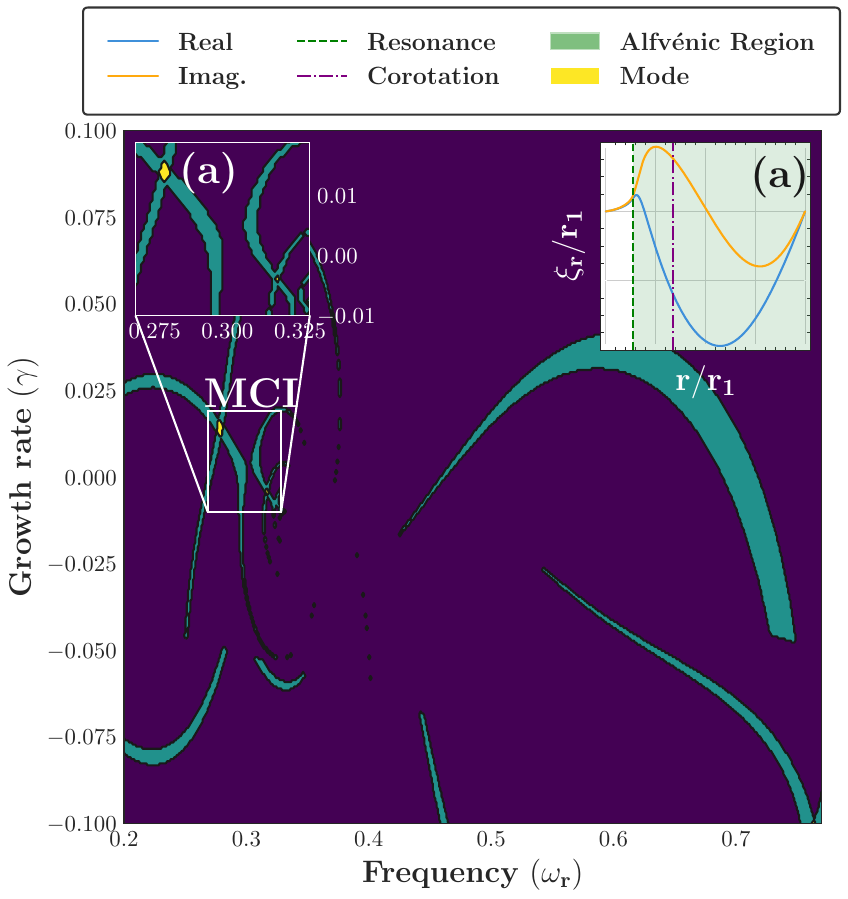}
     \caption{}
     \label{fig:d}
 \end{subfigure}

  \caption{MCI vs MRI: Locations of real and imaginary shooting tails and their overlap, which indicates a solution exists fitting the boundary conditions; Purple indicates neither the real nor the imaginary tail is within tolerance of zero; Teal indicates one tail is within tolerance; Yellow indicates both tails are within tolerance of zero, and there exists a solution. Shooting simulation ran for the solely vertical magnetic field of $V_A/(r_1\Omega_0) = 0.308$, $m = 1$, and $k_z = 1k_1$ $(r_1k_1 = \pi/4)$ for Rm $= \infty$ (Pm $= 0$) (a) and Rm $= 105$ (Pm $= 1$) (b). The left-hand set of solutions represents the low-frequency curvature mode (MCI), whereas the right-hand set of solutions represents the high-frequency mode (MRI). At high resistivities, only the low-frequency and global MCI mode remains unstable.}
  \label{fig:Scanplot-Ex}
\end{figure}

We solve the boundary-value ODE (Eq. \ref{eqn:shooting}) using the shooting method. For a given guess of the complex frequency $\omega$, we integrate the ODE from the inner boundary ($r_1$) to the outer boundary ($r_2$). An eigenfunction exists if and only if the calculated structure satisfies the boundary condition $\xi(r_2) = 0$. Numerically, we search for values $\omega = \omega_r + i\gamma$ where both the real and imaginary components of $\xi_r(r_2)$ (the ``tails") are simultaneously zero. Plotting these zero-crossings in the complex $\omega$ plane reveals the eigenvalues under specified parameters, as illustrated in Figure \ref{fig:Scanplot-Ex}. This figure also demonstrates how diffusion can eliminate certain solution branches present in the ideal limit.

Typically, the solutions fall into two distinct branches of modes: one associated with low frequencies and large radial scales (which we identify as the Magneto-Curvature Instability, MCI) and another with higher frequencies and more localized structures (identified as the Magneto-Rotational Instability, MRI). The exact frequencies, growth rates, and structure depend on system parameters (magnetic field configuration and strength, $m$, $k_z$, Rm, Pm, flow profile, etc). Our analysis generally focuses on the mode with the largest growth rate $(\gamma$) on each branch, which is most likely to be observed via experiments and simulations.

An important effect of diffusion becomes apparent when comparing the ideal (Rm $= \infty$, Fig. \ref{fig:Scanplot-Ex}a) and the non-ideal (finite Rm, Fig.\ref{fig:Scanplot-Ex}b) scans. In the ideal limit, the equations posses time-reversal symmetry, meaning every growth mode ($\gamma > 0$) has a corresponding decaying mode ($\gamma < 0$) with the same frequency ($\omega_r$). Diffusion breaks this symmetry. Furthermore, as we will explore using the effective potential (Sec. \ref{sec:res-pot}), diffusion tends to damp localized modes (like MRI)  at a higher rate than global modes (like MCI). This differential damping is evident in the eigenvalues scans (Fig.\ref{fig:Scanplot-Ex}b), where the MRI branch is suppressed at low Rm and the MCI remains unstable. Consequently, MCI (from the low-frequency branch) remains the only unstable mode at sufficiently low Rm.

\subsection{Resistive resonance condition} \label{sec:res-resonance}
A key concept for understanding mode structure and confinement is the Alfvénic resonance. In the presence of diffusion, these resonances occur at radial locations where the real part of the diffusion-modified Doppler-shifted frequency matches the Alfvén frequency:
\begin{equation}\label{eqn:reson}
   \; \mathrm{Re}\left[\omega_A^2 - \bar{\omega}_\eta\bar{\omega}_\nu\right] = 0
\end{equation}
These resonance points appear as finite singularities in that ODE formalism (Eq. \ref{eqn:shooting}) for any growing/decaying mode ($\gamma \neq 0$). In the effective potential formalism, these resonances manifest as potential barriers \cite{ebrahimi_generalized_2025} and will be further explored given diffusive effects in Section \ref{sec:res-pot}. This definition generalizes the ideal resonance condition ($\bar{\omega} = \pm \omega_A$) and is consistent with previous work in the appropriate limits \citep{matsumoto_magnetic_1995}, \citep{ebrahimi_nonlocal_2022}.

Expanding this condition (Eq. \ref{eqn:reson}) reveals how the resonance location depends on the mode's frequency ($\omega_r$), the flow configuration ($\Omega(r)$), the Alfvén frequency, the diffusivities ($\eta, \nu$), and the total wavenumber ($Q = k_r^2(r) + m^2/r^2+k_z^2$). 

\begin{equation}\label{eqn:resonance-condition}
    \omega_r - m \Omega(r) = \mathrm{Re}\left[\frac{i\left(\eta + \nu\right)Q}{2}\pm \sqrt{\omega_A^2 - \frac{1}{4}\left(\eta - \nu\right)^2Q^2}\right]
\end{equation}
 We have omitted the imaginary components, yet kept $i\left(\eta + \nu\right)Q$ as it is possible that $Q \in \mathbb{C}$ (see the TWKB approximation in Section \ref{sec:qr}) and thus contributes to the resonance condition. The presence of diffusion ($\eta,\nu$) shifts the resonance locations compared to the ideal case. The details of this shift depend on the relative magnitude of viscous to resistive diffusivities (Pm). Nonetheless, we will demonstrate via the effective potential formalism that diffusion generally acts to broaden potential barriers, contributing to the stabilization of modes.

\subsection{Resistive effective potential formalism}\label{sec:res-pot}

To gain deeper physical insight into mode confinement and the sources of instability free energy, we adapt the effective potential formalism, previously applied in ideal MHD \cite{curry_global_1996} \cite{pino_global_2008} \citep{ebrahimi_generalized_2025}, to our non-ideal global ODE (Eq. \ref{eqn:shooting}). By applying an integrating factor transform ($\xi_r(r^2) = u(r^2)\Psi(r^2)$ with $u(r^2) = \mathrm{exp} \left[-\frac{1}{2}\int dr^2 \left(\frac{f' + \frac{s}{2r}}{f} + \frac{1}{2r^2}\right) \right ]$), we recast the second-order ODE into a Schrödinger-like equation:
\begin{equation}\label{eqn:Potential}
    \Psi''(r^2) - U(r^2, \mathrm{mode})\Psi(r^2) = 0,
\end{equation}
with,
\begin{equation}
    U(r^2, \mathrm{mode}) = \frac{g}{4r^2f} + \frac{1}{2}\left(\frac{f'+ \frac{rs + f}{2r^2}}{f}\right)' + \frac{1}{4}\left(\frac{f'+\frac{rs+f}{2r^2}}{f}\right)^2
\end{equation}
Here, $\Psi$ represents the transformed radial mode structure, and $U(r^2, \mathrm{mode})$ is the complex effective potential. This potential $U$ depends on the original ODE coefficients ($f,s,g$) and thus on all physics parameters of the system, including the mode's complex frequency $\omega$ (which must be an eigenvalue found via shooting). We denote this dependence through the acronym ``mode" in the definition of the potential.

\begin{figure}
\centering
 \begin{subfigure}[b]{1\columnwidth}
    \includegraphics[width=\textwidth]{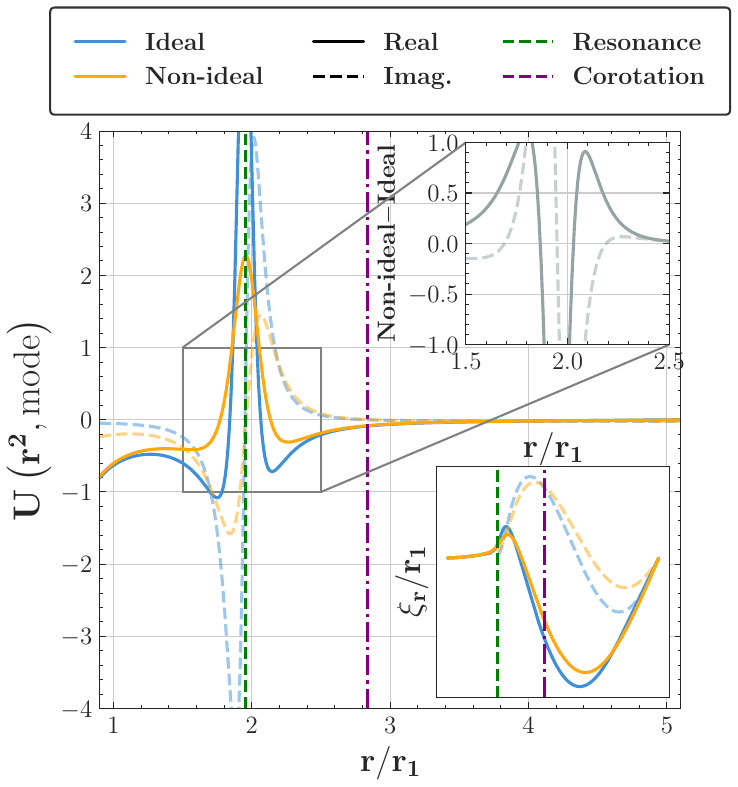}
     \caption{}
     \label{fig:c}
 \end{subfigure}
 \hfill
 \begin{subfigure}[b]{1\columnwidth}
    \includegraphics[width=\textwidth]{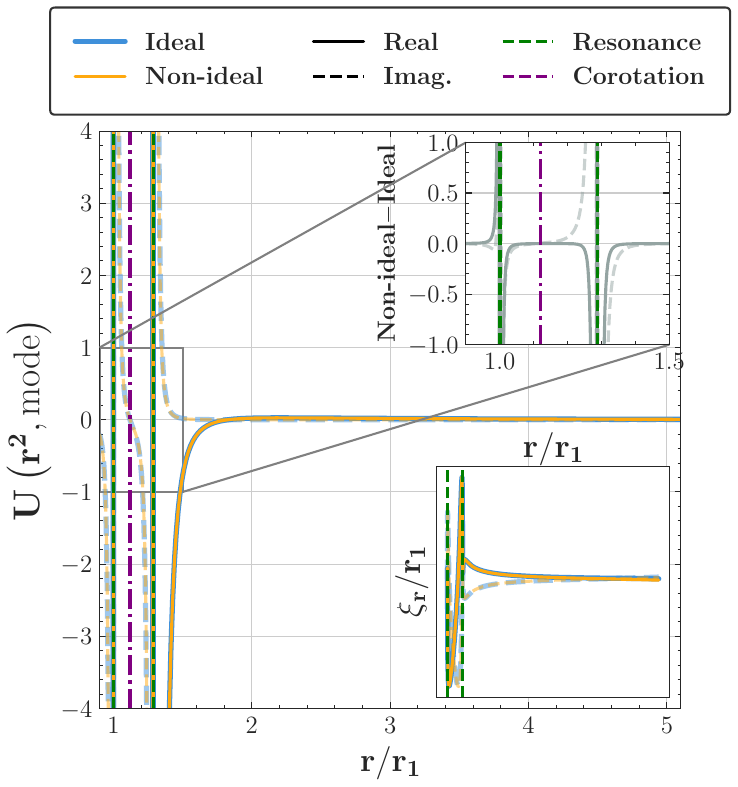}
     \caption{}
     \label{fig:d}
 \end{subfigure}

  \caption{Potentials for the non-axisymmetric ($m = 1$) MCI $1k_1$ (a) and MRI $1k_1$ (b) ($r_1k_1 = \pi/4$) modes at $V_A/(r_1\Omega_0) = 0.2$ as Rm is increased (Pm $= 1$). Non-ideal shooting for MCI and MRI are conducted at Rm $= 105$ and Rm $= 31623$ respectively. For both MRI and MCI, increasing diffusive effects causes the modes to become more global with respect to the resonances. This is demonstrated in the inset figures, which show the difference between ideal and non-ideal potentials.}
  \label{fig:Potential-NonIdeal}
\end{figure}
This formalism allows us to visualize how modes are spatially confined. Regions where Re$(U)$ is large and positive act as barriers, while regions where Re($U$) is negative act as potential wells where the mode amplitude ($\Psi$) becomes large. Figure \ref{fig:Potential-NonIdeal} illustrates these potentials for typical MCI and MRI modes with and without diffusive effects. A key observation is that finite diffusion ($\eta, \nu \neq 0$) tends to smooth out and broaden the potential wells, particularly about the Alfvénic resonances. This broadening makes the modes less tightly confined (more global) than their ideal counterparts, corroborating findings from the local analysis but now linking the structural change explicitly to the resonances. We expect MRI's inherent locality and resonance-localized nature to make the mode's instability more sensitive to diffusive contributions than the inherently global MCI modes.

\begin{figure}
    \centering
    \includegraphics[width=0.7\linewidth]{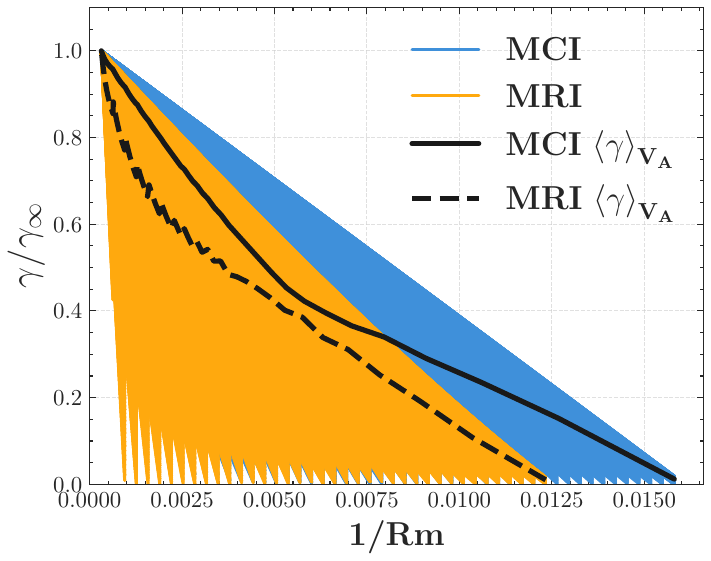}
    \caption{Growth rate scaling of MRI/MCI modes for a span of $V_A/(r_1\Omega_0)$ shows that MCI modes scale weakly with non-idealities. Finite Rm growth rates  $\gamma$ are normalized to ideal growth rates $\gamma_\infty$ to consider magnetic-field invariant scaling. Magnetic field average scaling ($\left<\gamma\right>_{V_A}$) is considered for both MRI and MCI modes. All values are from the non-ideal global shooting method with $k_z =2k_1 $ $(r_1k_1 = \pi/4)$ and the AR1 aspect ratio. Shooting was conducted with Pm $= 1$ to consider the largest non-ideal contribution to scaling.}
    \label{fig:res-scaling}
\end{figure}

Comparing the potentials and growth rates for MCI and MRI modes (e.g., Fig. \ref{fig:Potential-NonIdeal} and Fig. \ref{fig:res-scaling}) reinforces the idea that diffusion affects these modes differently. Figure \ref{fig:res-scaling} shows how the growth rates ($\gamma$, normalized by the ideal growth rate $\gamma_\infty$) decrease as diffusion effects increase (Rm decreases) over all magnetic fields where the mode is unstable. The MCI modes retain a larger fraction of their ideal growth rates than MRI at the same Rm. This ultimately results in MCI modes remaining unstable at high diffusivities (lower Rm) than MRI modes, which is showcased by the magnetic field averaged scalings for MRI and MCI in Figure \ref{fig:res-scaling}. This difference in scaling with diffusion strongly suggests that the first instability to appear (also known as the onset of instability) will generally be a global MCI mode, likely the one with the largest vertical wavelength ($k_z = 1k_1$), since diffusion scales with $k_z^2$.

\begin{figure}
\centering
 \begin{subfigure}[b]{0.49\columnwidth}
    \includegraphics[width=\textwidth]{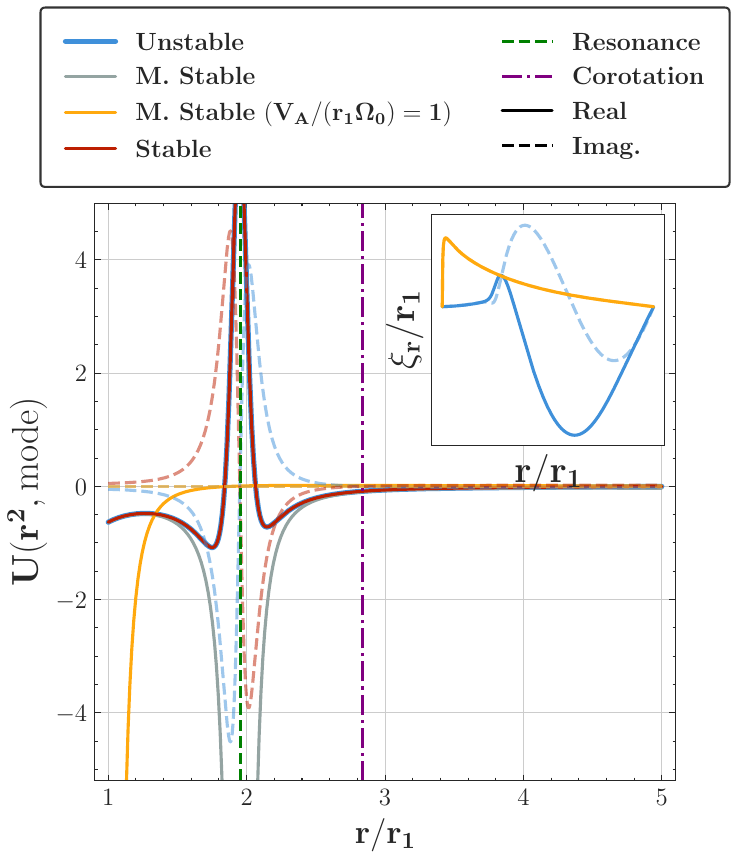}
     \caption{}
 \end{subfigure}
 \hfill
 \begin{subfigure}[b]{0.49\columnwidth}
    \includegraphics[width=\textwidth]{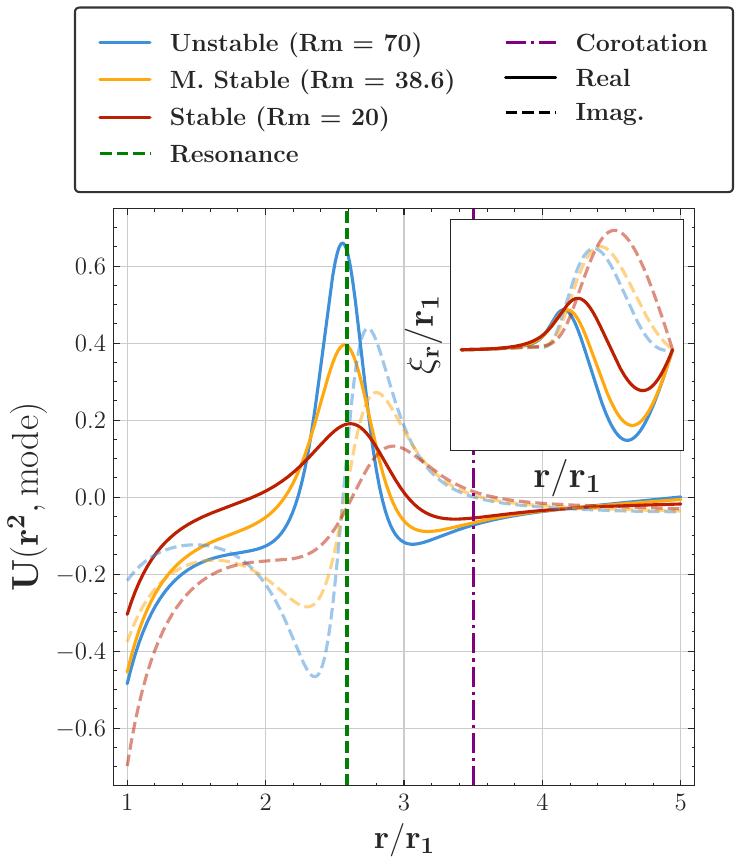}
     \caption{}
 \end{subfigure}
    \hfill
 \begin{subfigure}[b]{1\columnwidth}
    \includegraphics[width=\textwidth]{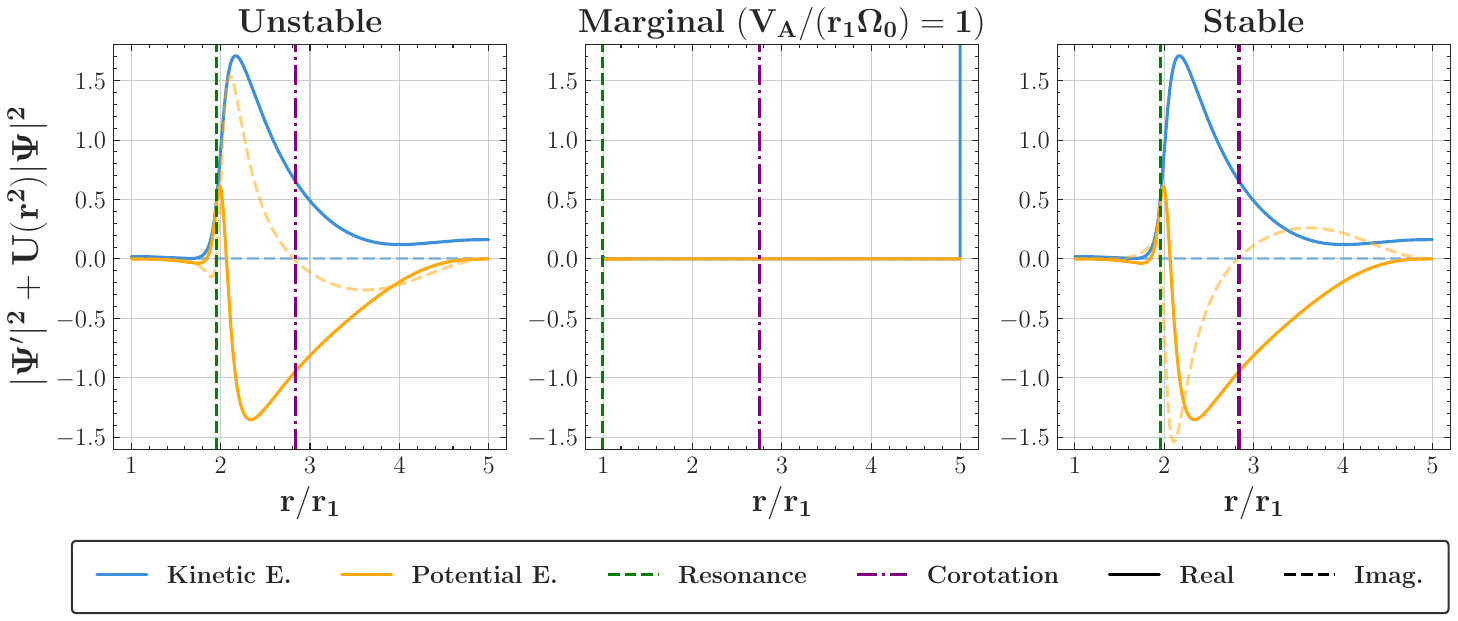}
     \caption{}
 \end{subfigure}
    \hfill
 \begin{subfigure}[b]{1\columnwidth}
    \includegraphics[width=\textwidth]{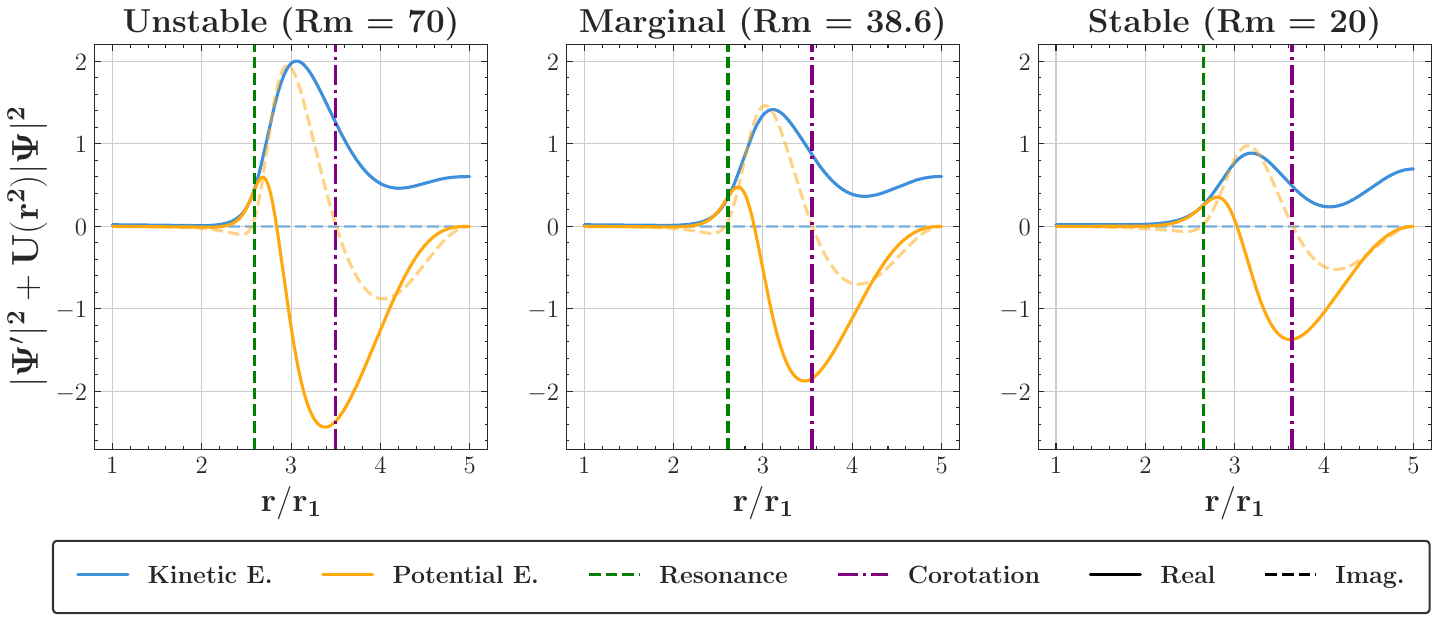}
     \caption{}
 \end{subfigure}
 
  \caption{Ideal (a,c) and non-ideal (b,d) potentials and energy densities for the global non-axisymmetric ($m = 1$) unstable ($\gamma > 0$), marginally stable (M. Stable) ($\gamma = 0$), and stable ($\gamma < 0$) MCI $1k_1$ ($r_1k_1 = \pi/4$) modes at $V_A/(r_1\Omega_0) = 0.2$ (stable/unstable), $V_A/(r_1\Omega_0) = 1$ (marginally stable) (a,c) and $V_A/(r_1\Omega_0) = 0.11$ (b,d) with $\mathrm{Rm} = \infty$ ($\mathrm{Pm} = 0$) (a,c) and as Rm (Pm $= 1$) varies (b,d). The plots of the potentials depict that all modes considered (Unstable, Marginally Stable, and Stable) are confined.}
  \label{fig:Stability-Char}
\end{figure}

\subsubsection{Energy Criterion}
The effective potential $U(r^2, \mathrm{mode})$ provides insights not only into spatial confinement but also into mode stability ($\gamma$). Examining the potential structure for known eigenvalues $\omega$ (found via shooting) helps understand these aspects (see Figure \ref{fig:Stability-Char}). In the ideal case, due to time reversal symmetry, the potential for a decaying mode is the complex conjugate of the potential for a growing mode ($U(r^2, \omega) = U^*(r^2,\omega^*)$). This means Re$(U)$ is the same for both, indicating confinement, while Im$(U)$ flips sign and relates to stability: Im$(U) > 0$ in the confinement region corresponds to $\gamma > 0$ (growing mode, unstable), Im$(U) < 0$ corresponds to $\gamma < 0$ (decaying mode, stable), and Im($U) = 0$ to $\gamma = 0$ (marginally stable). However, diffusion breaks this relationship with stability as demonstrated in Figure \ref{fig:Stability-Char}b. In the non-ideal case, the sign of Im$(U)$ no longer flips as the mode crosses marginal stability ($\gamma$). Establishing a criterion based on Im$(U)$ structure to determine the stability of non-ideal modes requires future investigation.

\begin{figure}
    \centering
    \includegraphics[width=1\linewidth]{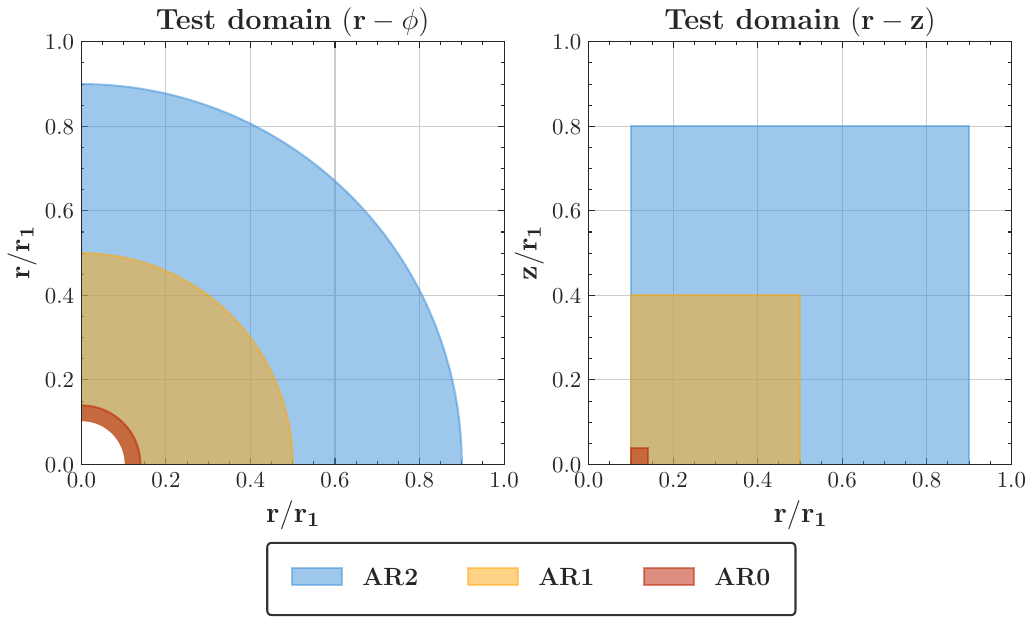}
    \caption{Slices of system geometry between different disk configurations. We define all domains such that disk thickness $\Delta z/\Delta r = 1/2$. Meanwhile, AR0, AR1, and AR2 exhibit aspect ratios $r_1/\Delta r = r_1/(r_2-r_1) \in \left\{5/2, 1/4, 1/8 \right\}$} 
    \label{fig:Ar-Config}
\end{figure}

\begin{figure} 
    \centering
    \includegraphics[width = 1\linewidth]{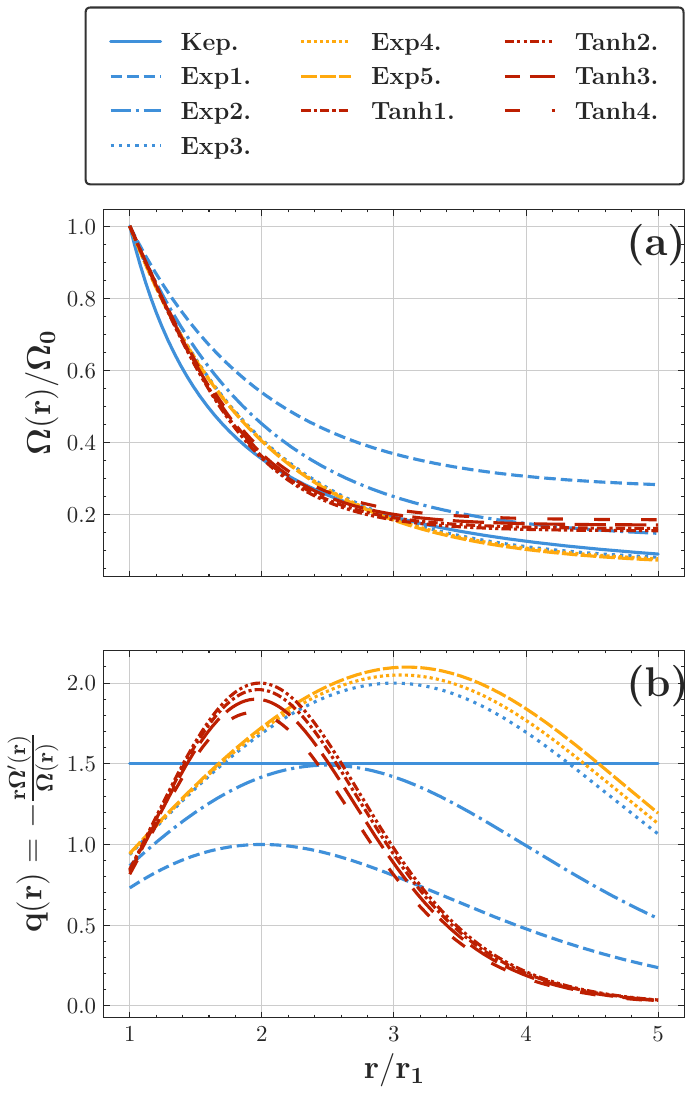}
  \caption{Variation of flow profile parameters. Depicts both the normalized rotational frequency of the plasma $\Omega(r)/\Omega_0$, as well as the flow shear parameter $q(r) = -\frac{r\Omega'(r)}{\Omega(r)}$. We consider flows unstable to axisymmetric (and thus non-axisymmetric) hydrodynamic perturbations ($1 \leq q(r) \leq 2$). Kep. is the standard Keplerian profile with $\Omega(r)/\Omega_0 = 1/r^{3/2}$. The exponential profiles (Exp.) are generated as $\Omega(r)/\Omega_0 = a\mathrm{exp}(1-r/r_1) + (1-a)$. Exp1, Exp2, and Exp3: $a = 0.7312$, $a = 0.8689$, and $a = 0.9366$ respectively. In addition, we consider hydrodynamically unstable configurations ($q(r) > 2$) generated via the exponential profile. Exp4, Exp5: $a = 0.9409,$ $a = 0.9448$. All configurations here are shown in the AR1 aspect ratio.}
  \label{fig:flowshearparam}
\end{figure}

We now transform Equation \ref{eqn:Potential} to define the form that a well-defined potential must obey. This approach is similar to \cite{curry_global_1996}, except under a different normalization, which makes the quantity we derive an energy. We begin by multiplying Equation \ref{eqn:Potential} by $\Psi^*(r^2)$, and then integrating by parts (noting that $\Psi$ at both boundaries is zero) to arrive at the following definition,
\begin{equation}
    E = \frac{\bigintss \biggl[ \big|\Psi'(r^2)\big|^2 + U(r^2)\big|\Psi(r^2)\big|^2 \biggr]dr^2}{\int \big|\Psi(r^2)\big|^2dr^2} = 0.
\end{equation}
For this integral to be zero, two conditions related to the potential $U(r^2)$ must be met:
\begin{enumerate}
    \item Real Component: The kinetic energy term $\bigl|\Psi'\bigr|^2$ is always positive. Therefore, the potential energy term $U(r^2)\bigl|\Psi|^2$ must contribute a negative real part overall. This implies that $\mathrm{Re}(U)$ must be sufficiently negative in the region where the mode exists to allow confinement ($\Psi \neq 0$). A purely positive potential cannot support a confined mode.
    \item Imaginary Component: The imaginary part of the integral, $\mathrm{Im}(E) \propto \int\mathrm{Im}(U)\bigl|\Psi\bigr|^2dr^2$ must also be zero. This requires that the imaginary part of the potential, $\mathrm{Im}(U)$, must either be identically zero (as in the marginally stable case), or it must change sign within the mode's confinement region such that the positive and negative contributions exactly cancel when weighted by $\bigl|\Psi\bigr|^2$. A potential where $\mathrm{Im}(U)$ is strictly positive or negative throughout the domain cannot support a confined eigenmode
\end{enumerate}
These criteria provide conditions for physically valid, confined solutions within the effective potential framework.

\begin{figure*}
\centering
 \begin{subfigure}{.7\textwidth}
     \includegraphics[width=\textwidth]{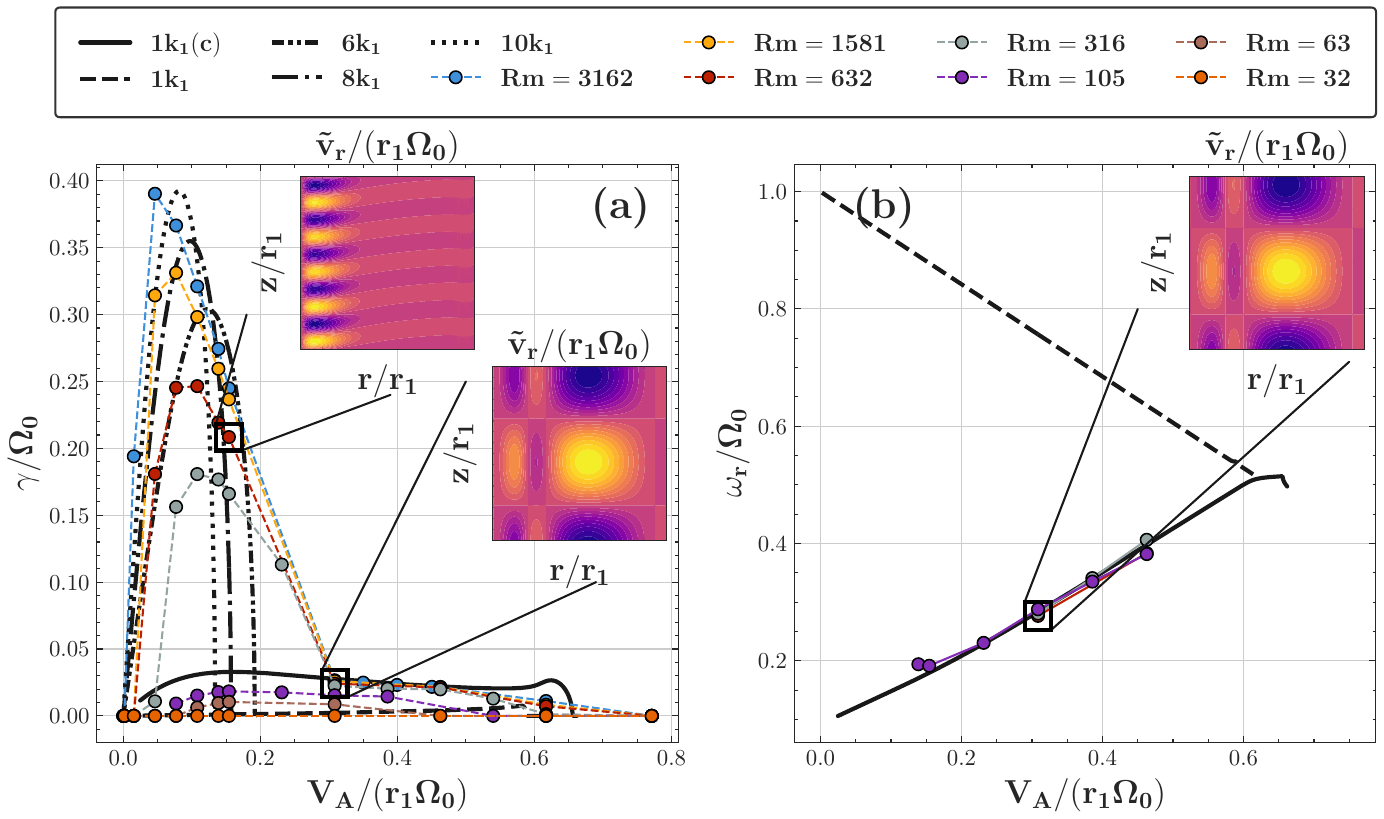}
 \end{subfigure}
 \bigskip

 \begin{subfigure}{.7\textwidth}
     \includegraphics[width=\textwidth]{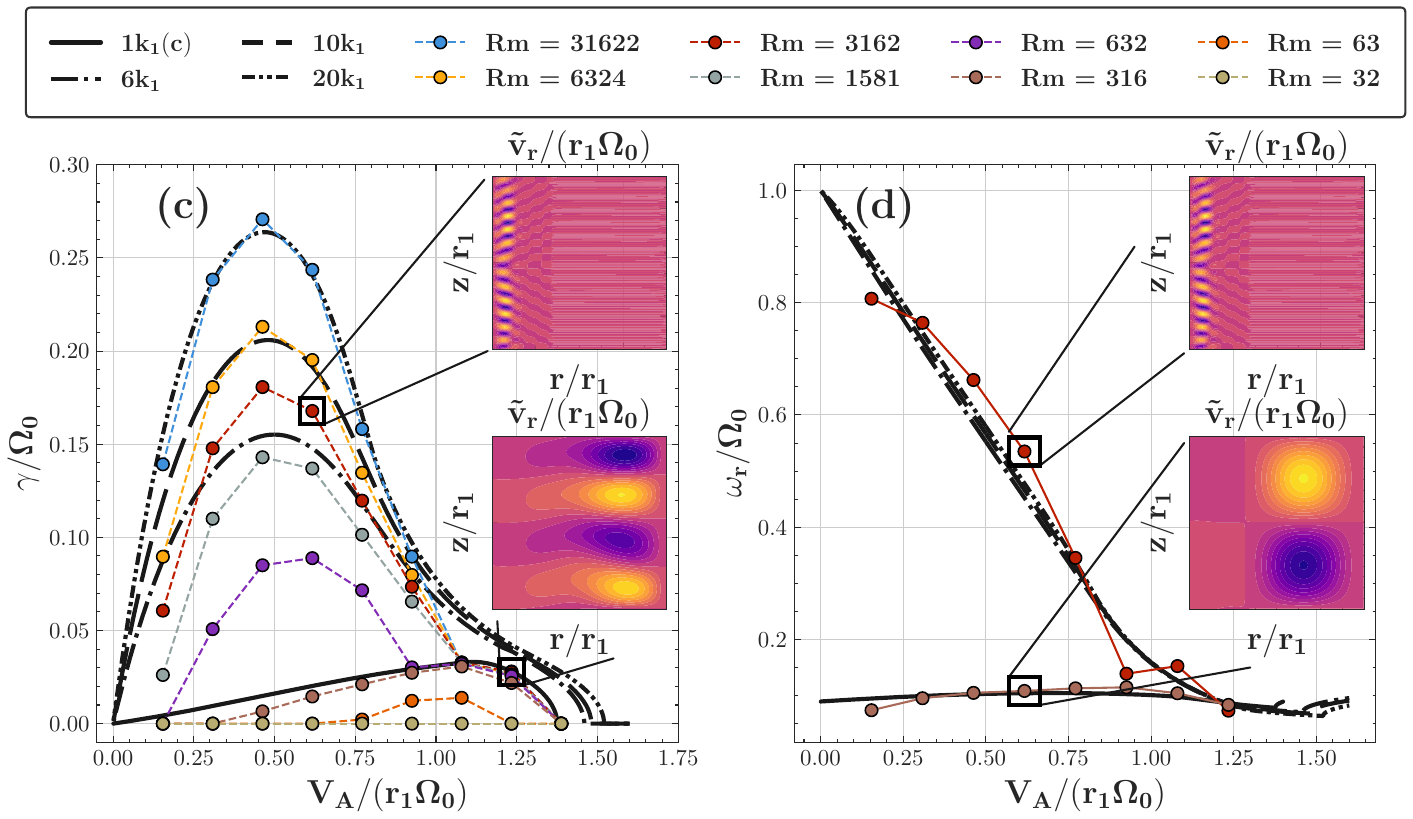}
     \label{fig:b}
 \end{subfigure}
     \caption{Growth rates and frequencies for pure vertical field (a,b) and azimuthal field (c,d) from both NIMROD simulations and the global shooting method for relevant vertical wave numbers. In both field configurations, there exists a $V_A/V_0$ where the system transitions from the localized MRI to global MCI mode. A transition $\mathrm{Rm}$ exists where the system becomes the MCI $1k_1$ mode for all $V_A/V_0$. Vertical wave numbers ($k_z$) for shooting curves were determined by looking at the mode structure from NIMROD, but do not capture all observed wave numbers for MRI. For a purely vertical field, ideal shooting is conducted for the $1k_1$ MCI mode (denoted as (c)) as well as MRI $1k_1$, $6k_1$, $8k_1$, and $10k_1$ modes. For a purely azimuthal field, ideal shooting is conducted for the $1k_1$ MCI mode (denoted as (c)) and MRI $6k_1$, $10k_1$, and $20k_1$ modes. Inset plots depict observed mode structures from NIMROD simulations showing that the MCI modes are global and low $k_z$, meanwhile the MRI modes are local and high $k_z$.}
    \label{fig:NIMROD-Comp}
\end{figure*}

\begin{figure*}
    \centering
    \begin{subfigure}[b]{1\columnwidth}
        \centering
        \includegraphics[width=\textwidth]{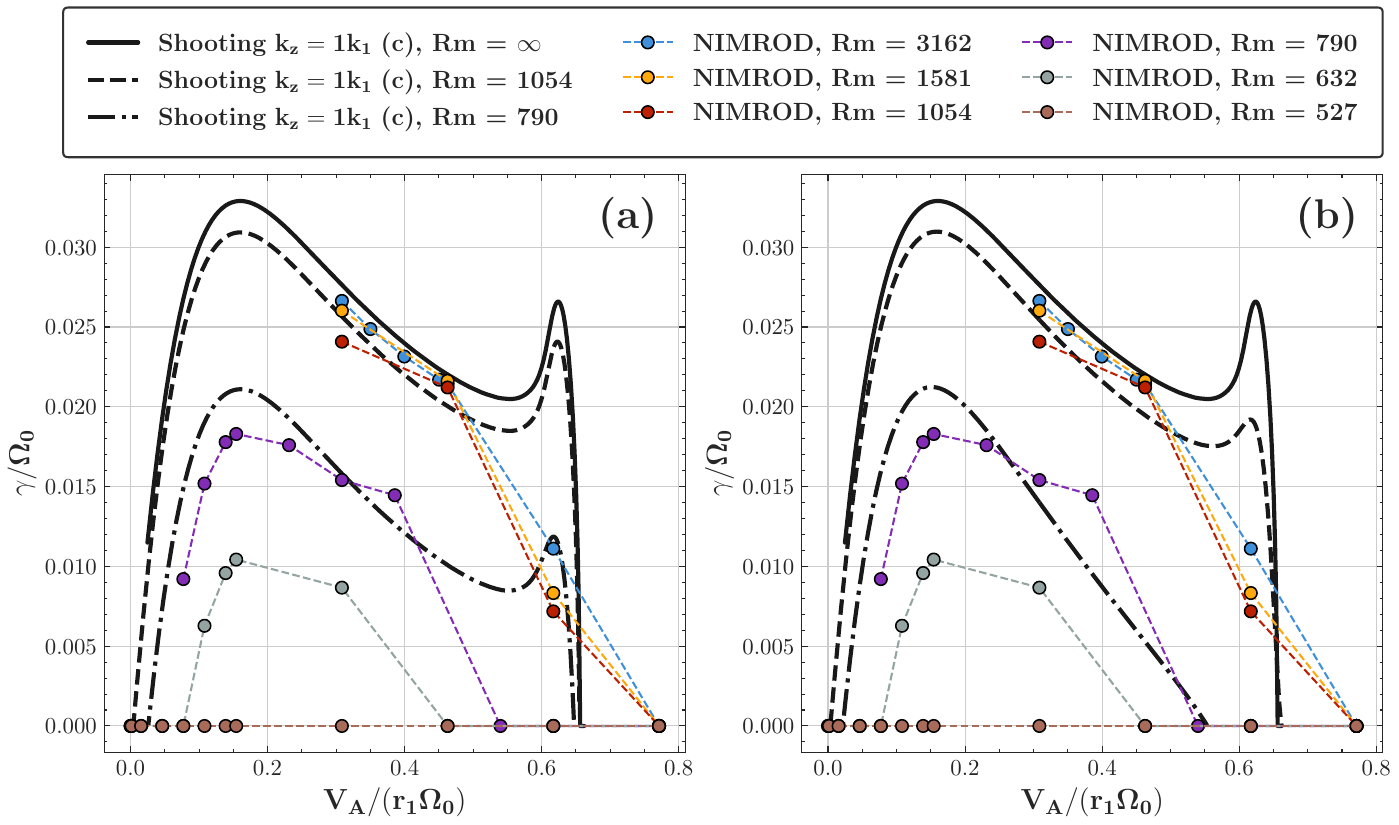}
     \label{fig:phase-a}
    \end{subfigure}
    \hfill
    \begin{subfigure}[b]{.95\columnwidth}
        \centering
        \includegraphics[width=\textwidth]{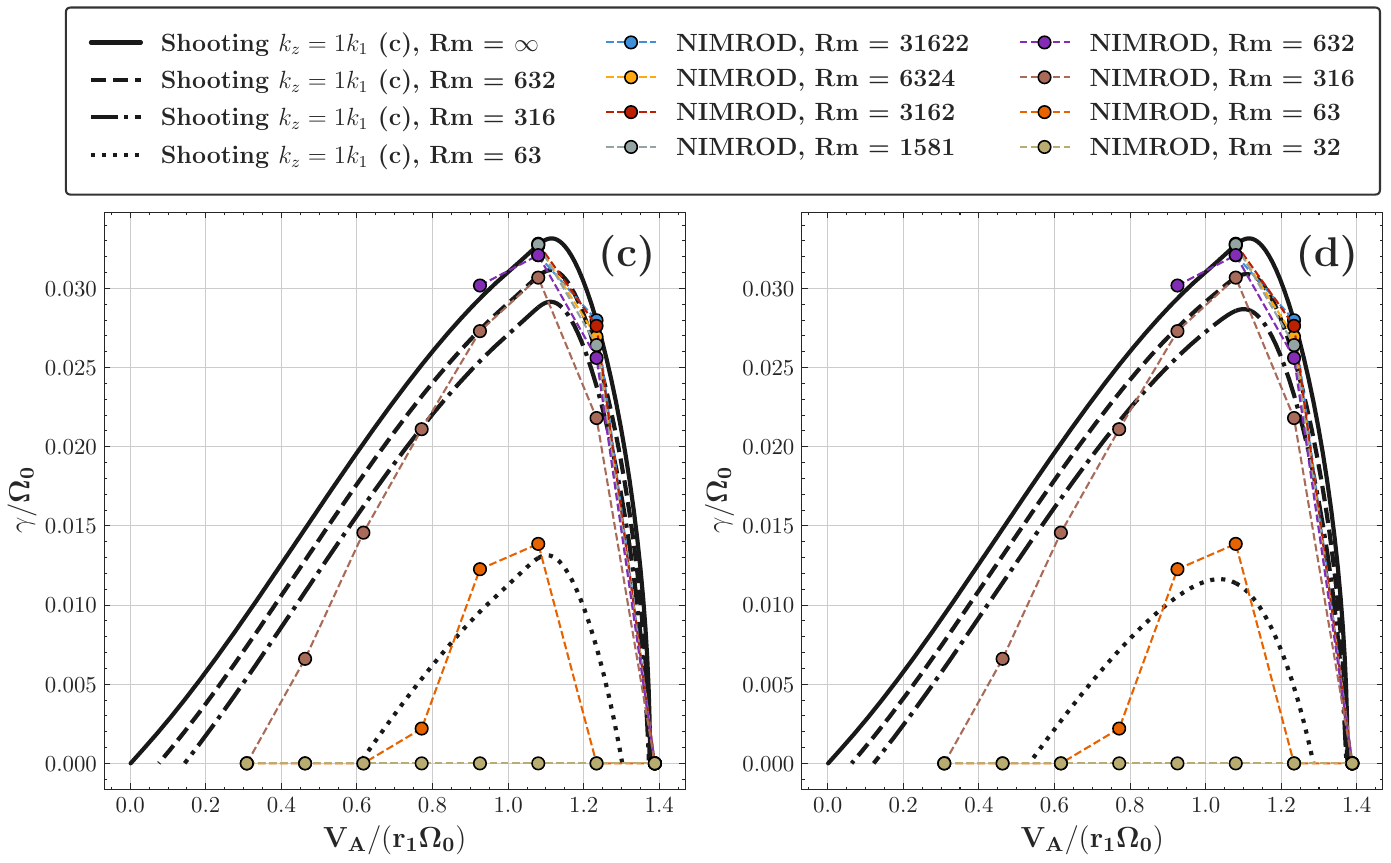}
     \label{fig:phase-b}
        \label{fig:2}
    \end{subfigure}
     \caption{Benchmarking of resistive shooting analysis on MCI $1k_1$ ($k_1 = \pi/4$) mode for both vertical (a,b) and azimuthal magnetic field (c,d) by comparing to growth rates from NIMROD simulations. In (a,c) we demonstrate the ``TWKB" method and in (b,d) we show the ``MWKB" method. All analysis was done with the AR1 aspect ratio, and Pm $ = 1$.}
    \label{fig:WKB-Qmagpi-Shooting}
\end{figure*}

\section{Global solutions under various scans}\label{sec:vf-res}

\begin{table*}
\begin{center}
\caption{Parameter inputs for global linear NIMROD simulations. Azim. refers to purely azimuthal field and Vert. purely vertical field. All simulations were ran with $r_1 = 0.1 \;m$, and $\Omega_0 = 10000\cdot(10)^{3/2} \; s^{-1}$. For AR0, low magnetic field cases $B_\phi \leq 30$ G were run with a finer temporal resolution $\mathrm{dt} = 10^{-9}$ s compared to $\mathrm{dt} = 10^{-8}$ s of all other Azim. cases. All Vert. cases were run with temporal resolution of $\mathrm{dt} = 10^{-9}$ s. Moreover, $\eta = 2 \; m^2/s$ for Vert. was ran with a resolution of $160 \times 160$ deg. 4 (m = 1).}
\label{table:AR0}
\centering
\begin{tabular}{ p{2.5cm} p{3cm} p{2cm} p{3cm} p{2cm} p{4cm}}
 \hline
 \hline
Aspect Ratio & Rm $= r_1^2\Omega_0/\eta$ & Pm $= \nu/\eta$ & S $= r_1V_A/\eta$& Config. & Resolution\\
 \hline
    AR0  & 6.3E2$-$3.2E4   & 1 &  9.8E0$-$4.4E2 &  Azim. & $80 \times 80$ deg. 4 (m = 1) \\
    AR1  & 3.2E1$-$3.2E4  & 1 &  4.9E-1$-$4.4E3 &   Azim. & $80 \times 80$ deg. 4 (m = 1) \\
    AR1  & 3.2E1$-$3.2E4   & 1 & 4.9E-1$-$4.4E3  &   Vert. & $120 \times 120$ deg. 4 (m = 1) \\
    AR2  & 6.3E2$-$3.2E4     & 1 & 4.9E-1$-$5.9E3  & Azim. & $160 \times 160$ deg. 4 (m = 1)\\
 \hline
\end{tabular} 
\end{center}
\end{table*}
This section investigates the behavior and scaling of Magneto-Rotational (MRI) and Magneto-Curvature (MCI) instabilities under variations in non-ideal effects (resistivity $\eta$, viscosity $\nu$), flow profiles $\Omega(r)$, and domain geometry. We utilize both linear initial-value simulations with the NIMROD code \citep{sovinec_nonlinear_2004} and the non-ideal global spectral shooting method developed in Section \ref{sec:methods}. First, we present NIMROD simulations across a range of magnetic Reynolds numbers (Rm) and Lundquist number (S) to map dominant instabilities and provide benchmarks (Sec. \ref{sec:NIMROD}). Subsequently, we introduce ``spectral diagrams" constructed using the shooting method to visualize regions of instability and mode dominance (Sec. \ref{sec:phase}). The influence of flow profile variations, including shear, vorticity, and its gradient, on mode structure, onset conditions, and the competition between MRI and MCI is examined (Sec. \ref{sec:flow}). Finally, the impact of varying domain geometry on mode scaling and dominance is explored using both NIMROD and spectral scans in Section \ref{sec:AR}.

\subsection{Resistive Scaling - NIMROD vs Non-ideal Shooting} \label{sec:NIMROD}
We begin by examining the non-axisymmetric ($m = 1$) stability landscape using linear initial-value NIMROD simulations for both purely azimuthal and purely vertical magnetic field configurations in the AR1 geometry (Fig. \ref{fig:Ar-Config}, Table \ref{table:AR0}), consistent with previous ideal studies \citep{ebrahimi_nonlocal_2022}. These simulations span across magnetic Reynolds number (Rm = $r_1^2\Omega_0/\eta$) and Lundquist number (S $= r_1V_A/\eta$), holding the Prandtl number fixed at Pm $= \nu/\eta = 1$, representing regimes with the largest cumulative non-ideal contributions. The primary goals are to map the dominant modes across the parameter space, identify transitions between MRI and MCI, and generate benchmark data for validating the non-ideal global approximations discussed in Sec. \ref{sec:qr}. Ideal spectral solutions derived from Eq. \ref{eqn:shooting} in the limit $\eta,\nu \rightarrow0$ are also computed for various vertical wavenumber ($k_z$) and serve as a baseline for understanding the mode behavior (shown as curves in Fig. \ref{fig:NIMROD-Comp}). We then extend the spectral solutions to finite Rm/Pm and provide the tools for determining the non-ideal onset of instability.

Two primary types of transitions between dominant modes are observed as Rm and the Lehnert number ($B_0 = V_A/(r_1\Omega_0)$) (proportional to $S/\sqrt{\mathrm{Rm}}$ for fixed $r_1, \Omega_0$) are varied (Fig. \ref{fig:NIMROD-Comp}). First, a transition driven by magnetic field strength can occur if the ideal growth rates of MRI and MCI modes cross. At low $V_A$, localized MRI modes may dominate, while at higher $V_A$, global MCI modes become most unstable. This type of transition is evident in the vertical field case (Fig. \ref{fig:NIMROD-Comp}a) but not observed for the ideal modes in the azimuthal field case within the parameter range studied (Fig. \ref{fig:NIMROD-Comp}c). Second, a transition driven by diffusivity occurs due to the different scaling of MRI and MCI growth rates with Rm (see Sec. \ref{sec:res-pot}). Even if MRI is dominant in the ideal limit or at high Rm, the system is expected to transition to MCI dominance as Rm is decreased, provided MCI is unstable. NIMROD simulations confirm this transition for both field configurations (Figs. \ref{fig:NIMROD-Comp}a,c). These transitions manifest in the simulations as abrupt changes in the mode's frequency and spatial structure, shifting from localized, high-$k_z$, high-frequency characteristics (MRI) to global, low-$k_z$, low-frequency features (MCI), as illustrated by the inset mode structures and frequency plots (Figs. \ref{fig:NIMROD-Comp}b,d).

The existence of the location of these transitions depends sensitively on the specific flow profile and domain geometry (explored further in Secs. \ref{sec:flow}, \ref{sec:AR}). However, comparing the ideal mode growth rates provides a practical predictive framework. An ideal field-driven transition exists if ideal MRI and MCI growth rates cross. If instead the ideal MRI growth rate envelope (is always larger than) MCI, no field-driven transition occurs. However, a diffusivity-driven transition at low Rm to MCI may exist, provided the MCI is unstable. Conversely, if MCI modes envelope MRI, there will exist no transitions (diffusivity-driven or field-driven) and MCI dominates throughout (see Sec. \ref{sec:AR}).

We now proceed to demonstrate that the non-ideal global shooting method, incorporating the approximation for diffusive terms, successfully captures these complex dynamics observed in NIMROD. This validation sets the stage for using the computationally efficient shooting method to construct comprehensive ``spectral diagrams" for predicting instability boundaries and dominant mode characteristics.

\begin{figure*}
    \centering
    \includegraphics[width=\textwidth]{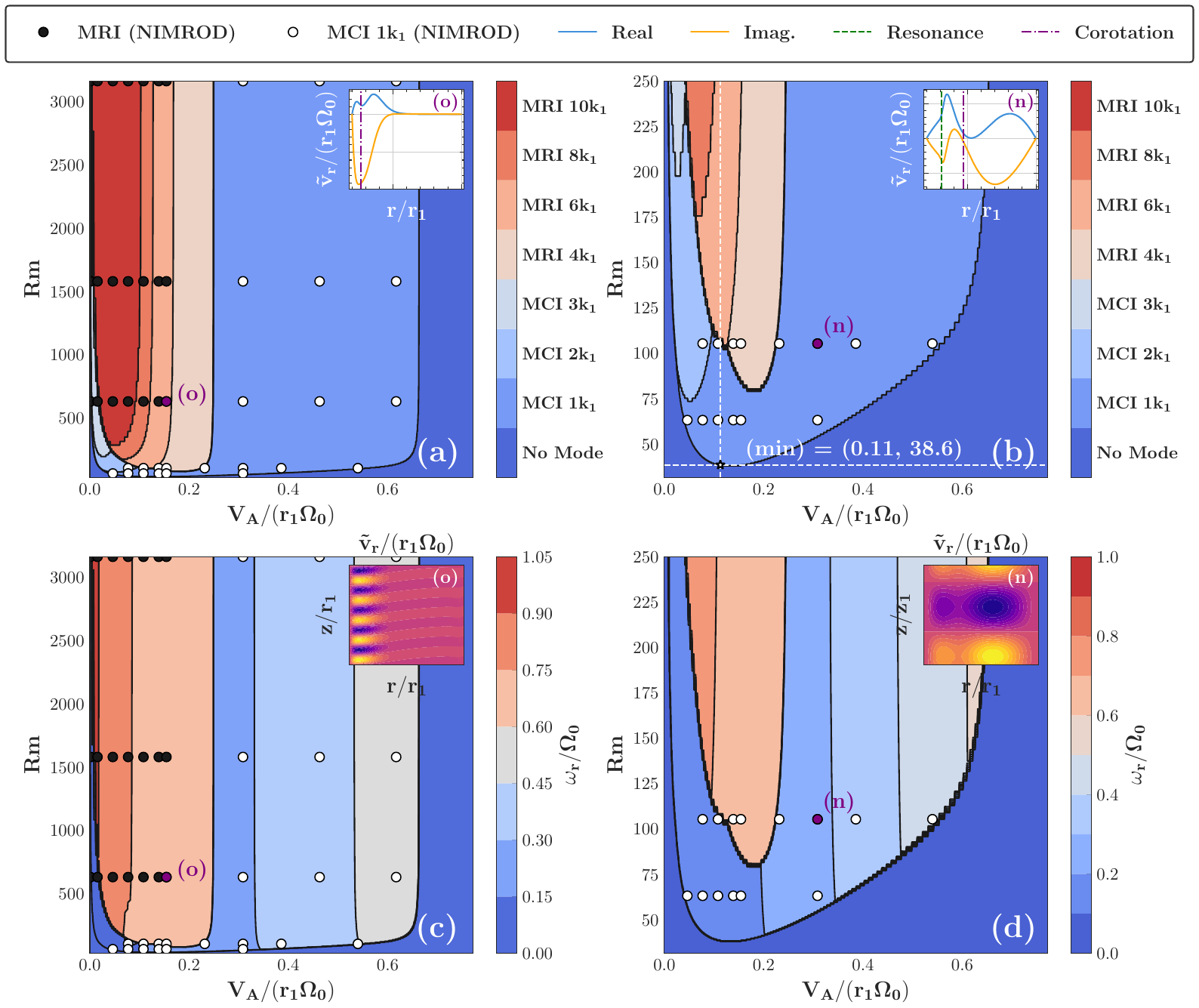}
 \caption{Vertical field dominant mode scans over Rm and $V_A/(r_1\Omega_0)$. (a) shows growth rate results for high Rm, and (b) shows low Rm. At high Rm and low fields, high $k_z$ MRI modes dominate, whereas at low Rm and high fields, low $k_z$ MCI modes dominate. (c) illustrates frequencies $(\omega_r)$ at given Rm and $V_A$ for the dominant mode, and (d) shows the same relationship for low Rm. All eigenvalue analysis was done via the resistive global shooting method for non-axisymmetric ($m = 1$) modes with $k_1 = \pi/4$, $k_r^2(r) = \pi^2/r^2$ (MWKB), and $r_2 - r_1 = 4r_1$ (AR1). We show even $k_z$ for MRI up to $k_z = 10k_1$ for simplicity, however, more modes do coexist and are seen via NIMROD simulations. Both shooting and NIMROD were tested for Pm $= 1$ and the Keplerian profile. Inset plots show mode structure from the non-ideal global shooting method (a,c), and NIMROD simulations (b,d).}
\label{fig:res-shooting-phase}
\end{figure*}

To assess the accuracy of the non-ideal global shooting method and the underlying TWKB and MWKB approximation (Sec. \ref{sec:qr}), we compare their predictions against the NIMROD simulation results across varying parameters. Since individual MRI modes ($k_z \geq 1k_1$) often dominate only in narrow regions of parameter space (see Sec. \ref{sec:phase}), the MCI $1k_1$ mode, which typically occupies a larger domain and represents the longest radial wavelength (posing the strongest for the short-wave approximations), is used for detailed benchmarking. Figure \ref{fig:WKB-Qmagpi-Shooting} compares the growth rates computed using both approximations against NIMROD data for the MCI $1k_1$ mode at Pm $= 1$.

For the vertical field configuration (Figs. \ref{fig:WKB-Qmagpi-Shooting}a,b), both approximations reproduce the NIMROD growth rates reasonably well at low and intermediate field strengths ($V_A/(r_1\Omega_0$), where modes tend to be more localized. At high field strengths, where the MCI modes become strongly global, deviations appear: TWKB tends to under-damp (overestimate $\gamma$), while MWKB tends to over-damp (underestimate $\gamma$) compared to NIMROD. However, since the instability onset occurs at intermediate field strengths, both methods generally capture the onset parameters accurately.

For the azimuthal field configuration (Figs. \ref{fig:WKB-Qmagpi-Shooting}c,d), both approximations show good agreement with NIMROD across the entire range of field strengths tested, even for strongly global modes. Further investigation is needed to determine if this superior performance for an azimuthal field is a general feature.

Given that the MCI mode represents a limit case for the short-wavelength approximations, the reasonable agreement here suggests that the dynamics of the generally more localized MRI modes should also be adequately captured by either method. As discussed in Section \ref{sec:qr}, the optimal choice between TWKB and MWKB can depend on the domain configuration. Overall, these results validate the non-ideal global shooting method as a reliable tool for exploring instability characteristics.

\subsubsection{Spectral diagrams and the Instability Onset}\label{sec:phase} 
A significant advantage of the spectral shooting method is its ability to efficiently compute the entire spectrum of unstable modes (including subdominant ones) for any given set of parameters. This capability allows for the construction of ``spectral diagrams," analogous to phase diagrams, which map the character of the dominant instability across parameter space. Such diagrams provide valuable insights into the stability landscape, mode competition, and instability onset conditions.

Figure \ref{fig:res-shooting-phase} presents an example spectral diagram for the vertical field case (AR1 geometry, Keplerian flow, Pm $= 1$), computed using the MWKB approximation. The diagram plots the dominant mode type (identified by its vertical wavenumber $k_z$) and its corresponding frequency $\omega_r$ as a function of Rm and Lehnert number ($V_A/(r_1\Omega_0)$). The shooting results (colored regions) are overlaid with points indicating the dominant mode identified in corresponding NIMROD simulations. There is excellent agreement between the two methods regarding the region of dominance: high-$k_z$, high frequency MRI modes prevail at high Rm and low $V_A$, while the low-$k_z$, low frequency MCI mode dominates at low Rm and high $V_A$. The discontinuous jumps in frequency associated with transitions between these regimes are captured. Furthermore, the mode structures computed by the shooting method match those observed in NIMROD (inset plots).

This analysis corroborates the finding from Section \ref{sec:res-pot} that the instability onset (the threshold for any mode to become unstable as Rm or $V_A$ is increased from zero) typically occurs via the global, low-frequency MCI $1k_1$ mode for configurations with sufficient curvature. The spectral diagram method provides a direct visualization of this onset boundary.

These diagrams also have potential implications for the experimental identification of modes. For instance, if Rm can be varied experimentally (e.g., by changing temperature or rotating rate $\Omega_0$), a discontinuous jump in the observed frequency could signify a transition from MCI dominance near onset to MRI dominance at higher Rm, provided the ideal MRI is unstable under those conditions.

The spectral diagram approach is a versatile tool. We employ it further in Section \ref{sec:flow} to investigate how flow profiles modify instability domains, mode characteristics, and onset criteria, demonstrating its utility for systematic parameter exploration relevant to astrophysical and laboratory plasmas.

\subsection{Flow Configuration} \label{sec:flow}

\begin{figure*}
\centering
\includegraphics[width=2\columnwidth]{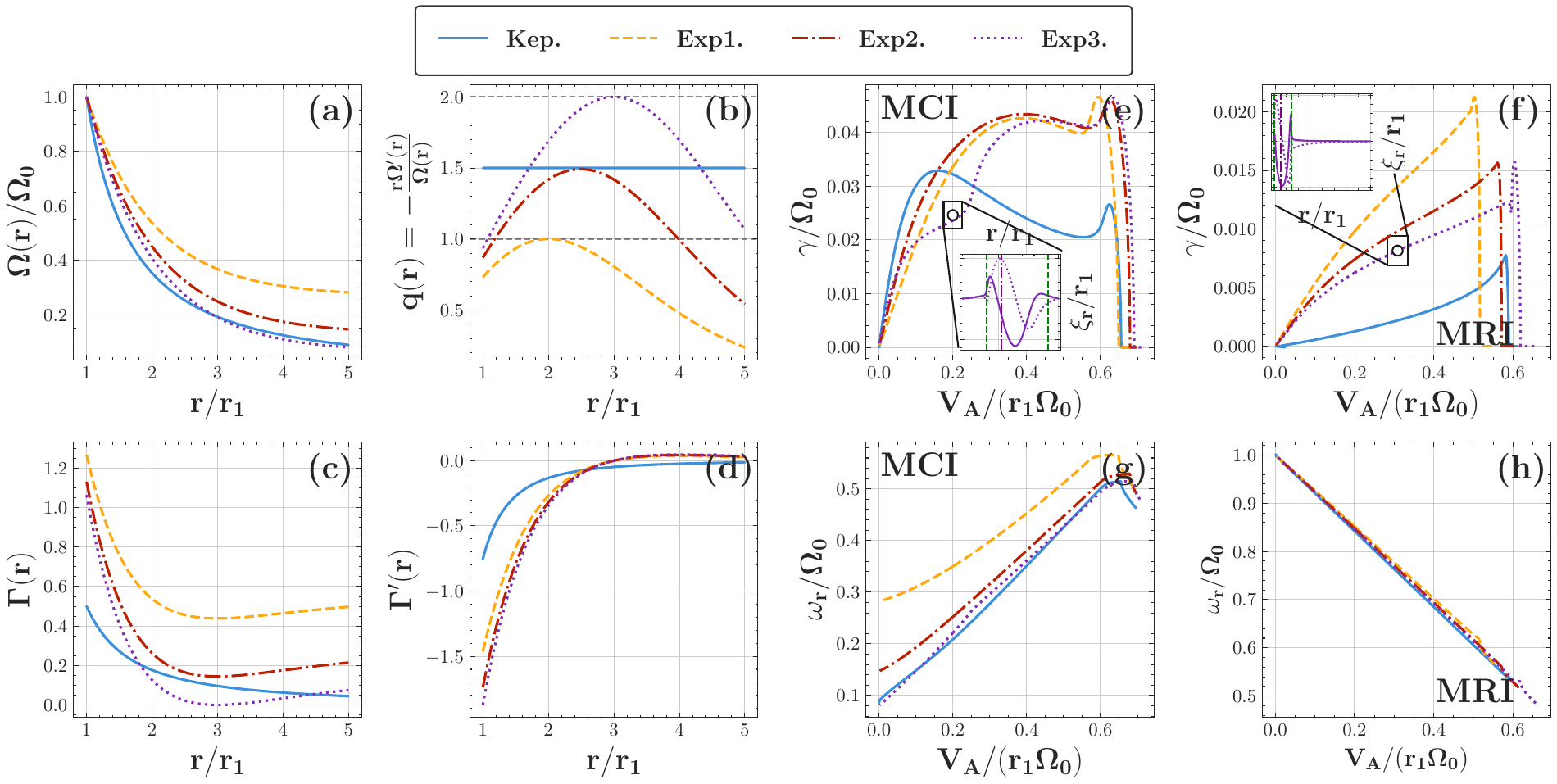}

  \caption{Flow rate configuration (a,b,c,d), growth rates (e,f), and frequencies (g,h) for the Keplerian and Exponential flow configuration. We define the Exponential profile as $\Omega(r)/\Omega_0 = a\mathrm{exp}(1-r/r_1) + (1-a)$. Exp1: $a = 0.7312$; Exp2: $a = 0.8689$; Exp3: $a = 0.9366$. All shooting (e,f,g,h) was conducted for the non-axisymmetric ($m = 1$), $k_z = 1k_1$ ($k_1 = \pi/4)$ modes with a purely vertical magnetic field. We define Exp1 such that Max($q(r)$) $= 1,$ Exp2 such that Max($q(r)$ $ = 3/2$, and Exp3 such that Max($q(r)$ $ = 2$. The exponential profile can be tuned for Max($q(r)$ $ = 5$ ($a = 1$). However, we chose a maxima of $q(r) = 2$ to include only MHD perturbations. While the MRI modes are uniformly damped by increased relative flow curvature, the MCI modes persist at large vorticities.}
  \label{fig:vorticity-configuration}
\end{figure*}

\begin{figure}
    \centering
    \includegraphics[width=1\linewidth]{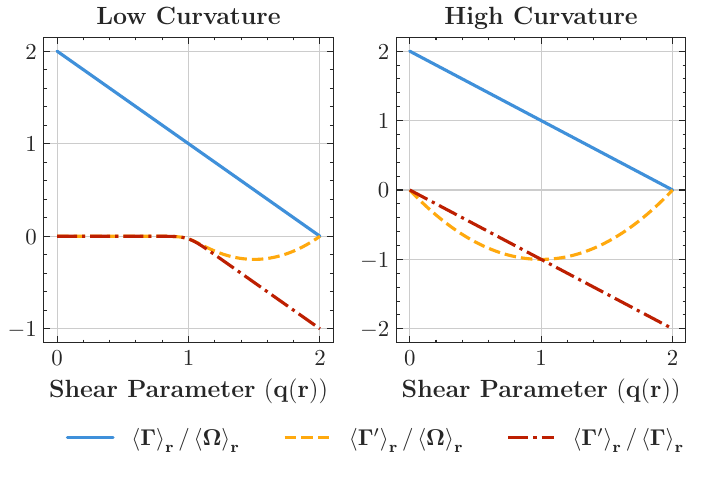}
    \caption{Scaling of relative vorticity ($\left<\Gamma\right>_r/\left<\Omega\right>_r$), relative vorticity gradient ($\left<\Gamma'\right>_r/\left<\Omega\right>_r$), and relative flow curvature ($\left<\Gamma'\right>_r/\left<\Gamma\right>_r$) as curvature of the domain changes. Low Curvature corresponds to $r_2/r_1 \rightarrow \infty$, whereas High Curvature corresponds to $r_2/r_1 \rightarrow 1$. Scaling emphasizes that vorticity gradients become an increasingly important contribution in highly curved domains.}
    \label{fig:vorticity-scaling}
\end{figure}

\begin{figure*}
\centering
\includegraphics[width=1.9\columnwidth]{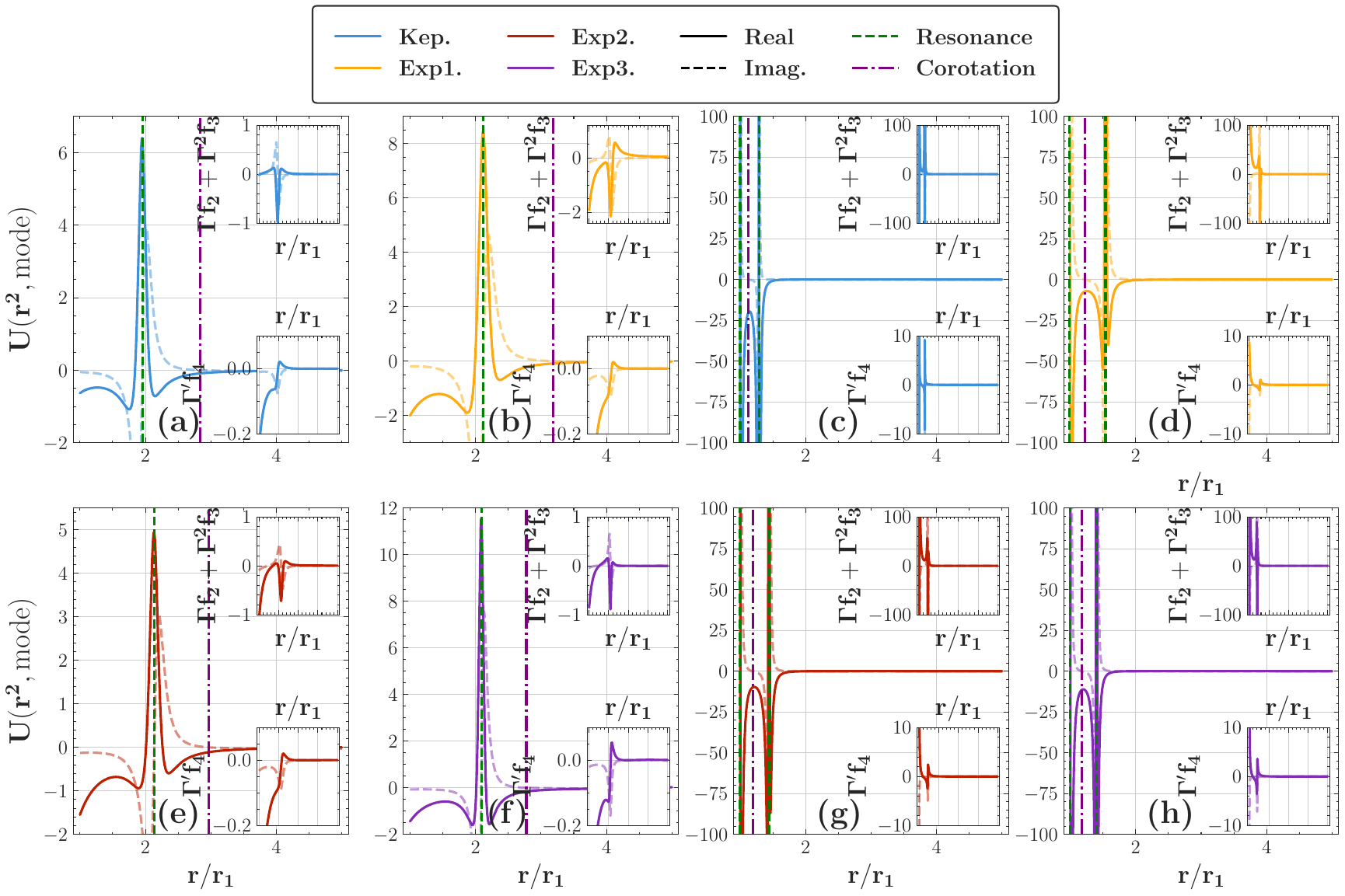}
 
\caption{Vertical field potentials for MCI $1k_1$ (a,b,e,f) and MRI $1k_1$ (c,d,g,h) modes at $V_A/(r_1\Omega_0) = 0.2$ for four flow configurations (Kep (a,c), Exp1 (b,d), Exp2 (e,g), and Exp3 (f,h)). Upper right inset figures show vorticity ($\Gamma$) contribution to potential, and lower right inset figures show vorticity gradient ($\Gamma'$) contribution.}
  \label{fig:Vorticity-Potential}
\end{figure*}

\begin{figure*}
\centering
 \begin{subfigure}[b]{0.48\columnwidth}
     \includegraphics[width=\textwidth]{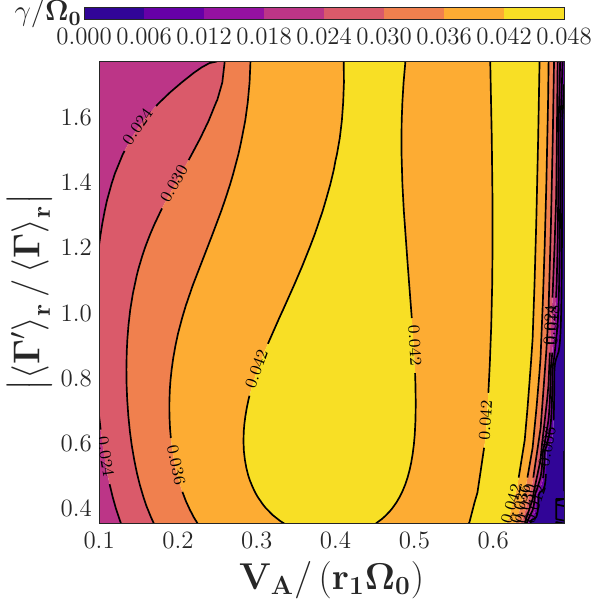}
     \caption{MCI, Relative Flow Curvature, $q'(r)\neq0$}
 \end{subfigure}
 \hfill
  \begin{subfigure}[b]{.48\columnwidth}
     \includegraphics[width=\textwidth]{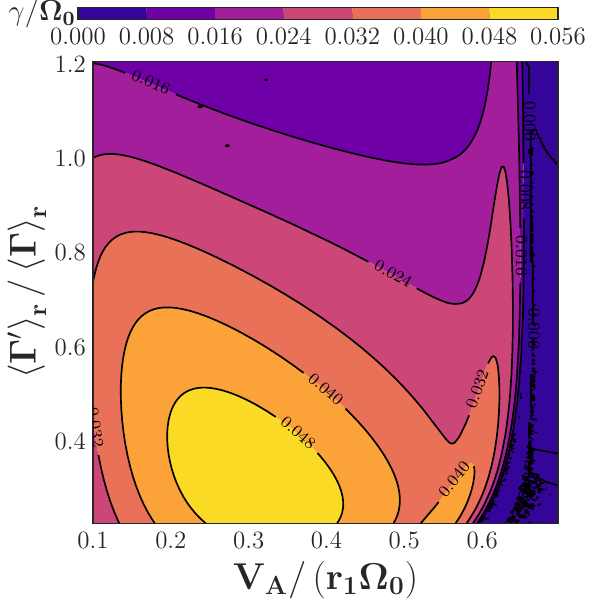}
     \caption{MCI, Relative Flow Curvature, $q'(r)=0$}
 \end{subfigure}
 \hfill
  \begin{subfigure}[b]{.48\columnwidth}
     \includegraphics[width=\textwidth]{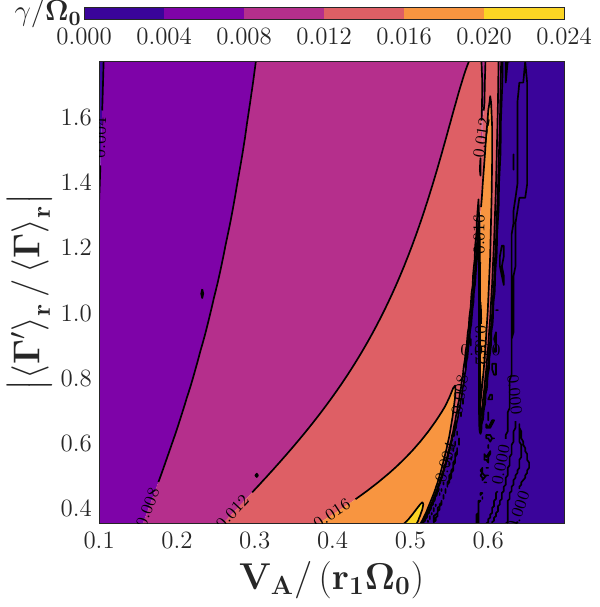}
     \caption{MRI, Relative Flow Curvature, $q'(r)\neq0$}
 \end{subfigure}
 \hfill
  \begin{subfigure}[b]{.48\columnwidth}
     \includegraphics[width=\textwidth]{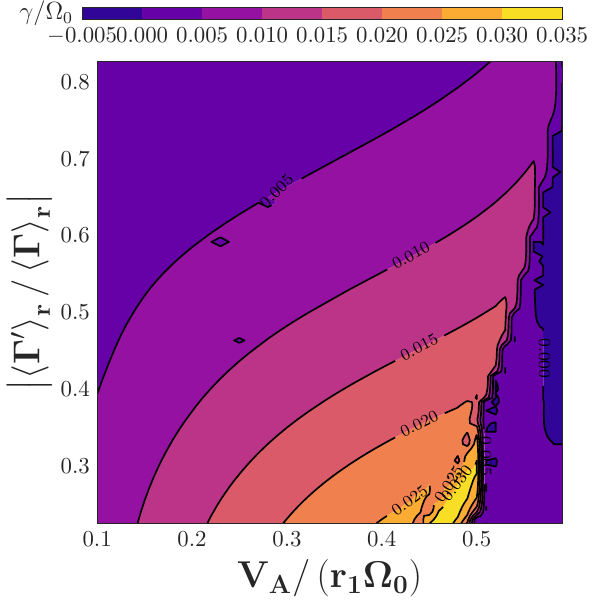}
     \caption{MRI, Relative Flow Curvature, $q'(r)=0$}
 \end{subfigure}

 \begin{subfigure}[b]{.458\columnwidth}
     \includegraphics[width=\textwidth]{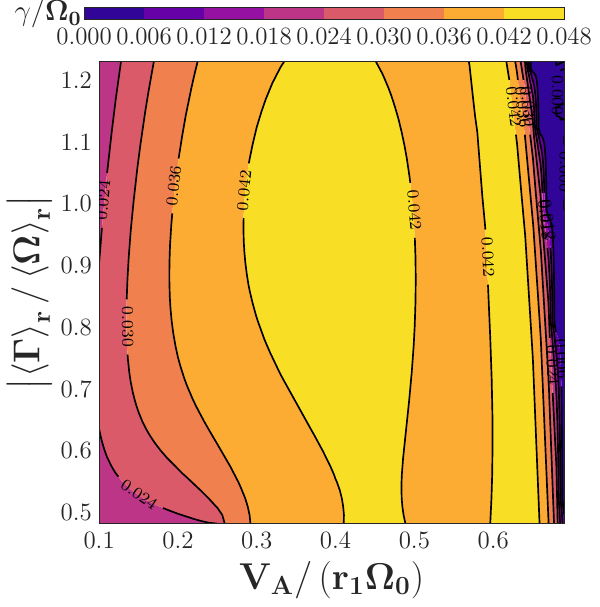}
     \caption{MCI, Relative Vorticity, $q'(r) \neq 0$}
 \end{subfigure}
 \hfill
  \begin{subfigure}[b]{.48\columnwidth}
     \includegraphics[width=\textwidth]{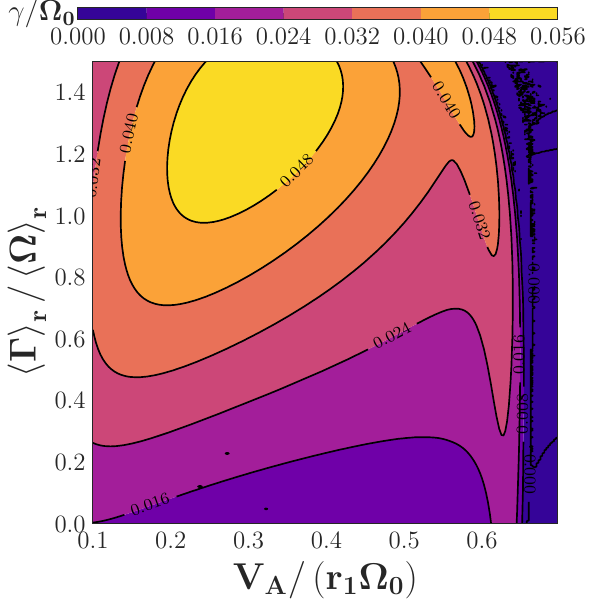}
     \caption{MCI, Relative Vorticity, $q'(r)=0$}
 \end{subfigure}
  \hfill
  \begin{subfigure}[b]{.48\columnwidth}
     \includegraphics[width=\textwidth]{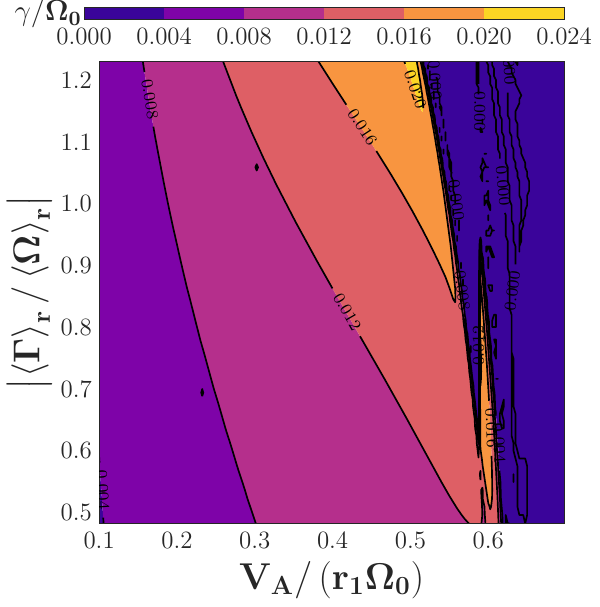}
     \caption{MRI, Relative Vorticity, $q'(r)\neq0$}
 \end{subfigure}
  \hfill
  \begin{subfigure}[b]{.48\columnwidth}
     \includegraphics[width=\textwidth]{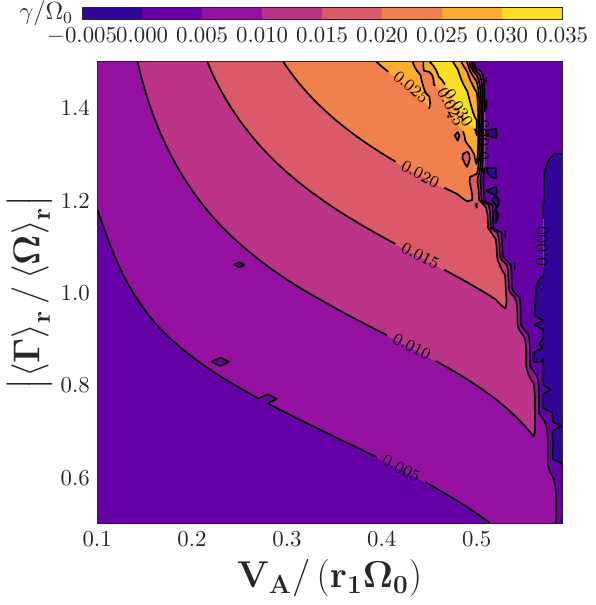}
     \caption{MRI, Relative Vorticity, $q'(r)=0$}
 \end{subfigure}
\caption{Growth Rate scan of most unstable non-axisymmetric (MCI (a), MRI (b)) modes for $k_z = 1k_1$ as relative flow curvature ($R_\Gamma$) and magnetic field strength are varied for a purely vertical field. In (a), (c), (e), and (g), we utilize the Exp. profile ($\Omega(r)/\Omega_0 = a\mathrm{exp}(1-r/r_1) + (1-a)$) with non-uniform $q(r)$ ($q'(r) \neq0$), and vary $\left<\Gamma'\right>_r/\left<\Gamma\right>_r$ and $\left<\Gamma\right>_r/\left<\Omega\right>_r$ by changing $a$. We span from the Exp1 to the Exp3 profile to only study the dynamics of MHD perturbations. Similarly, in (b)(d)(f)(h), we consider we utilize profiles with uniform $q(r)$ ($\Omega(r)/\Omega_0 = 1/r^q$, spanning between $q = 1/2$ to $q = 3/2$ (Keplerian). MCI modes remain dominant for both small vorticity and large vorticity gradients, with the domain of instability narrowing at large relative flow curvatures. Conversely, the MRI modes are destabilized by vorticity gradients and remain unstable only for low relative flow curvatures.}
  \label{fig:Vorticity-Scan}
\end{figure*}

\begin{figure*}
\centering
\includegraphics[width=1.8\columnwidth]{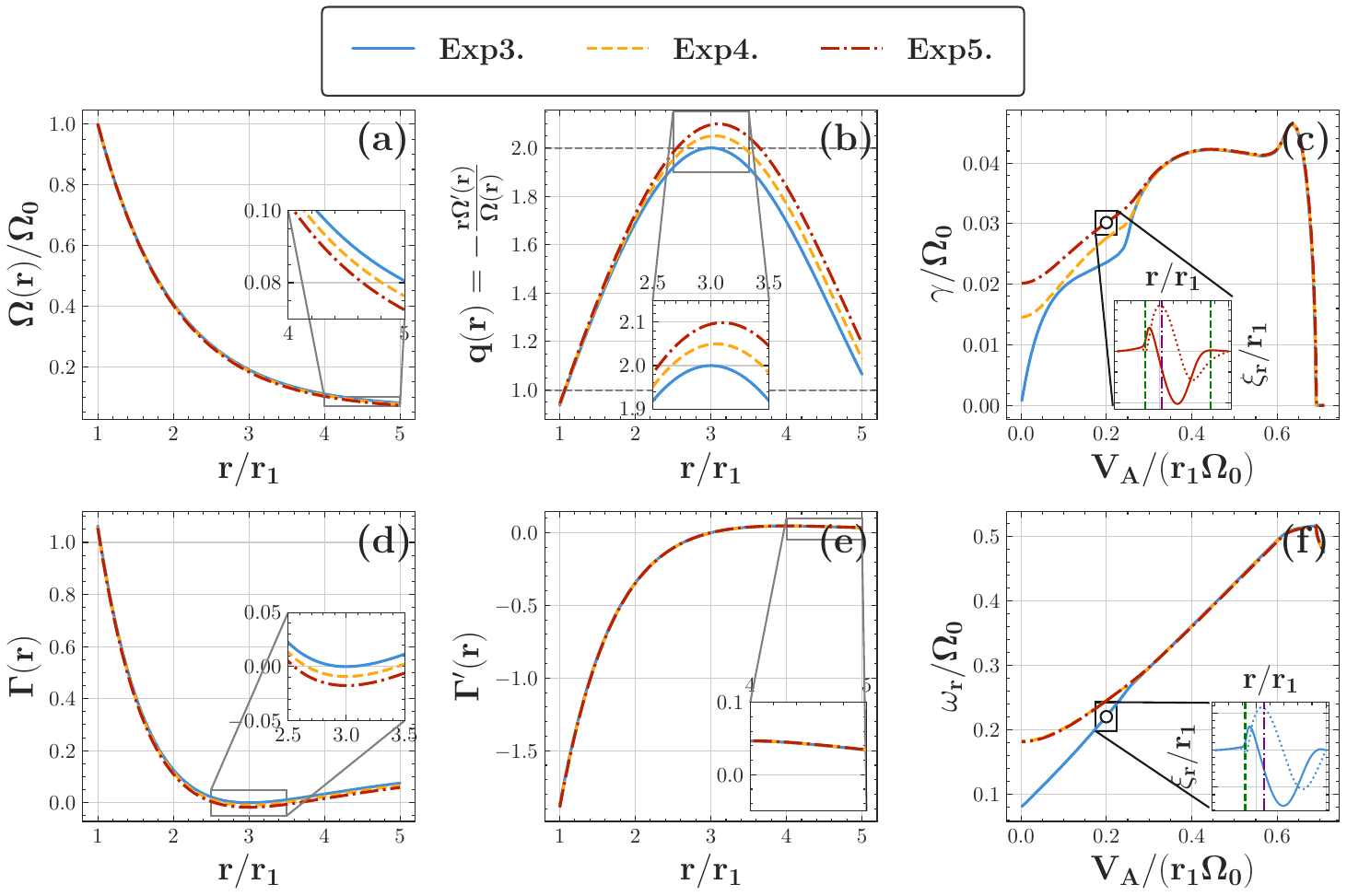}

  \caption{Flow rate configuration (a,b,d,e), growth rates (c), and frequencies (f) for the Exponential flow configuration as $q(r)$ crosses two shows bifurcation of MCI mode into hydrodynamic branch. We define the Exponential profile as $\Omega(r)/\Omega_0 = a\mathrm{exp}(1-r/r_1) + (1-a)$. Exp3: $a = 0.9366$; Exp4: $a = 0.9409$; Exp5: $a = 0.9448$. All shooting (c,f) was conducted for the non-axisymmetric ($m = 1$), $k_z = 1k_1$ ($k_1 = \pi/4)$ modes with a purely vertical magnetic field.}
  \label{fig:hyd1-configuration}
\end{figure*}

\begin{figure*}
\centering
\includegraphics[width=1.8\columnwidth]{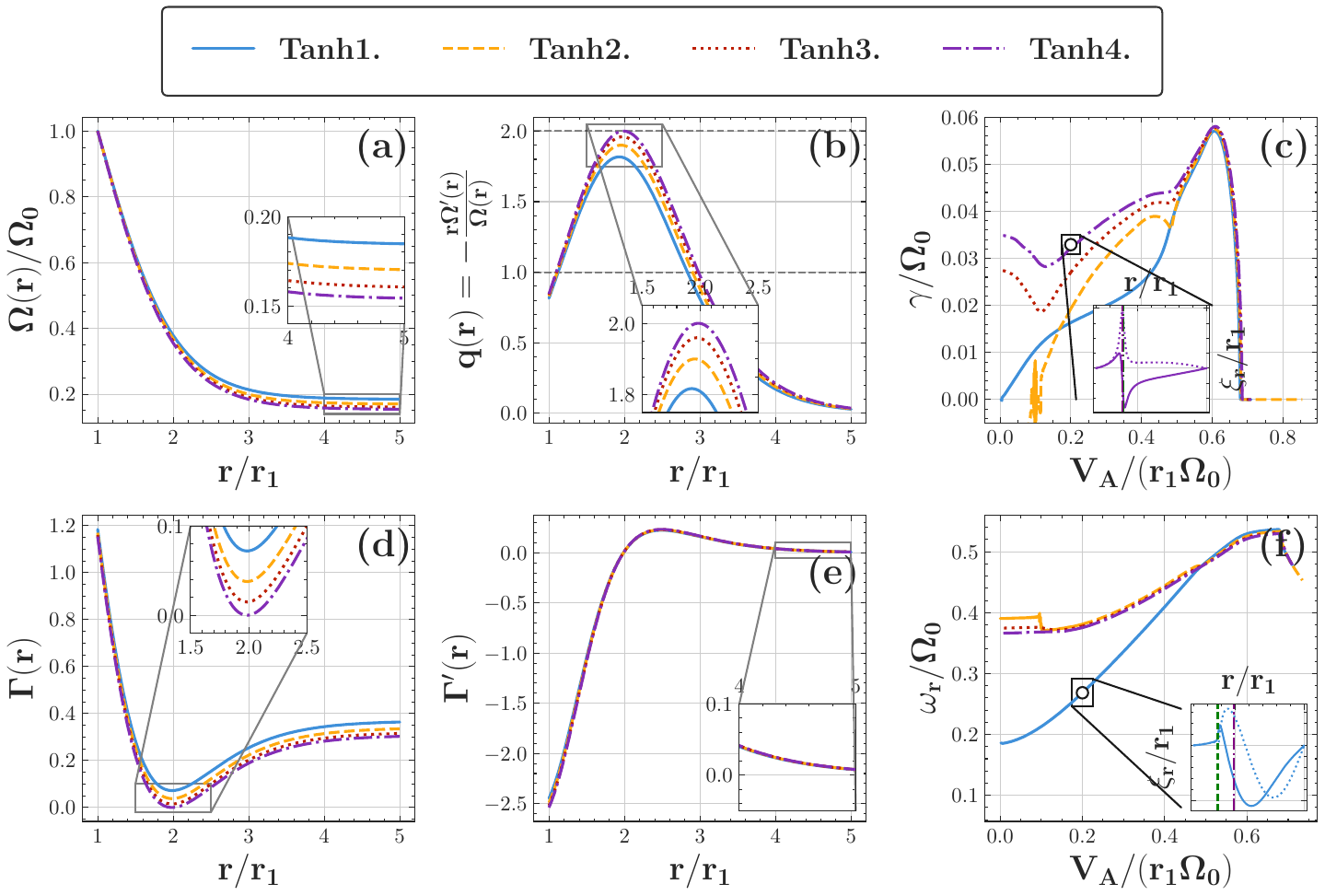}

  \caption{Flow rate configuration (a,b,d,e), growth rates (c), and frequencies (f) for the Tanh flow configuration as a vorticity well creates a hydrodynamic branch for $q < 2$. We define the Tanh profile as $\Omega(r)/\Omega_0 = a\mathrm{exp}(1-r/r_1) + 1$. Tanh1: $a = 0.8158$; Tanh2: $a = 0.8303$; Tanh3: $a = 0.8400$; Tanh4: $a = 0.8463$. All shooting (c,f) was conducted for the non-axisymmetric ($m = 1$), $k_z = 1k_1$ ($k_1 = \pi/4)$ modes with a purely vertical magnetic field.}
  \label{fig:hyd2-configuration}
\end{figure*}

While Keplerian flow ($\Omega(r) \propto 1/r^{3/2}$) is a standard benchmark motivated by astrophysical context and its stability to axisymmetric hydrodynamic perturbations (per the Rayleigh criterion), realistic flows can deviate significantly. This section investigates how variations in the background flow profile $\Omega(r)$, specifically altering the vorticity $\Gamma = (1/r)d(r^2\Omega)/dr$ and its radial gradient $\Gamma'$, influence the stability, structure, and competition of non-axisymmetric ($m = 1$) MHD modes, particularly MRI and MCI. We explore regimes both stable and unstable to hydrodynamic (HD) perturbations, including non-axisymmetric instabilities that can arise when the flow is Rayleigh-stable ($q(r) = -\frac{r\Omega'}{\Omega} < 2$). We demonstrate that MHD onset conditions are sensitive to the flow profile, although the global MCI $1k_1$ mode typically remains the first to become unstable. Furthermore, we analyze the bifurcation of the MCI mode into a hydrodynamic branch in HD-unstable flows and develop a criterion to predict the MHD onset threshold based upon resonance dynamics.

\subsubsection{Vorticity and the Flow Curvature Effect} \label{sec:vorticity}
We now investigate the influence of vorticity, $\Gamma(r) = (\nabla \times\mathbf{v}_0)_z = (1/r)d(r^2\Omega)/dr$, and its radial gradient $\Gamma'$ on MHD stability. By introducing three additional profiles with tunable shear and vorticity (Exp1-3, see Fig. \ref{fig:vorticity-configuration}), we reveal that both MRI and MCI modes are significantly more unstable under these flows than the Keplerian profile, even in configurations with $q(r)$ close to the Keplerian profile (cf. Kep and Exp3). To understand the underlying mechanisms, we recast the effective potential formalism (see Sec. \ref{sec:res-pot}) for a purely vertical field, explicitly separating terms involving $\Gamma, \Gamma'$ to demonstrate their individual destabilizing effects:

\begin{equation}
    U(r^2, \mathrm{mode}) = f_1 + \Gamma f_2 + \Gamma^2 f_3 + \Gamma' f_4
\end{equation}
with,
\begin{widetext}
\begin{equation} \label{eqn:hyd-pot}
\begin{split}
    f_1 =   \frac{k_z^2r^2+m^2-1}{4r^4} + \frac{k_z^2}{4r^2}\left(\frac{k_z^2r^2-2m^2}{k_z^2r^2+m^2}\right)  - \frac{\Omega^2\left(k_z^2+m^2 \right)\omega_A^2}{r^4\left( \omega_A^2-\bar{\omega}^2\right)^2}  + \frac{m\Omega\bar{\omega}}{r^4\left(\omega_A^2-\bar{\omega}^2\right)}\left(\frac{2k_z^2r^2+m^2}{k_z^2r^2+m^2}\right)  
    \\ - \frac{\Omega\bar{\omega}}{r^4}\left(\frac{k_z^2r^2\left(m^2+1\right) + m^4}{k_z^2r^2+m^2}\right), \; \; 
    f_2 = \frac{m^2\bar{\omega}^2\Omega}{r^4\left(\omega_A^2-\bar{\omega}^2\right)^2} + \frac{\Omega\left(k_z^2r^2+2m^2\right)}{2r^4\left(\omega_A^2-\bar{\omega}^2\right)} - \frac{m\bar{\omega}k_z^2}{2r^2\left(k_z^2r^2+m^2\right)\left(\omega_A^2-\bar{\omega}^2\right)}, \\  \; \; 
    f_3 = -\frac{m^2\omega_A^2}{4r^4\left(\omega_A^2-\bar{\omega}^2\right)^2}, \; \mathrm{and} \; f_4 = \frac{m\bar{\omega}}{4r^3\left(\omega_A^2-\bar{\omega}^2\right)} 
\end{split}
\end{equation}
\end{widetext}
The coefficients $f_1, f_2, f_3, f_4$ depend on the system parameters and the mode's growth rate/frequency ($\gamma, \omega_r$). The term $f_1$ contains contributions independent of vorticity and its gradient, including curvature and flow energy. Meanwhile, $f_2,f_3$ capture the linear and quadratic vorticity effects, whereas $f_4$ captures the effect of $\Gamma'$.

Figure \ref{fig:Vorticity-Potential} illustrates the decomposition (Eq. \ref{eqn:hyd-pot}) for representative MCI and MRI modes using Keplerian and exponential flows (Exp1-Exp3, defined in Fig. \ref{fig:vorticity-configuration}). While the non-vorticity term $f_1$ generally provides the largest contribution to the potential well/barrier structure, the terms involving $\Gamma$ ($\Gamma f_2 + \Gamma^2 f_3$) and $\Gamma'$ ($\Gamma' f_4$) can significantly modify the potential, influencing mode localization and stability. The sign and magnitude of these contributions depend intricately on the mode type (via $\omega$, $\bar{\omega}$), field strength (via $\omega_A$), and flow properties ($\Omega, \Gamma, \Gamma'$).

To gain qualitative insight, we consider the expected signs of $\bar{\omega}$ and ($\omega_A^2-\bar{\omega}^2$) in different regimes (Table \ref{table:flow-scaling}). For typical low-frequency MCI modes ($\bar{\omega} < 0$) at low field strength ($\omega_A^2<\bar{\omega}^2$), both $f_2, f_3$ terms (related to $\Gamma$) and the $f_4$ term (related to $\Gamma'$) tend to b e positive (assuming $\Gamma > 0$, $\Gamma' < 0$ as in typical disk profiles), contributing to potential barriers (stabilizing). Conversely, for typical high-frequency MRI modes ($\bar{\omega} > 0$) at low field, these terms tend to be negative, deepening the potential well (destabilizing). The denominators change sign at intermediate fields where $\bar{\omega}^2 \approx \omega_A^2$, reversing some trends. At high fields ($\omega_A^2 > \bar{\omega}^2$), the low-field characteristics are reversed for MRI and MCI, respectively, as mode frequency uniformly decreases/increases for MRI/MCI. Therefore, at high fields ($\omega_A^2 > \bar{\omega}^2$), MCI modes ($\bar{\omega} > 0$) tend to be stabilized by $\Gamma$, but destabilized by $\Gamma'$. Conversely, MRI modes at high fields ($\bar{\omega} < 0$) tend to be destabilized by $\Gamma$ and stabilized by $\Gamma'$. These heuristics align with the low-field examples in Figure \ref{fig:Vorticity-Potential}.

\begin{table}[!h]
\begin{center}
\caption{Scaling of field-dependent MCI and MRI terms to show scaling of vorticity (gradient) terms with different field strengths. (C) and (R) refers to MCI and MRI modes, respectively. Meanwhile, (1) and (-1) refer to destabilizing and confining, respectively. Finally, we consider three field strengths: Low, Intermediate (Interm.), and High.}
\label{table:flow-scaling}
\centering
\begin{tabular}{ p{1.2cm} p{1cm} p{2cm} p{2cm} p{1cm}}
 \hline
 \hline
Field Strength & $\left<\bar{\omega}\right>_r$ & $\left<\omega_A^2-\bar{\omega}^2\right>_r $  & $\left<\Gamma f_2 + \Gamma^2 f_3\right>_r$ & $\left<\Gamma' f_4\right>_r$\\
 \hline
    Low (C/R)  & -1/1    & -1 & -1/1 & -1/1 \\
    Interm. (C/R)  & -1/1    & 1/-1 & 1 & 1\\
    High (C/R)  & 1/-1   & 1 & 1/-1 & -1/1 \\
 \hline
\end{tabular} 
\end{center}
\end{table}

Further insight comes from estimating vorticity and vorticity gradient scaling characteristics by expanding arbitrary flow configurations as power law profiles: $\Omega(r)/\Omega_0 = 1/r^{q}$. Vorticity and its gradient can then be expanded as, 
\begin{equation} \label{eqn:vort-qscale}
    \Gamma = \frac{\left(2-q\right)\Omega_0}{r^q}, \; \Gamma' = -q\frac{\left(2-q\right)\Omega_0}{r^{q+1}}
\end{equation}
We can utilize these scalings to construct the relative vorticity $\left<\Gamma\right>_r/\left<\Omega\right>_r$, relative vorticity gradient $\left<\Gamma'\right>_r/\left<\Omega\right>_r$, and relative flow curvature $\left<\Gamma'\right>_r/\left<\Gamma\right>_r$, which describe the scaling of each contribution as the underlying flow configuration is varied:
\begin{equation}
\begin{split}
    \frac{\left<\Gamma\right>_r}{\left<\Omega\right>_r} = 2-q, \; \; 
    \frac{\left<\Gamma'\right>_r}{\left<\Omega\right>_r} = (2-q)(1-q)\frac{r_2^{-q}-r_1^{-q}}{r_2^{1-q}-r_1^{1-q}}, \\
    \frac{\left<\Gamma'\right>_r}{\left<\Gamma\right>_r} = (1-q)\frac{r_2^{-q}-r_1^{-q}}{r_2^{1-q}-r_1^{1-q}}
\end{split}
\end{equation}

The relative vorticity indicates that configurations with lower shear (smaller $q$) will have higher vorticity contributions to confinement of the mode. Importantly, the scaling with vorticity is a domain-independent quantity. Conversely, the relative vorticity gradient and relative flow curvature are intrinsically domain-dependent quantities, as they are inherently curvature-dependent terms. This is because the definition of vorticity gradient inherently considers the curvature of the flow configuration ($\Gamma'\propto \nabla^2_r\Omega$ in cylindrical coordinates). We can tabulate these scaling relations taking into account domain configuration (see Fig. \ref{fig:vorticity-scaling}). We find that the contribution of vorticity gradients and relative flow curvature is maximized in domains with large curvature ($r_2/r_1\rightarrow 1$), and minimized in domains with low curvature ($r_2/r_1\rightarrow\infty$). This suggests that the destabilizing effect of $\Gamma'$ on intermediate-field MCI might be enhanced in strongly curved geometries, potentially contributing to MCI dominance in such domains (cf. Sec. \ref{sec:AR}). We include the relative vorticity gradient and relative flow curvature to decouple $\Gamma, \Gamma'$ scaling when varying parameters underlying the flow configuration. 

To verify these scalings numerically, we performed ideal spectral scans using the shooting method, varying the flow profile using both exponential profiles (non-uniform $q$) and power-law profiles (uniform $q$) (see Fig. \ref{fig:Vorticity-Scan}). The results broadly confirm the theoretical expectations. We confirm that MCI/MRI growth rates generally increase with higher relative vorticity, and are suppressed by relative vorticity gradients. Typically, MRI modes scale relatively uniformly with variation of parameters, whether the underlying flow has uniform or non-uniform $q(r)$ (cf. Figs \ref{fig:Vorticity-Scan}c,g and Figs.\ref{fig:Vorticity-Scan}d,h). However, MCI is much more sensitive to the uniformity of the profile (cf. Figs \ref{fig:Vorticity-Scan}a,e and Figs.\ref{fig:Vorticity-Scan}b,f). Since the non-uniformity in the profile $q(r)$ can be equivalently described as flow curvature, this reinforces the concept that MCI is a curvature-driven instability, whereas MRI is relatively agnostic due to its local nature. We thus argue that profiles with non-uniform $q$ will generally be more unstable, as onset will still occur at the MCI mode (see Sec. \ref{sec:onset}). These findings emphasize the sensitivity of MHD stability to the detailed flow structure. Investigating more complex profiles, such as those that exhibit local minima in the vorticity (as observed in recent simulations \citep{wang_observation_2024}), is a promising direction of future work.

\subsubsection{Hydrodynamic Modes} \label{sec:hyd}
When the background flow is unstable to hydrodynamic (HD) perturbations, MHD modes can interact and merge with HD modes. The MCI branch, in particular, smoothly connects to an HD instability branch as the magnetic field strength approaches zero ($V_A\rightarrow 0$) if the flow permits HD instability. Though we typically consider HD stable flows, we will demonstrate that these hybrid HD-MCI modes exhibit a resonance-localized mode structure, and given a vorticity well, can occur at shear parameters below the Rayleigh criterion.

We begin by studying configurations unstable to axisymmetric hydrodynamic perturbations. Utilizing Equation \ref{eqn:vort-qscale}, we see that as $q$ crosses two at some point $r^*$ in the domain, $\Gamma$ changes sign at $r^*$. From the potential, we see at this location  $r^*$, vorticity switches signs. At this location, we also have a switching of sign in $\Gamma'$, which acts as a restoring force to confine hydrodynamic modes. It is generally true, however, that any power law profile will not have $\Gamma$ crossing zero. We thus must consider profiles with non-uniform $q(r)$, such as profiles generated via the tanh or exponential functions, or those considered by \cite{ebrahimi_generalized_2025}.

In Figure \ref{fig:hyd1-configuration}, we perturb the HD stable Exp3 profile to form the Exp4 and Exp5 profiles, which are both HD unstable. We find that marginally stable profiles (e.g., Exp3) will exhibit knees in the profile that bifurcate upon perturbation of $q$. We observe this behavior in the Exp4 and Exp5 profiles, which become increasingly hydrodynamically unstable as $q(r)$ increases. It was also recently observed that profiles with vorticity wells can exhibit hydrodynamic instability for $q(r) < 2$ as shown by \cite{ebrahimi_generalized_2025} and illustrated here using the Tanh profiles (see Fig. \ref{fig:hyd2-configuration}). Physically, $\Gamma'$ acts to confine the modes near the vorticity minimum, thus creating HD instability given a sufficiently deep well. Again, the spectral results show the low-field MCI branch bifurcating and connecting to the $V_A/(r_1\Omega_0) = 0$ HD instability branch (Tanh3, Tanh4 in Fig. \ref{fig:hyd2-configuration}).

A consequence of bifurcation is that in HD-MHD unstable flows, the low-field MCI modes will tend to be resonance localized, inheriting the localization properties of the HD mode to which it connects. This is because hydrodynamic modes are localized about the Corotation point (the point in the regime where $\bar{\omega} = 0$) (see inset of Fig. \ref{fig:hyd2-configuration}c). In the absence of magnetic fields, there are no Alfvénic resonances. Since the resonance condition is continuous, the resonances must bifurcate from the Corotation point as the magnetic field strength becomes non-zero. We will utilize this bifurcation argument in Section \ref{sec:onset} to provide a tool to calculate MHD onset in the presence of HD instability. We can observe such a resonance localized MCI mode illustrated in the inset figure of Figure \ref{fig:hyd1-configuration}c. 

\subsubsection{MHD Onset - Critical Rm} \label{sec:onset}

\begin{figure*}
    \centering
    \begin{subfigure}[b]{1\columnwidth}
        \centering
        \includegraphics[width=\textwidth]{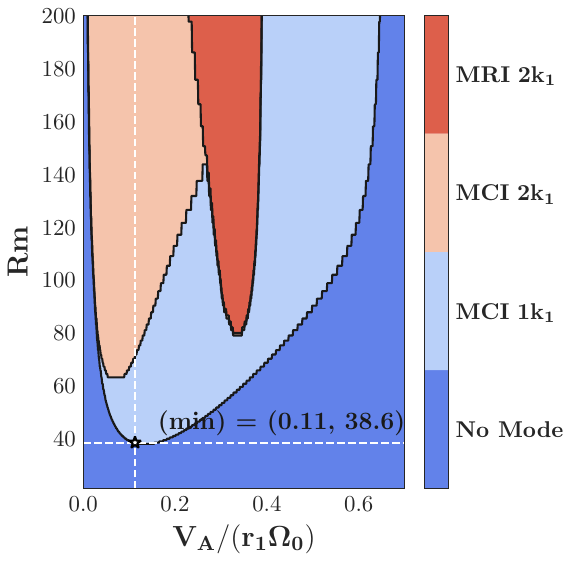}
     \caption{}
    \end{subfigure}
    \hfill
    \begin{subfigure}[b]{1\columnwidth}
        \centering
        \includegraphics[width=\textwidth]{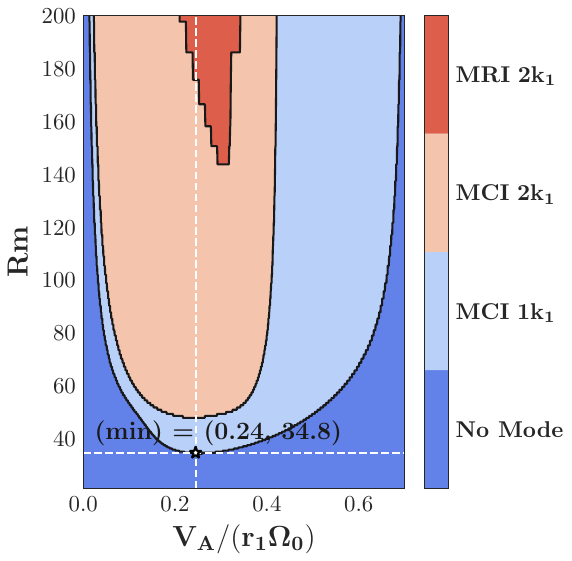}
     \caption{}
    \end{subfigure}
\caption{Vertical field dominant mode at low Rm for Kep (a) and Exp3 (b) determined via global resistive shooting method. All eigenvalue analysis was done for non-axisymmetric ($m = 1$) modes with $k_1 = \pi/4$, $k_r^2(r) = $ MWKB, and $r_2 - r_1 = 4r_1$ (AR1). We show that at low Rm, only low $k_z$ ($\leq 2k_1$) MRI and MCI modes are unstable. Shooting is run with Pm $= 1$. Moreover, both the onset phase boundary and onset parameters vary between configurations, yet both onsets occur at MCI $1k_1$.}  \label{fig:phase-res-flow}

\end{figure*}

The critical parameters (e.g., minimum Rm and $V_A$) for the onset of MHD instability depend on the background flow configuration. Using the spectral diagram method (see Sec. \ref{sec:phase}), we compare the instability boundaries at low Rm for the Keplerian flow versus the Exp3 profile (Fig. \ref{fig:phase-res-flow}). While both profiles are hydrodynamically stable and exhibit onset via the MCI $1k_1$ mode, the location and shape of the onset boundary in the (Rm-$V_A$) plane differ significantly, underscoring the sensitivity of the instability domains to flow configuration.

 In flows that are already hydrodynamically unstable (when $V_A = 0$, $\gamma > 0$), the concept of MHD onset requires careful definition. As shown in Sec. \ref{sec:hyd}, the MCI mode bifurcates into an HD branch as magnetic fields vanish. We propose viewing the emergence of MHD-HD dynamics as a bifurcation process involving the Alfvénic resonances. Alfvénic resonances must vanish in the HD limit, and only the Corotation point remains in the domain. For the Corotation point to be the only solution of the resonance condition (see Eq. \ref{eqn:resonance-condition}), the roots must be degenerate, such that:
$$\mathrm{Re}\left(\sqrt{4\omega_A^2-\left(\eta-\nu\right)^2Q^2} \right) = 0$$
This degeneracy is generally true without magnetic fields, irrespective of non-idealities. However, transition parameters $\eta, \nu$ should exist at finite fields that retain this degeneracy.

\begin{figure}
    \centering
    \includegraphics[width = 0.49\textwidth]{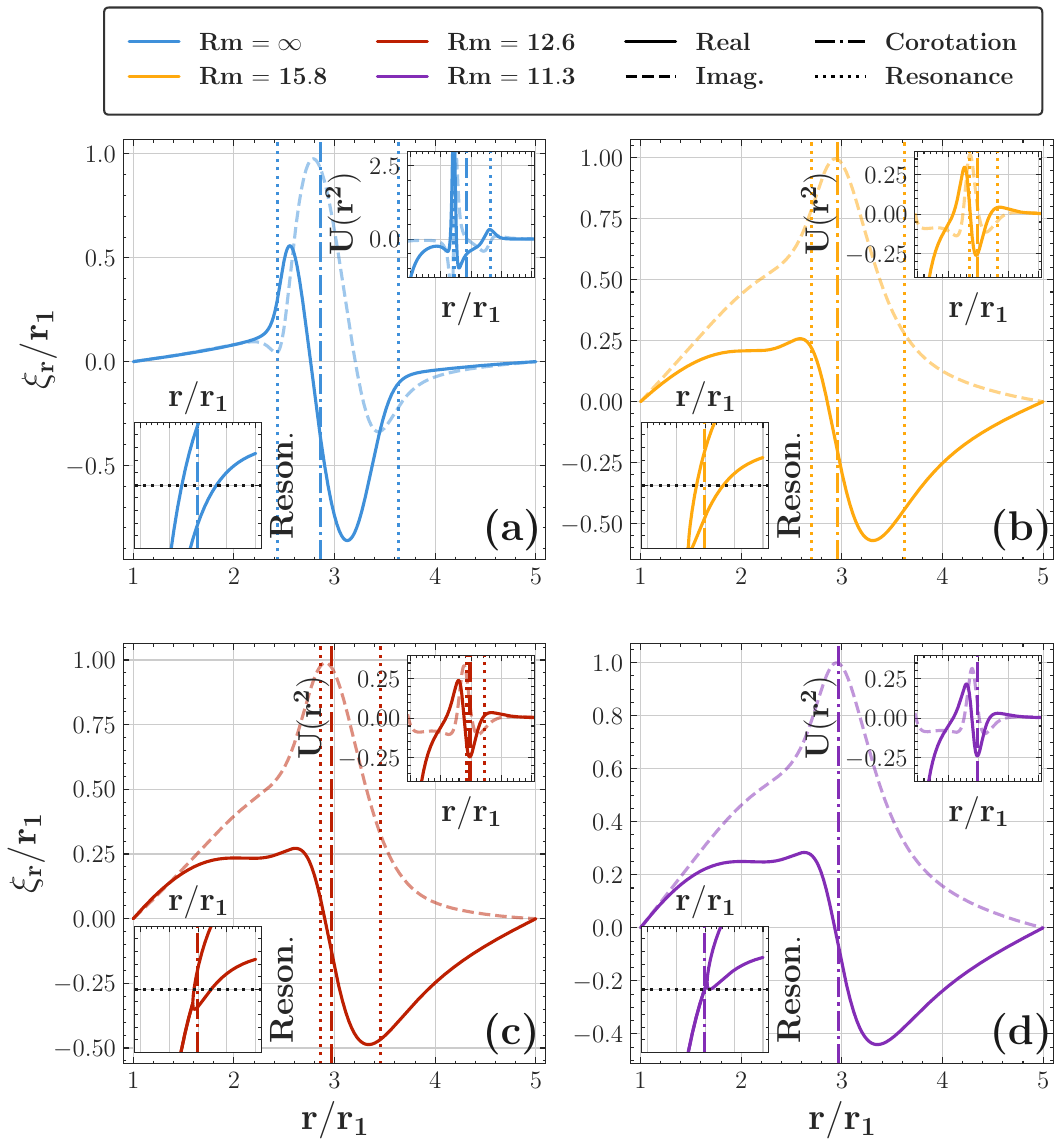}
 \caption{Vertical field shooting resonances for the Exp5 configuration shows onset of the MHD mode as Rm is increased (Rm $= 11.3$) for the $k_z = 1k_1$ $(k_1 = \pi/4)$ non-axisymmetric ($m = 1$) MCI mode at $V_A/(r_1\Omega_0) = 0.1$. Four resistivities are chosen to show that the Alfvénic resonances bifurcate from the Corotation point, with (a) Rm $= \infty$, (b) Rm $= 15.8$, (c) Rm $= 12.6$, and (d) Rm $= 11.3$ at $\mathrm{Pm} = 0$. The upper right inset plots show that the potentials move from those destabilized by the Alfvénic resonances to a well about the Corotation point. The lower left inset plots show the resonance condition (see Eq. \ref{eqn:resonance-condition}), plotting both branches and resonances attributed to the equation's roots. Equation \ref{eqn:MHD-HD-Onset} predicts an onset at $\mathrm{Rm}_H = 12.4$, whereas an onset at $\mathrm{Rm} = 11.3$ is observed via shooting.}
\label{fig:Res-Convergence}
\end{figure}

As $Q^2$ and $\omega_A^2$ are radially varying quantities, we must estimate the radial location where the resonance condition must degenerate. Since We estimate this location as the Corotation point $r_c \equiv \left\{r\in\left[r_1,r_2\right]\big | \omega_r-m\Omega(r) =0\right\}$, as the Alfvénic resonances converge about $r_c$. Without this choice, the onset criterion would be a flow and ideal MHD mode independent quantity ($\omega_r$). We know, however, from simulations \citep{wang_observation_2024} and Section \ref{sec:onset} that MHD onset is intrinsically a flow-dependent parameter. Thus, the choice of $r_c$ is fitting to evaluate the MHD onset, such that the onset criterion can be estimated as,

\begin{equation} \label{eqn:MHD-HD-Onset}
    \mathrm{Rm}_H = \frac{\left(1-\mathrm{Pm}\right)\left(k_r^2(r_c)+m^2/r_c^2 + k_z^2\right)}{2\omega_A(r_c)}.
\end{equation}

Figure \ref{fig:Res-Convergence} illustrates the bifurcation process for the HD-unstable Exp5 profile from direct global shooting. As Rm decreases towards the onset value found via shooting (Rm $\approx 11.3$), the Alfvénic resonances are shown to merge on the Corotation point $r_c$. The predicted onset from Eq. \ref{eqn:MHD-HD-Onset} using the ideal frequency is $\mathrm{Rm}_H \approx 12.4$, in reasonable agreement. The discrepancy arises from the shift in $\omega_r$ (and thus $r_c$) between the ideal limit and the onset point, though iteration via shooting could improve this prediction. This analysis provides a useful method for estimating MHD onset thresholds in the presence of HD instability, linking it directly to the fundamental resonance structure. Developing analogous resonance-based onset criteria for HD-stable flows remains an area of future investigation.

\subsection{Disk Configuration and space curvature} \label{sec:AR}

\begin{figure}

 \begin{subfigure}[b]{0.49\columnwidth}
     \includegraphics[width=\textwidth]{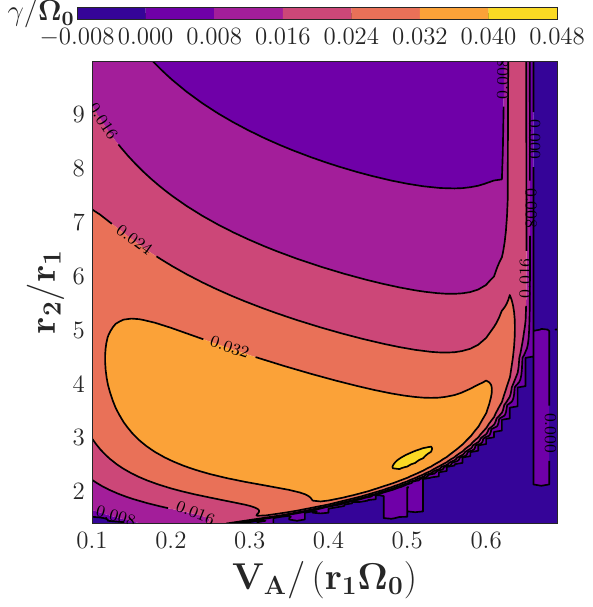}
     \caption{MCI, Curvature}
     \label{fig:c}
 \end{subfigure}
\hfill
 \begin{subfigure}[b]{0.49\columnwidth}
     \includegraphics[width=\textwidth]{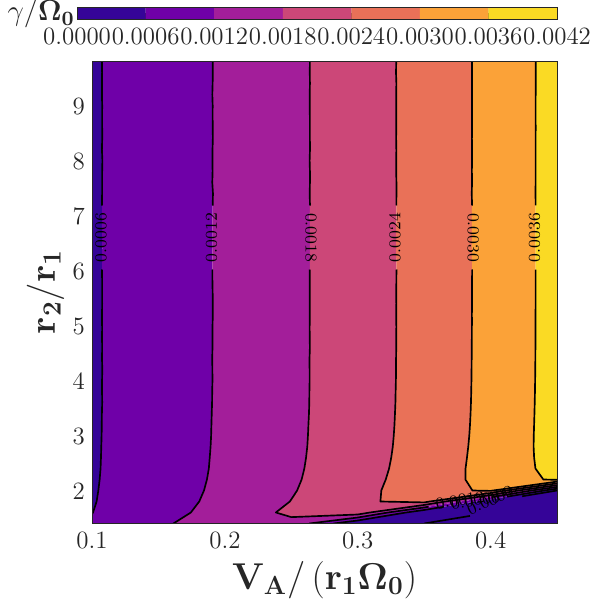}
     \caption{MRI, Curvature}
     \label{fig:c}
 \end{subfigure}
\hfill
 \begin{subfigure}[b]{0.49\columnwidth}
     \includegraphics[width=\textwidth]{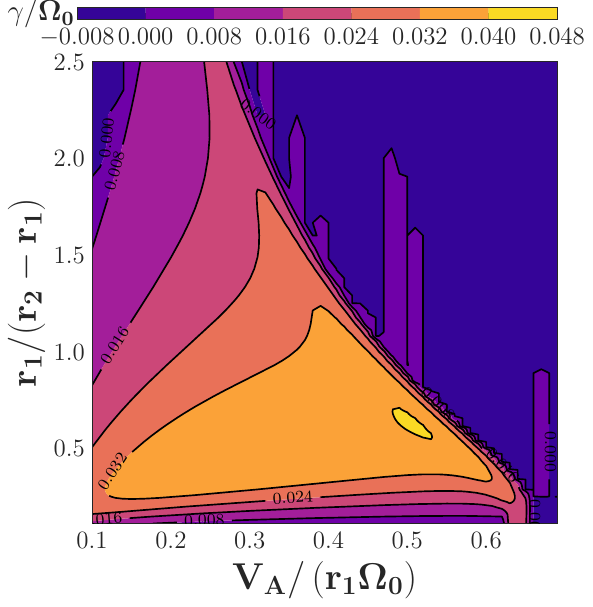}
     \caption{MCI, Aspect Ratio}
     \label{fig:c}
 \end{subfigure}
\hfill
 \begin{subfigure}[b]{0.49\columnwidth}
     \includegraphics[width=\textwidth]{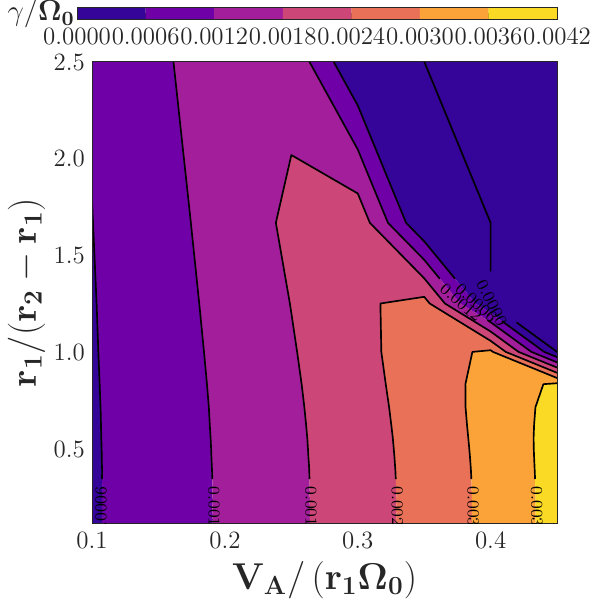}
     \caption{MRI, Aspect Ratio}
     \label{fig:c}
 \end{subfigure}

 \begin{subfigure}[b]{0.49\columnwidth}
     \includegraphics[width=\textwidth]{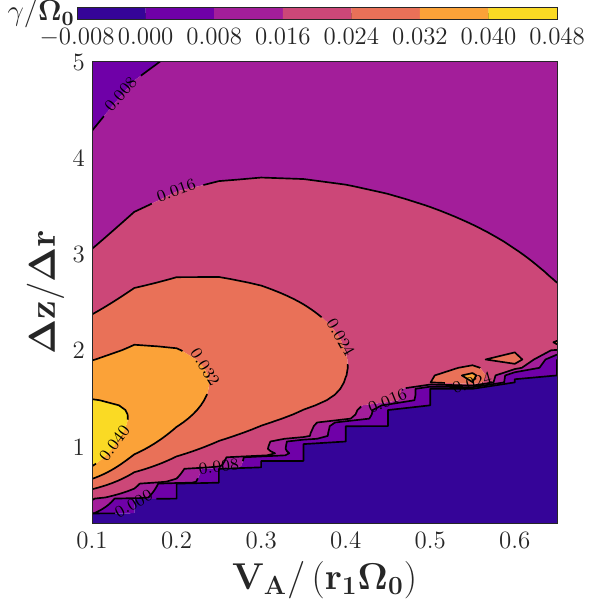}
     \caption{MCI, Relative Thickness}
     \label{fig:c}
 \end{subfigure}
\hfill
 \begin{subfigure}[b]{0.49\columnwidth}
     \includegraphics[width=\textwidth]{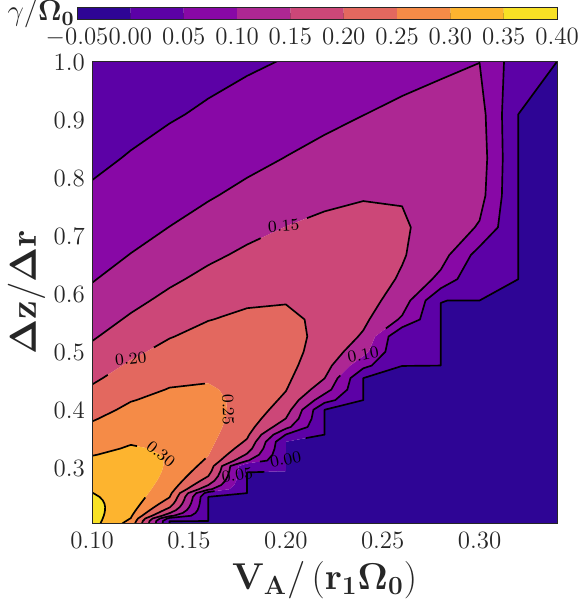}
     \caption{MRI, Relative Thickness}
     \label{fig:c}
 \end{subfigure}

  \caption{Variation of MCI (a,c,e) and MRI (b,d,f) mode growth rates with disk configuration from ideal global shooting method with the Keplerian profile. In (a)(b)(c)(d) we vary disk curvature and aspect ratio by tuning $r_2/r_1$ and $r_1/(r_2-r_1)$ respectively for $k_z/k_1 = 1$. As disk curvature increases ($r_2/r_1\rightarrow1$), the MCI mode becomes increasingly dominant and vice versa. In contrast, curvature has little effect on the growth rates of the MRI mode. In (e), (f) we vary disk thickness by tuning $\Delta z/\Delta r$ ($r_2/r_1 = 5)$. We find MCI modes dominate in intermediate disk thickness ($\Delta z/\Delta r\sim 1$), whereas MRI dominates in thin disk ($\Delta z/\Delta r \ll 1$).}   \label{fig:Disk-Scaling}
\end{figure}


The geometry of the differentially rotating plasma significantly influences MHD mode stability. This has already been hinted at in Section \ref{sec:vorticity}, which established that vorticity gradients $\Gamma'$ are intrinsically linked to the curvature of the flow profile (and hence the domain). Here, we investigate the distinct effects of the physical boundaries and dimensions, specifically the global curvature (related to $r_2/r_1$), the aspect ratio ($r_1/(r_2-r_1)$, and the relative disk thickness ($\Delta z/\Delta r$). Increased geometric curvature strongly favors the Magneto-Curvature instability, reinforcing its characterization as a ``curvature mode," while the MRI is less sensitive to global curvature. Furthermore, we will demonstrate MCI dominates in vertically moderate disks ($\Delta z/\Delta r \sim 1$), whereas MRI prevails in thin disks ($\Delta z/\Delta r \ll 1$). We explore these dependencies using the ideal global shooting method with a Keplerian profile ($\Omega(r) \propto 1/r^{3/2}$) and a purely vertical magnetic field, varying one geometric parameter at a time for MRI and MCI individually (see Fig. \ref{fig:Disk-Scaling}). We vary disk thickness, aspect ratio, and curvature for this span while keeping the inner radius fixed. 

First, we vary the ratio $r_2/r_1$ (implicitly varying the aspect ratio as well), effectively changing the global curvature (Fig. \ref{fig:Disk-Scaling}a,b). As $r_2/r_1$ decreases towards 1 (increasing curvature), the MCI growth rates significantly increase, dominating over MRI. This supports the notion that geometric curvature provides free energy for the MCI. At very high curvature ($r_2/r_1$ very close to 1), MCI growth rates slightly decrease, potentially due to boundary effects or reduced radial variation limiting access to flow energy. At low curvature, MCI modes approach stability ($r_2/r_1\rightarrow \infty$, which corroborates the expectation that MCI vanish in the Cartesian limit. In contrast, the MRI growth rates show only a weak dependence on $r_2/r_1$, consistent with its more localized nature.

Next, we vary aspect ratio (Fig. \ref{fig:Disk-Scaling}c,d). Smaller aspect ratios (larger radial gaps) generally allow modes access to a larger reservoir of free energy from the mean flow shear. Accordingly, MRI and MCI growth rates tend to increase with aspect ratio. However, unlike MRI, which uniformly favors smaller aspect ratios, MCI is also dominant for larger aspect ratios. This is because varying aspect ratio, while keeping $r_1$ fixed, also intrinsically varies spatial curvature. Thus, sufficiently small aspect ratios destabilize the MCI as curvature vanishes. Therefore, MCI dominates for intermediate aspect ratios, as there is a tradeoff between energy from curvature effects and accessing the energy from the mean flows.

Finally, we investigate the effect of disk thickness ($\Delta z/\Delta r$) for both MRI and MCI modes (see Fig. \ref{fig:Disk-Scaling}e,f) by varying $z_2,z_1$ while keeping $r_1,r_2$ fixed. We find that MRI is uniformly preferred in thinner disks ($\Delta z/\Delta r \ll 1$), whereas MCI is preferred in intermediate thickness disks ($\Delta z/\Delta r\sim 1$), while being stabilized for sufficiently thick disks $(\Delta z/\Delta r \gg 1$).

These ideal MHD results suggest that MCI is most dominant in geometrically moderately thick, highly curved domains with an intermediate aspect ratio (intermediate radial gap). Conversely, MRI is favored in thin, less curved domains with a small aspect ratio (large radial gap).

\begin{figure*}
\begin{subfigure}[b]{1\columnwidth}
     \includegraphics[width=\textwidth]{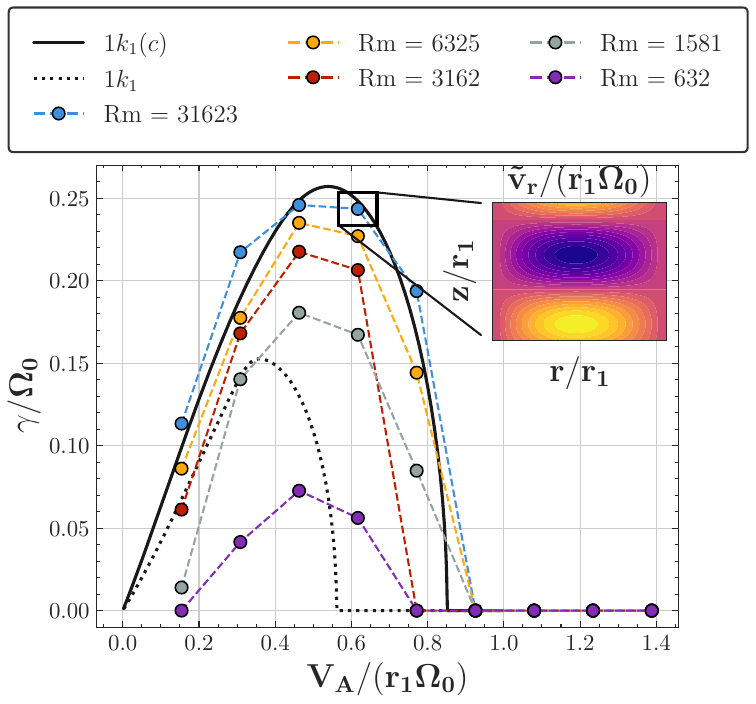}
     \caption{}
     \label{fig:c}
 \end{subfigure}
 \hfill
 \begin{subfigure}[b]{1\columnwidth}
     \includegraphics[width=\textwidth]{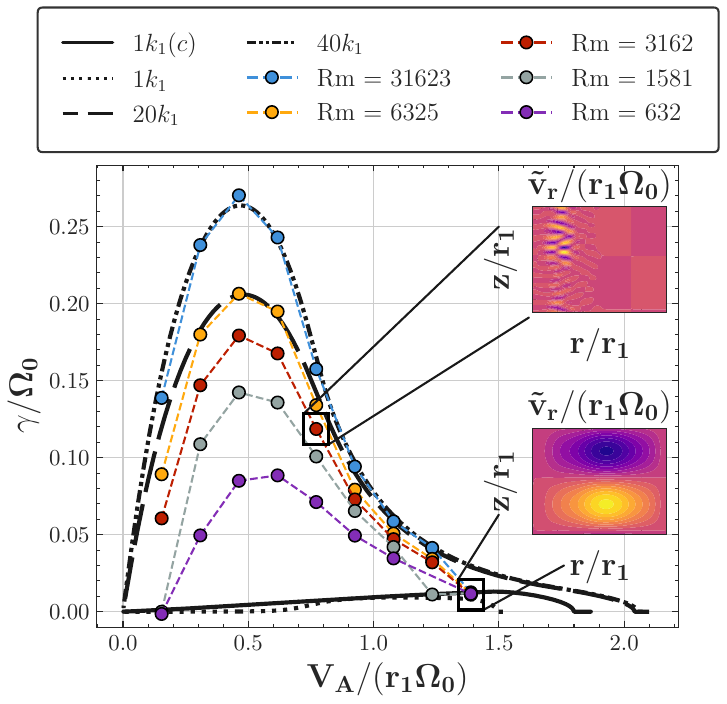}
     \caption{}
     \label{fig:c}
 \end{subfigure}
 
  \caption{Azimuthal field growth rates from both NIMROD simulations and the ideal global shooting method for the AR0 aspect ratio (a) and the AR2 aspect ratio (b). The AR0 aspect ratio has the global $1k_1$ MCI mode dominant for the entire regime due to the increased curvature of the domain. Conversely, the AR2 aspect ratio has the MRI mode dominant for more of the regime than previous simulations (AR1, see Fig. \ref{fig:NIMROD-Comp}) due to decreased curvature. Wave numbers ($k_z$) for shooting curves were determined by looking at the mode structure from NIMROD, but do not capture all observed wave numbers for MRI. Inset figures show mode structures from NIMROD simulations. Finally, definitions for the AR0, AR2 aspect ratios are given in Figure \ref{fig:Ar-Config}.}
  \label{fig:AR-Comp-NIMROD}
\end{figure*}

We corroborate these spectral results via a sweep of linear NIMROD simulations for two additional aspect ratios (AR0 \& AR2 with a purely azimuthal magnetic field. Figure \ref{fig:AR-Comp-NIMROD} compares growth rates from NIMROD with corresponding ideal shooting results (using dominant $k_z$ observed in NIMROD). Despite the change to an azimuthal field, we demonstrate that these scaling trends hold at finite resistivity: the high-curvature AR0 geometry shows MCI dominance across the entire range of $V_A$ simulated, while the low-curvature AR2 geometry exhibits a transition from MRI to MCI dominance at a significantly higher $V_A$ compared to the intermediate aspect ratio AR1 case (cf. Fig. \ref{fig:AR-Comp-NIMROD}). This confirms that increased geometric curvature robustly promotes MCI dominance.

These findings have important implications for laboratory experiments, which typically operate in high curvature, intermediate aspect ratio, and moderate thickness regimes. PPPL's MRI experiment, for example, operates in a regime with $r_2/r_1 = 3$, $r_1/(r_2-r_1) = 1/2$, and $\Delta z/\Delta r = 2$ (\cite{wang_identification_2022}, \cite{wang_observation_2022}, \cite{wang_observation_2024}). Our analysis indicates that such parameters strongly favor the MCI (low-frequency branch of non-axisymmetric modes) as the dominant instability, especially at moderate to strong magnetic fields. Observing MRI (high frequency localized branch) in such configurations might require accessing regimes of very high magnetic Reynolds number (Rm) at low magnetic field strengths, where MRI growth rates might exceed those of MCI. Identifying transitions between MCI and MRI could be achieved by observing discontinuous changes in the dominant mode frequency, provided the target mode (e.g., MRI) has a sufficiently large ideal growth rate in the relevant parameter space to overcome damping and competing modes. spectral diagrams, as constructed in Section \ref{sec:phase}, can help delineate the parameter regimes where different modes are expected to dominate.

\section{Summary and Conclusions} \label{sec:conc}

Understanding the linear stability of differentially rotating magnetized disks is crucial for explaining the drivers of turbulence and angular momentum transport in systems like accretion disks. We investigated the global linear dynamics of non-axisymmetric ($m = 1$) MHD instabilities in such disks, threaded with vertical or azimuthal magnetic fields, exploring the dependence on diffusivity, flow configuration, and disk geometry. Utilizing a combination of local WKB analysis, a non-ideal global spectral method (solved via shooting), and linear initial-value simulations (NIMROD), we established that the two primary $m = 1$ non-axisymmetric modes -- the Magneto-Rotational Instability (MRI) and the Magneto-Curvature Instability (MCI) -- dominate in different regimes of parameter space due to their distinct mode characteristics (structure, frequency, and configuration sensitivity). MRI modes are typically high-frequency, localized radially and vertically (high $k_z$), whereas MCI modes are low-frequency, global, and have low $k_z$. A key finding is that at low magnetic Reynolds number (Rm), the global structure of MCI allows it to persist. At the same time, MRI is preferentially stabilized due to the broadening of its structure around its Alfvénic resonances (see Fig. \ref{fig:Potential-NonIdeal}). Consequently, we argue that in disks with finite curvature, the onset of MHD instability typically occurs via the global MCI $1k_1$ mode (e.g., Fig. \ref{fig:res-shooting-phase}), consistent with recent simulation results showing onset via a global low-frequency mode \cite{wang_observation_2024}. The detailed dependence of this stability landscape, including onset conditions and mode competition, on flow profiles, field structure, and disk geometry was systematically explored using these methods and is summarized below.

Our analysis began with a local WKB approach (Sec. \ref{sec:local}), which indicated that non-ideal effects shift dominant instabilities towards larger spatial scales (lower $k_z$, $k_r$, $m = 1$) compared to the ideal limit. Recognizing the limitations of local analysis, we developed a non-ideal global stability method (Sec. \ref{sec:methods}). This involved deriving a 1D second-order ordinary differential equation (ODE) governing the radial mode structure by applying a targeted approximation to the diffusive terms \cite{zou_analysis_2020}, solved via shooting to find the eigenvalues and eigenfunctions. To understand free energy contributions, we extended the effective potential formalism \cite{ebrahimi_generalized_2025} to the non-ideal regime, revealing that diffusion broadens the potential structure around Alfvénic resonances (Sec. \ref{sec:res-pot}, Fig. \ref{fig:Potential-NonIdeal}). This broadening preferentially stabilizes the localized MRI, explaining the persistence of the global MCI at lower Rm.

The global spectral method was validated against linear initial-value NIMROD simulations (Sec. \ref{sec:NIMROD}). The MWKB (``Modified WKB") and TWKB (``Traditional WKB") approximations for the diffusive terms yielded reasonable agreement with NIMROD growth rates across various parameters, particularly capturing onset trends, stability boundaries, and mode characteristics (mode structure, frequencies, etc) despite some differences in growth rates when MCI is most global (Sec. \ref{sec:qr}, Figs. \ref{fig:NIMROD-error},\ref{fig:WKB-Qmagpi-Shooting}, and \ref{fig:res-shooting-phase} (insets)). Leveraging the efficiency of the spectral method, we introduced ``spectral diagrams" (Sec. \ref{sec:phase}), which map the dominant instability (largest growth rate) (MCI or MRI with specific $k_z$) across parameter space (e.g., Rm vs. $V_A$, Figs. \ref{fig:res-shooting-phase}/\ref{fig:phase-res-flow}). These diagrams successfully delineated regions of mode dominance consistent with NIMROD simulations and provide a valuable tool for predicting instability boundaries, onset parameters, and observable characteristics like mode frequency, or growth rate. Notably, they predict distinct frequency shifts associated with the transition between instability branches (MRI to MCI and vice versa).

Systematic parameter scans and analysis of the effective potential formalism (see reformulation in Eq. \ref{eqn:hyd-pot}) were conducted to clarify the influence of flow configuration -- specifically vorticity ($\Gamma$ and its radial gradient $\Gamma'$-- on MHD stability (Sec. \ref{sec:flow}). Both MRI and MCI exhibit complex dependency on vorticity ($\Gamma$) and its gradient ($\Gamma'$), as suggested by a heuristic analysis of the potential terms (summarized in Table \ref{table:flow-scaling}). At low field strengths ($\omega_A^2 \ll \bar{\omega}^2$), vorticity ($\Gamma$ terms) tend to stabilize MCI while destabilizing MRI; concurrently, vorticity gradients ($\Gamma'$ term) act to destabilize the MRI and confine the MCI. At intermediate field strengths ($\omega_A^2 \sim \bar{\omega}^2$), both vorticity and its gradient generally act to destabilize both modes. At high field strengths ($\omega_A^2 \gg \bar{\omega}^2$), vorticity tends to destabilize MCI while confining MRI; conversely, vorticity gradients tend to stabilize MCI while destabilizing MRI. These complex, field-dependent interactions highlight the sensitivity of mode competition and stability to the detailed structure of the flow profile beyond simple shear parameters. We perform spectral scans over profiles with non-uniform $q(r)$ (Exponential configuration, defined in Fig. \ref{fig:vorticity-configuration}) and uniform $q(r)$ (power-law profiles), finding that the observed scaling generally matches our heuristic analysis (see Fig. \ref{fig:Vorticity-Scan}). In this figure, it is also apparent that profiles with non-uniform $q(r)$ appear more resilient under perturbation of flow characteristics compared to the uniform $q(r)$ profiles, likely due to additional free energy contributions associated with vorticity gradients ($\Gamma'$) when $q'\neq0$. Nonetheless, we do find that flows with larger vorticity (and thus larger shear) will exhibit onset at lower Rm, and even a small perturbation in the profile can have a significant impact on the onset parameters (see Fig. \ref{fig:phase-res-flow}).

We also explored the interaction with hydrodynamic (HD) instability (Sec. \ref{sec:hyd}), finding that the MCI branch typically bifurcates into an HD mode at low magnetic fields when the flow is HD unstable (see Fig. \ref{fig:hyd1-configuration}). This includes non-axisymmetric HD instability driven by vorticity wells, even for flows stable by the axisymmetric Rayleigh criterion ($q(r) < 2$) (see Fig. \ref{fig:hyd2-configuration}. At finite fields, these hybrid HD-MCI modes exhibit a superposition of both global MCI and Corotation localized (points where the Doppler shifted frequency equals zero) mode structure (see Fig. \ref{fig:hyd1-configuration}), leading to resonance-localized MCI modes as field strength increases. As Alfvénic resonances must vanish in the absence of magnetic fields, we argue that MHD onset can be viewed as a bifurcation process of the resonances about the Corotation point as magnetic Reynolds number (Rm) increases. With this perspective, we provide a criterion to estimate the magnetic Reynolds number (Rm) threshold for MHD onset in the presence of HD instability (Eq. \ref{eqn:MHD-HD-Onset}) and demonstrate the bifurcation process (see Fig. \ref{fig:Res-Convergence}), finding reasonable agreement with the theoretical estimate.

Disk geometry was also found to be critical (Sec. \ref{sec:AR}). MRI modes dominate in thin disks ($\Delta z$/$\Delta r \ll 1$) with low global curvature ($r_2/r_1 \gg 1$) and small aspect ratio (large radial gaps). In contrast, MCI modes dominate in intermediate thickness disks ($\Delta z/ \Delta r \sim 1$ with high global curvature $(r_2/r_1\rightarrow 1)$ and intermediate aspect ratio (moderate radial gap) (Fig. \ref{fig:Disk-Scaling}). This trend was qualitatively confirmed by NIMROD simulations in different aspect ratio geometries with azimuthal fields (Fig. \ref{fig:AR-Comp-NIMROD}). These findings suggest that MCI likely dominates typical laboratory experiments operating in highly curved, moderately thick domains.

While the resistive approximations have quantitative limitations, this work demonstrates they capture essential scaling behaviors and mode properties, providing a computationally efficient tool for exploring parameter space and guiding simulations or experiments. Methods for refining these approximations remain a path for future research. Given the successful application to resistive MHD, extending these spectral and potential methods to incorporate additional physics, such as Hall MHD effects, which are significant in specific plasma regimes, represents another promising direction. Similarly, applying these efficient tools to analyze stability in more complex, non-uniform flow profiles, such as those exhibiting vorticity minima \cite{wang_observation_2024}, is warranted. This linear stability analysis provides a foundation for future studies focusing on the non-linear evolution, turbulence, and associated transport driven by these instabilities.

\begin{acknowledgments}
The authors thank Matthew Pharr, Yin Wang, Hantao Ji, Jeremy Goodman, and Erik Gilson for insightful discussions. Simulations were performed on the cluster at the Princeton Plasma Physics Laboratory. This work was supported by the National Science Foundation under Award No. NSF 2308839 and DOE under Grant No. DE-AC02-09CH11466.
\end{acknowledgments}

\bibliographystyle{apsrev4-2}
\bibliography{apssamp}

\appendix
\section{Convergence On Ideal MHD Limit} \label{sec:conv}
This section shows that Equation \ref{eqn:shooting} converges on Equation 16 of \cite{ebrahimi_nonlocal_2022} in the ideal limit. First, recall the 1D second-order equation boundary valued equation for $\xi_r(r)$, where $\xi_r(r) = -r\tilde{v}_r/(i\bar{\omega}_\eta)$:
    \begin{equation}
        \frac{d}{dr}\left(f\frac{d\xi_r}{dr}\right) + s\frac{d\xi_r}{dr} - g\xi_r = 0
    \end{equation}
   where the coefficients $f$, $s$, and $g$ are defined below. Taking the ideal limit, we note that much simplification can be done simply by taking $\bar{\omega}_\eta \rightarrow \bar{\omega}_\nu$. 

\begin{equation}
\begin{split}
    f =\frac{r\left(\omega_A^2 - \bar{\omega}_\nu\bar{\omega}_\eta\right)}{k_z^2r^2 + m^2}, \; \; \; s = \underbrace{\frac{m(\bar{\omega}_\nu - \bar{\omega}_\eta)}{k_z^2r^2 + m^2}r\Omega'}_{\text{= 0, since $\bar{\omega}_\eta \rightarrow \bar{\omega}_\nu$}}, \\
    g = \frac{\mathrm{d}}{\mathrm{dr}}\Biggl\{\frac{m}{\left(k_z^2r^2 + m^2\right)}\Biggl[\underbrace{\left(\bar{\omega}_\eta-\bar{\omega}_\nu\right)r\Omega'}_{\text{= 0, since $\bar{\omega}_\eta \rightarrow \bar{\omega}_\nu$}} + \left(2\Omega\bar{\omega}_\eta+\omega_A\omega_c\right)\Biggr]\Biggr\} \\
    + \frac{\left(\omega_A^2 + \omega_c^2- \bar{\omega}_\nu\bar{\omega}_\eta\right)}{r} + \underbrace{\frac{\omega_s^2}{r}\Biggl(\frac{\omega_A^2 - \frac{k_z^2r^2\bar{\omega}_\eta^2 + m^2\bar{\omega}_\eta\bar{\omega}_\nu}{k_z^2r^2+m^2}}{\omega_A^2 - \bar{\omega}_\nu\bar{\omega}_\eta}\Biggr)}_{\text{=$\frac{\omega_s^2}{r}$, since $\bar{\omega}_\eta \rightarrow \bar{\omega}_\nu$}} \\ 
    - \frac{k_z^2r}{k_z^2r^2 + m^2 }\frac{\left(2\Omega\bar{\omega}_\eta + \omega_A\omega_c\right)^2}{\omega_A^2 - \bar{\omega}_\nu\bar{\omega}_\eta}
     + \underbrace{\frac{k_z^2r^2\left(\bar{\omega}_\eta-\bar{\omega}_\nu\right)}{k_z^2r^2+m^2}\frac{\omega_A\omega_c\Omega'}{\omega_A^2 - \bar{\omega}_\nu\bar{\omega}_\eta}}_{\text{= 0, since $\bar{\omega}_\eta \rightarrow \bar{\omega}_\nu$}}
\end{split}
\end{equation}

If we then take the final limit $\bar{\omega}_\eta \rightarrow \bar{\omega}_\nu \rightarrow \bar{\omega}$, we arrive at equation 16 except in $r$ instead of $x$ coordinates, and they converge in change of coordinates.

\begin{equation}
\begin{split}
    f =\frac{r\left(\omega_A^2 - \bar{\omega}^2\right)}{k_z^2r^2 + m^2}, \; \; \; s = 0, \\
    g = \frac{m}{2}\frac{\mathrm{d}}{\mathrm{dr}}\Biggl\{\frac{\Omega\bar{\omega}+\frac{1}{2}\omega_A\omega_c}{\left(k_z^2r^2 + m^2\right)}\Biggr\} - \frac{\left(\bar{\omega}^2 - \omega_A^2 - \omega_c^2 - \omega_s^2 \right)}{r} \\ + \frac{1}{4}\frac{k_z^2r}{k_z^2r^2 + m^2 }\frac{\left(\Omega\bar{\omega} + \frac{1}{2}\omega_A\omega_c\right)^2}{\bar{\omega}^2 - \omega_A^2}
\end{split}
\end{equation}

\end{document}